%% file: main.tex
\documentclass[11pt]{article}
\usepackage{jheppub}
\usepackage{graphicx}
\usepackage[space]{grffile}
\usepackage{latexsym}
\usepackage{amsfonts,amsmath,amssymb}
\usepackage{url}
\usepackage[utf8]{inputenc}
\usepackage{hyperref}
\hypersetup{colorlinks=false,pdfborder={0 0 0}}
\usepackage{textcomp}
\usepackage{longtable}
\usepackage{multirow,booktabs}
\usepackage{amsfonts}
\usepackage{datetime}

\usepackage{tikz}
\usetikzlibrary{%
    decorations.pathreplacing,%
    decorations.pathmorphing%
}
\usepackage{mathtools}
\usepackage{enumitem}
\usepackage{breqn}
\usetikzlibrary{calc}
\newcommand{\kt}{\langle k_T\rangle}

\newcommand{\lsim}{\mathrel{\hbox{\rlap{\lower.75ex \hbox{$\sim$}} \kern-.3em \raise.4ex \hbox{$<$}}}}
\newcommand{\gsim}{\mathrel{\hbox{\rlap{\lower.75ex \hbox{$\sim$}} \kern-.3em \raise.4ex \hbox{$>$}}}}

\begin{document}

\hfill{\tt CERN-TH-2020-007, BNL-213623-2020-FORE}

\title{Far-forward neutrinos  at the Large Hadron Collider}

\author[a]{Weidong Bai,}
\author[b]{Milind Diwan,}
\author[c]{Maria Vittoria Garzelli,}
\author[d]{Yu Seon Jeong,}
\author[a]{Mary Hall Reno}

\affiliation[a]{Department of Physics and Astronomy, University of Iowa, Iowa City, IA 52242, USA}
\affiliation[b]{Brookhaven National Laboratory, USA}
\affiliation[c]{Universit\`a degli Studi di Firenze, Dipartimento di Fisica e Astronomia \& INFN, Firenze, Italy}
\affiliation[d]{Theoretical Physics Department, CERN, 1211 Geneva 23, Switzerland}

\emailAdd{weidong-bai@uiowa.edu}
\emailAdd{diwan@bnl.gov}
\emailAdd{garzelli@mi.infn.it}
\emailAdd{yuseon.jeong@cern.ch}
\emailAdd{mary-hall-reno@uiowa.edu}

\date{\today}
\abstract{
We present a new calculation of the energy distribution of high-energy neutrinos from the decay of charm and bottom hadrons produced at the Large Hadron Collider (LHC). In the kinematical region of very forward rapidities, heavy-flavor production and decay is a source of tau neutrinos that leads to thousands of { charged-current} tau neutrino events in a 1 m long, 1 m radius lead neutrino detector at a distance of 480 m from the interaction region. 
In our computation, next-to-leading order QCD radiative corrections are accounted for in the production cross-sections. Non-perturbative intrinsic-$k_T$ effects 
are approximated by a simple phenomenological model introducing a Gaussian $k_T$-smearing of the parton distribution functions, which might also mimic perturbative effects due to multiple initial-state soft-gluon emissions.  
The transition from partonic to hadronic states is described by phenomenological fragmentation functions. 
To study the effect of various input parameters, theoretical predictions for $D_s^\pm$ production are compared with LHCb data on 
double-differential cross-sections in transverse momentum and rapidity. 
The uncertainties related to the choice of the input parameter values, ultimately affecting the predictions of the tau neutrino event distributions, are discussed.
We consider a 3+1 neutrino mixing scenario  
to illustrate the potential for a neutrino experiment to constrain
the 3+1 parameter space using tau neutrinos and antineutrinos.
We find large theoretical uncertainties in the predictions of the neutrino fluxes in the far-forward region. Untangling the effects of tau neutrino oscillations into sterile neutrinos and distinguishing a 3+1 scenario from the standard scenario with three active neutrino flavours, will be challenging due to the large theoretical uncertainties from  QCD.}
\dedicated{\today}
\maketitle

\flushbottom

\input{intro}

\input{geo}

\input{production-14TeV}

\input{results-14TeV-comb} 
\input{newphysics} 
\input{conclusions}

\input{appendix}

\acknowledgments
This work is supported in part by Department of Energy grants DE-SC-0010113, DE-SC-0012704, and the Korean Research Foundation (KRF) through the CERN-Korea Fellowship program. The authors would like to express a special thanks to the Mainz Institute for Theoretical Physics (MITP) of the Cluster of Excellence PRISMA+ (Project ID 39083149) for its hospitality and support. We thank the Rencontres du Vietnam/VietNus 2017 workshop at 
the International Center for Interdisciplinary Science and Education (ICISE), QuiNhon, Vietnam 
for support and discussions in the early stage of this work, and we thank C. Giunti and M.L. Mangano for comments and discussions. M.V.G. is grateful to the II Institute for Theoretical Physics of the University of Hamburg for hospitality during the completion of this work.

\bibliography{LHC}

\end{document}

%% file: intro.tex
\section{Introduction}  

Since the discovery of oscillation 
properties of neutrinos, their fundamental roles in shaping the universe have become an important line of inquiry \cite{pdg:2016}.
 A high-luminosity neutrino program with a flux of neutrinos produced {at an energy scale of few GeV} is the focus of the large experimental neutrino physics community working on the Deep Underground Neutrino Experiment (DUNE) \cite{Acciarri:2015uup,Acciarri:2016crz,Abi:2018dnh}. With a proton beam energy of 80-120 GeV, pions are produced with the highest multiplicity in proton-nucleon interactions. Thus the beams at DUNE are predominantly muon neutrino (and muon antineutrino) beams, generated by charged pion decays. 
Future measurements of muon neutrino disappearance and electron neutrino appearance, and of the differences between neutrino and antineutrino rates, will allow the extraction of elements of the Pontecorvo-Maki-Nakagawa-Sakata (PMNS) mixing matrix with better precision than they are currently known. The mixing of tau neutrinos with muon and electron neutrinos will be determined indirectly through disappearance and appearance probabilities since at these energies, tau neutrinos are not directly produced in the beam and the tau mass threshold severely suppresses the number of tau neutrino charged-current events at the far detector. 

Recently, attention has turned to opportunities of measuring the interactions 
of highly energetic tau neutrinos produced by $pp$ collisions at the LHC \cite{Park:2011gh,Feng:2017uoz,Ariga:2018pin,Ariga:2019ufm,Buontempo:2018gta,Beni:2019gxv,Beni:2019pyp}. Tau neutrino beams would allow for
direct tests of lepton universality and to explore tau neutrino PMNS mixing in the traditional three-neutrino mixing paradigm and in scenarios including additional exotic neutrinos.  
It was already recognized 
several decades ago that through the production and prompt decays of the $D_s^\pm$ and
$B$ mesons to taus and tau neutrinos, hadron colliders produce large fluxes of tau neutrinos in the forward direction
\cite{DeRujula:1984pg,Winter:1990ry,DeRujula:1992sn,Vannucci:1993ud}. Electron and muon neutrinos and antineutrinos
from $D$ and $B$ meson decays will also be produced.
Charm and bottom quark production in  
the Standard Model mostly occurs in quark-antiquark pairs, leading to an approximately equal number of hadrons and antihadrons\footnote{Differences in the forward region between the total number of hadrons and antihadrons from quark-antiquark pair production arise from the recombination of  forward final-state heavy quarks with partons from the initial-state protons non participating to the hard scattering, considering the fact that the parton distribution functions of the up and down quarks in the nucleon differ from those of the antiquarks. Baryon-antibaryon asymmetries, in principle, would show a larger effect than for meson-antimeson asymmetries. We show below that the $\Lambda_c$ gives a small contribution to neutrino production. The $D_s^\pm$ has no valence component, and experimental studies by LHCb show a $D_s^+-D_s^-$ asymmetry below 1\% \cite{Aaij:2018afd}}, so there will be approximately equal fluxes of neutrinos and antineutrinos from heavy flavor. In our discussion below, we will refer to both particles and antiparticles as neutrinos.
As described in 1984 by De Rujula and Ruckl \cite{DeRujula:1984pg}, using Feynman scaling arguments, a quark-gluon string model and empirical extrapolations based on collider data available at the time, a few thousand tau neutrino events in future $pp$ and $p\bar{p}$ colliders could be detected with a 2.4 ton detector placed 100 m distant from the interaction point along the tangent to the accelerator arc. An estimation of the neutrino flux using \textsc{Pythia} \cite{Sjostrand:2019zhc} tuned to Tevatron data and rescaled to a $\sqrt{s}=14$ TeV center-of-mass energy gives qualitatively consistent event rates \cite{Park:2011gh}. 

In the last few years, there is renewed interest in far-forward neutrino production and detection \cite{Feng:2017uoz,Ariga:2018pin,Ariga:2019ufm,Beni:2019gxv,Beni:2019pyp,Buontempo:2018gta,Abreu:2019yak,Abreu:2020ddv}. 
The ForwArd Search ExpeRiment at the LHC (\textsc{Faser}), primarily dedicated to searches for light, extremely weakly interacting particles \cite{Ariga:2019ufm,Ariga:2018pin,Feng:2017uoz},  with phase 1 approved and under construction, will be sensitive to tau neutrinos if the detector mass is sufficiently large. The location of the detector is projected to be 480 m from the ATLAS interaction point along the colliding beam axis. The \textsc{Faser} collaboration uses a pseudorapidity cut $\eta>6.87$ on particle momenta (here, neutrinos) in its second phase to determine if they enter
a detector of radius 1.0 m. Other evaluations for this baseline use $\eta>6.7$ \cite{Buontempo:2018gta}.
For a half-cylinder, 2-meter long lead detector covering 
pseudorapidities $\eta>6.7$, Beni et al. \cite{Beni:2019gxv} used \textsc{Pythia} 8 \cite{Sjostrand:2019zhc,Sjostrand:2014zea} to find $\sim 8,700$ tau neutrino events for a
$3,000$ fb$^{-1}$ integrated luminosity at the LHC.   Other configurations for detectors, for example, in $\eta$ ranges of $8-9.5$ and $7.4-8.2$ for XSEN \cite{Beni:2019pyp} and in the range $7.2-8.7$ for SND$@$LHC
\cite{Ahdida:2020evc} are also under consideration. The prototype \textsc{Faser}-$\nu$ has $\eta>8.6$
\cite{Abreu:2019yak,Abreu:2020ddv}.
We adopt a minimum pseudorapidity $\eta>6.87$ in this work for definiteness. Our results for $\eta>6.7$ are qualitatively similar. 

A source of high-energy tau neutrinos opens the possibility of new tests of the Standard Model that can not be done at DUNE. Measurements of LHC forward tau neutrino interactions can be used for direct tests of lepton flavor universality in charged current  interactions with much higher statistics than achieved by \textsc{Donut} \cite{Kodama:2000mp,Kodama:2007aa} and \textsc{Opera} \cite{Pupilli:2016zrb,Agafonova:2018auq}.
Both Super-Kamiokande and IceCube have reported signs of tau neutrino appearance in their datasets, but the statistics is still very limited \cite{Li:2017dbe, Aartsen:2019tjl}. Measurements of the interaction cross-sections of muon neutrinos from heavy-flavor decays at the LHC 
with the nucleons/nuclei of the target will help to close the gap between direct neutrino 
cross-section measurements for
$E_\nu<370$ GeV \cite{pdg:2016} and the IceCube Collaboration's determination { of the averaged cross-section for neutrino plus antineutrino deep-inelastic scattering with nucleons} for $E_\nu=6.3-980$ TeV \cite{Aartsen:2017kpd} (see also, e.g., ref. \cite{Bustamante:2017xuy}). LHC forward neutrinos will provide the first opportunity for direct neutrino and antineutrino cross-section measurements for neutrino energies up to $E_\nu\lesssim 2$ TeV.

In this paper, we perform a new evaluation of the $D$ and $B$ meson contributions 
to the $pp$ differential cross section as a function
of tau neutrino and muon neutrino energy for
$\eta> 6.87$. Differently from previous evaluations  
which are limited to leading order/leading logarithmic accuracy, our evaluation accounts for the effects of next-to-leading order (NLO) QCD radiative corrections to the heavy-quark hadroproduction  \cite{Nason:1987xz,Nason:1989zy,Mangano:1991jk} and neutrino deep-inelastic-scattering cross-sections. 
The effects due to the intrinsic transverse momentum of initial state partons confined in the protons are accounted for by a simple model adding $k_T$-smearing effects to the standard  collinear parton distribution functions (PDF) as an input for the calculation. The same model might also mimic the effects of the resummation of logarithms related to initial-state soft gluon emissions in an approximate and purely phenomenological way.
Furthermore, we include a description of the fragmentation of partons into heavy mesons, relying on widely used phenomenological fragmentation functions.  

Our main focus is on tau neutrinos which come predominantly from $D_s^\pm$ decays.
To study the effects of different choices of the values of various parameters entering our computation, we compare our theoretical
predictions with the LHCb data on $D_s^\pm$ production
\cite{Aaij:2015bpa} in the rapidity range
of $2 < y < 4.5$. These same parameters give predictions in similar agreement with experimental data for
other charm hadron distributions at LHCb. 

As discussed below,
the effect of the transverse momenta of initial state partons  can significantly
impact the predictions for forward tau neutrino event rates. 
In general, the LHCb and other charm data give hints of the need for higher-order effects in the description of single-inclusive open D-hadron production, beyond NLO  and the limited logarithmic accuracy of the parton shower implementations and of the analytical resummations
of various kinds of logarithms presently available. Power suppressed non-perturbative terms might also play a relevant role, considering the smallness of the charm quark mass, still larger but not too large with respect to $\Lambda_{QCD}$. At present, it is not clear if the discrepancies between theoretical predictions and experimental data at low transverse momenta can be completely cured within the collinear factorization framework, or if it is necessary to go beyond this scenario. 
Considering the unavailability of rigorous perturbative and non-perturbative QCD theoretical calculations accurate enough to reproduce the shape of the experimental transverse-momentum distributions, we incorporate initial-state partonic $k_T$-smearing effects in a purely phenomenological way in the QCD calculation of $D_s^\pm$ production for the phase space covered by LHCb and assess their impact on the predicted number of neutrino charged-current interaction events.

The elements of the PMNS mixing matrix are least constrained in the tau sector, compared to the other flavor sectors. We demonstrate how sterile neutrino mass and mixing parameters can begin to be constrained by measuring oscillations of LHC tau neutrinos, and we discuss the challenges to pushing these constraints to sterile neutrino masses of order~$\sim 20$~eV. 

Heavy-flavor decays are not the only sources of forward neutrinos, but they dominate the forward tau-neutrino flux. The contributions from $pp\to W,Z$ production followed by $W,Z$ leptonic decays are negligible for $\eta \gsim 6.5$ \cite{Beni:2019gxv}. 
On the other hand, for muon and electron neutrinos, charged pion ($\pi^\pm$) and  kaon  ($K^\pm$) decays to $\nu_\mu$ and $K_L\to \nu_e$ decays ($K^0_{e3})$ are most important. 
For our evaluation of muon and electron neutrino oscillations, we use parametrizations of light meson distributions based on \textsc{Pythia} distributions \cite{Koers:2006dd}.

We begin in section \ref{sec:general} with an overview of the far-forward geometry used for our discussion here. In section \ref{sec:production}, we present our $D$ and $B$ hadron production results. 
Energy distributions of charged-current interaction events generated in a forward detector by neutrinos from heavy-flavor production and decay are shown in section \ref{sec:results} for both tau neutrinos and muon neutrinos. In our estimation, 
we also account both for muon neutrinos from the decays of charged pions and kaons and for electron neutrinos from kaon decays. 
Section \ref{sec:newphysics} shows an application to the study of tau neutrino oscillations, considering a 3+1 oscillation framework with three active neutrinos and one sterile neutrino. 
We conclude in section \ref{sec:conclusions}.
Appendix A collects formulas for the decay distributions to neutrinos.

%% file: geo.tex
\section{Overview of forward neutrino detection geometry}
\label{sec:general}

A forward detector along a line tangent to the LHC beam line necessitates calculations in 
high-pseudorapidity regimes. A  detector with radius of 1.2 m placed at 480 m from the LHC interaction point is used for the evaluations in ref. \cite{Beni:2019gxv}. This corresponds to a neutrino pseudorapidity of $\eta>6.7$ for detection. The \textsc{Faser}2 proposal \cite{Ariga:2019ufm} has $R=1.0$ m, corresponding to $\eta> 6.87$, which we use for the results shown below. Other smaller detectors like XSEN \cite{Beni:2019pyp} and SND$@$LHC \cite{Ahdida:2020evc} are discussed in the recent literature. A prototype \textsc{Faser}-$\nu$ 
at even higher $\eta$,\footnote{The \textsc{Faser}-$\nu$ experiment has recently been approved. The \textsc{Faser}-$\nu$ detector will be installed in front of \textsc{Faser}. It is expected to be fully operational and collect data during the LHC Run-III.} with 
a 25 cm $\times$ 25 cm cross sectional area
and length of 1.35 m of tungsten interleaved with emulsion detectors,
will have a few tens of tau neutrino events with an integrated luminosity of 150 fb$^{-1}$ delivered during Run 3 of the LHC \cite{Abreu:2019yak}.  
A detector radius of $1.0$ meters, for a comparable target mass, increases the number of events by a factor of $\sim 50$ when scaling by cross-sectional area alone. 

The LHC interaction region is a very  compact source of tau neutrinos. Most of the tau neutrinos come from $D_s\to \nu_\tau \tau$ decay. The $\tau\to \nu_\tau X$ decay is also a prompt process. The characteristic size of the region where tau neutrinos are produced by $D_s$ decays is of order 
$\gamma c \tau_{D_s} 
\simeq E_{D_s}/m_{D_s}\cdot 150\ \mu$m, so of order of $1.5-15$ cm for $E_{D_s}=200$ GeV$-2$ TeV. 
The tau decay length $c\tau=87.11\ \mu$m, multiplied by the $\gamma$-factor for the same energy range, gives a size of $0.98-9.8$ cm.
Similarly, $\gamma c \tau_{B^+} = E_{B^+}/m_{B^+}\cdot 496\ \mu$m produces a size of $1.9-19$ cm for the same energy range.
Thus, for tau neutrinos produced along the beam pipe, the longitudinal distance is a few to 20 cm. The transverse size is $17\ \mu$m~\cite{Bruning:2004ej} from the 
proton bunch size. The compact source means the assumed detector radius of 1 m and distance of 480 m
from the interaction point translates to a maximum
angle relative to the beam axis for the tau neutrino
three-momentum of $\theta_{\rm max}=2.1$ mrad ($\eta_{\rm min}=6.87$). This same constraint applies to the 
momenta of muon and electron neutrinos from heavy-flavor decays.

While the focus of this paper is on heavy-flavor production of neutrinos in the forward region, consideration of oscillations in a 3+1 mixing framework also requires an estimate of the number of electron and muon neutrinos from both 
heavy-flavor decays and light-meson decays. In sec. 4.2, we make an estimation of the production of electron neutrinos and muon neutrinos from light-meson decays. The light-meson decay lengths are long compared to heavy-meson decay lengths, so the detector and magnets near the interaction point play a role.
In ref. \cite{Abreu:2019yak}, an evaluation of the number of
$\nu_\mu+\bar{\nu}_\mu$ events in a detector of
25 cm $\times$ 25 cm cross sectional area finds that most of the events below 1 TeV come from charged pion and kaon decays that occur within 55 m of the interaction point and stay within the opening of the front quadrupole absorber with inner radius of 17 mm. This corresponds to light-meson momenta lying within
1 mrad from the beam axis. A more detailed discussion of this point appears in section 4.2.

Heavy-meson production at small angles with respect to the beam axis receives dominant contributions from the transverse momentum region below few GeV. 
For example, 
for $E_{D_s}=1$ TeV, approximating the neutrino
direction by the $D_s$ direction means that the $p_T$ of the $D_s$ meson must be smaller than 2.1 GeV, and even smaller for lower energies. 
Non-perturbative effects related to the intrinsic $k_T$ of the partons confined in the initial state nucleons are important at such low transverse momenta. Additionally, perturbative effects related to the appearance of large high-energy logarithms 
together with those due 
to initital-state multiple soft-gluon emissions, are potentially relevant in the $p_T$ range [0, 15] GeV of the LHCb data considered in this paper, covering low to intermediate $p_T$ values. 
 
In our evaluation of differential cross sections for open charm and bottom production, we include Gaussian 
$k_T$-smearing to better match LHCb data, using a purely phenomenological approach. In this approach the non-perturbative effects are reabsorbed in the $\langle k_T\rangle $ smearing model, which may also mimic part of the all-order perturbative effects in a rough way.  We found that a better description of the LHCb data, given the renormalization and factorization scale choices discussed below, is provided in our approach by a $\langle k_T \rangle$ value higher than the naive estimate due to Fermi motion and even larger than the upper end of the range of typical non-perturbative $\kt$ values of $\sim 1-2$ GeV reported in the literature \cite{Apanasevich:1998ki,Mangano:1998oia,Miu:1998ju,Balazs:2000sz,Skands:2010ak}. 
This gives hints, on the one hand, that non-perturbative physics aspects not yet well understood might be particularly relevant for the process we are studying. On the other hand, it hints that the contribution from the Sudakov resummation of double logarithmic perturbative terms related to the emission of an arbitrarily large number of soft gluons, missing in fixed-order calculations, may be large. 
Since our focus is on $\nu_\tau+\bar{\nu}_\tau$ production, we use the LHCb data for $D_s$ production at $\sqrt{s}=13$ TeV for rapidities in the range  of $2.0-4.5$ \cite{Aaij:2015bpa}.

%% file: production-14TeV.tex
\section{Forward heavy-flavor production and decay at the LHC}
\label{sec:production}

We evaluate the single-particle 
inclusive heavy-flavor energy and angular distributions. Our approach relies on perturbation theory (pQCD) in the collinear factorization framework. In particular, we include NLO QCD corrections to the heavy-quark production cross sections \cite{Nason:1987xz,Nason:1989zy,Mangano:1991jk}. 

As noted above, at very high rapidities, 
the effect of relatively small transverse momenta of the initial state partons
can affect the acceptance of neutrino
events. In shower Monte Carlo event generators like \textsc{Pythia}, 
the transverse momentum distribution of the produced heavy quarks 
is affected, on the one hand, by 
the effects of multiple soft and collinear gluon emissions, accounted for with a limited logarithmic accuracy by the introduction of Sudakov form factors resumming the relative main logarithmic contributions, and, on the other hand, by the inclusion of a small intrinsic transverse momentum, related to the confinement of partons in finite-size nucleons and the uncertainty principle.
An alternative to the collinear factorization
approach is to use unintegrated parton distribution functions that
have a transverse momentum $k_T$ dependence in addition to the usual 
longitudinal momentum fraction $x$ dependence \cite{Catani:1990eg,Collins:1991ty}.  
In both collinear and $k_T$ factorization, further higher-order effects can also modify the transverse momentum distribution of heavy-flavor hadroproduction in the low $p_T$ region. In particular, the effect of the resummation of high-energy logarithms, not yet implemented into a publicly available code, and the joint resummation of these and other logarithms could also play a relevant role, which deserves future dedicated investigations, but is beyond the scope of this paper.  

In the present absence of calculations capable of fully reproducing the experimental shape of the transverse momentum distributions of charm mesons at small transverse momenta at the LHC, our approach here is phenomenological and uses Gaussian smearing of the outgoing charm quark. With the NLO QCD calculation of
$g_{\rm NLO}(q_T,y)=d^2\sigma({\rm NLO})/dq_T\, dy$ for the quark, we evaluate the Gaussian smeared
one-particle inclusive charm-quark distribution $g(p_T,y)=d^2\sigma/dp_T\, dy$ according to 
\begin{equation}
g(p_T,y) = \int d^2 \vec{k}_T\, f(\vec{k}_T,\langle k_T^2\rangle) g_{\rm NLO}(q_T=|\vec{p}_T-\vec{k}_T|,y)\ ,
\end{equation}
where $f(\vec{k}_T,\langle k_T^2\rangle)$ is a two-dimensional Gaussian function normalized to unity, 
\begin{equation}
  \label{eq:gaussian}
\int \, d^2 \vec{k}_T\, f(\vec{k}_T,\langle k_T^2\rangle) = \int \, d^2 \vec{k}_T\,  \frac{1}{\pi\langle k_T^2\rangle}\exp[{-k_T^2/\langle k_T^2\rangle}]=1\ .
\end{equation}
The smearing 
generates a shift of the outgoing heavy-quark momentum vector $\vec{q}_T$ by $\vec{k}_T$
after the hard
scattering, before fragmentation. 
The single Gaussian $f(\vec{k}_T,\langle k_T^2\rangle)$ is equivalent to starting
with two Gaussian functions, one for each incoming parton ($k_{iT}$), making a change of variables to $\vec{k}_T= (\vec{k}_{1T}+\vec{k}_{2T})/2$ for one of the integrals, and integrating over the other
parton's $k_T$,
\begin{equation}
    d^2 \vec{k}_{1T}\, f(\vec{k}_{1T},2\langle k_T^2\rangle) \,
    d^2 \vec{k}_{2T}\, f(\vec{k}_{2T},2\langle k_T^2\rangle) 
    \to d^2 \vec{k}_{T}\, f(\vec{k}_{T},\langle k_T^2\rangle)\ .
\end{equation}
This definition of the effective
$\vec{k}_T$ variable distributes the transverse momentum smearing equally to the one-particle transverse momentum and to the recoil hadrons \cite{Apanasevich:1998ki}.
The quantity
$\langle k_T^2\rangle$ is related to the average
magnitude of $\vec{k}_T$, $\langle k_T\rangle$, by
\begin{equation}
\langle k_T\rangle^2=\langle k_T^2\rangle\pi/4\ .
\end{equation}
As we discuss below, we set the $\langle k_T^2\rangle$ value based on comparisons with
double-differential distributions in rapidity and transverse momentum
for forward $D_s$ production measured by LHCb at $\sqrt{s}=13$ TeV~\cite{Aaij:2015bpa}.

Charm quark fragmentation to mesons is accounted for by using fragmentation functions of the Peterson form
\cite{Peterson:1982ak}.
For $c\to  D^{0}$, $D^{+}$ and $D_{s}^{+}$, we take fragmentation
fractions 0.6086, 0.2404 and 0.0802, respectively \cite{Lisovyi:2015uqa}.
The parameter $\epsilon$ in the Peterson fragmentation function that
describes the hardness of the meson spectrum relative to the heavy quark is taken to be
$\epsilon$ = 0.028, 0.039 and 0.008 for $D^0$, $D^+$ and $D_s^+$,
respectively, and $\epsilon = 0.003$  for $B$ production \cite{pdg:2019}. 
The fragmentation fraction for $B$'s is $b\to B^+=b\to B^0=0.362$ from ref. \cite{Aaij:2019pqz}.

We use the NLO nCTEQ15 parton distribution function (PDF) grids 
\cite{Kovarik:2015cma}, available for free proton and nuclear targets, 
for our evaluations here. We use their best fit set for free proton targets
in our evaluation of charm production in $pp$ collisions. As noted below, we use the NLO nCTEQ15 nuclear
PDF set for lead targets in our neutrino cross section calculation.
We take a charm pole mass value
$m_c=1.3$ GeV consistent with the choice of PDFs. A $b$-quark pole mass of 4.5 GeV is used here, also consistent with the PDFs.
We use renormalization and factorization scales ($\mu_R, \mu_F$) which are factors $(N_R,N_F)$ of the
transverse mass
$m_T = \sqrt{m_Q^2 + p_T^2}$, where $p_T$ is the magnitude of the
transverse momentum of the heavy quark $Q=c,\ b$.
Conventionally, renormalization and factorization scales are chosen in a range with factors $N_R,N_F= 0.5-2$ in $(\mu_R, \mu_F)= (N_R,N_F)m_T$, with $N_R=N_F=1$ set as conventional scale
factors \cite{Cacciari:2012ny}.
The values $N_R = 1.0$ and $N_F= 1.5$ lie in the standard scale factor uncertainty range and are used
as our default parameters as discussed below, but we also show predictions obtained with the standard conventional choice $N_R=N_F=1.0$ multiplying $m_T$ for $(\mu_R,\mu_F)$ used in most, (see, e.g., ref. \cite{Cacciari:2012ny}), though not all (see, e.g., ref. \cite{Benzke:2017yjn,Zenaiev:2019ktw}), of the literature on heavy-flavor production. In particular, in the following we will discuss predictions obtained with the following configurations:
\begin{itemize}[nosep]
    \item $N_R$ = 1, $N_F$ = 1 (conventional central scale choice), with
    $\langle k_T \rangle$ = 0.7 GeV,
    \item $N_R$ = 1, $N_F$ = 1.5, (alternative central scale choice, used as default in this paper, as better motivated in the following),
    with $\langle k_T \rangle$ = 0.7 GeV,  
    $\langle k_T \rangle$ = 0 GeV and
    $\langle k_T \rangle$~=~2.2~GeV.
\end{itemize}

The scale input, the PDFs, the fragmentation functions and the non-perturbative transverse momenta
all influence the predicted heavy-flavor energy and rapidity distributions. 
Since our focus is on tau neutrino production, LHCb data on forward $D_s$ production \cite{Aaij:2015bpa} are used to set $\langle k_T\rangle$ for selected
$(\mu_R,\mu_F)$ combinations,
after fixing the PDF and fragmentation function details.
The $D_s$ data cover the range $0< p_T<14$ GeV$\,$\footnote{The LHCb measurements reported in Ref.~\cite{Aaij:2015bpa} extend to a $p_T$ value of 15 GeV for $D^\pm$, $D^0$, and $\bar{D}^0$, while, for the less abundant $D_s^\pm$, the experimental results in the largest $p_T$ bin are not reported.}  
and $2.0<y<4.5$. While varying the input parameters,
a $\chi^2$ is computed, which combines the double-differential predictions, 
the binned data for 71 data points and the experimental uncertainties.  
As additional constraint, 
the integrated cross section evaluated	
with our Monte Carlo integration program is also required to be within the experimental error bars of the measured cross section for the same kinematic region.
Fitting the LHCb {data} for $D_s$ production with $p_T<14$ GeV, by minimizing the $\chi^2$ with $\mu_R=m_T$, $\mu_F=1.5\,m_T$ and 
varying $\langle k_T\rangle$, leads to  $\langle k_T\rangle=2.2\pm0.7$ GeV ($\langle k_T^2\rangle = 6.2$ GeV$^2$). This represents a reasonably good fit, with $\chi^2/$DOF$=2.8$ and the cross section within $10\%$ of the experimentally measured cross section in this kinematic region. 
With this parameter choice,
the theoretical total cross section for $\sigma_{c\bar{c}}$ is within 2\%  
of the central value of the LHCb estimate reported in ref. 
\cite{Aaij:2015bpa}. 

\begin{figure}[t]
  \begin{centering}
  \includegraphics[scale=0.48]{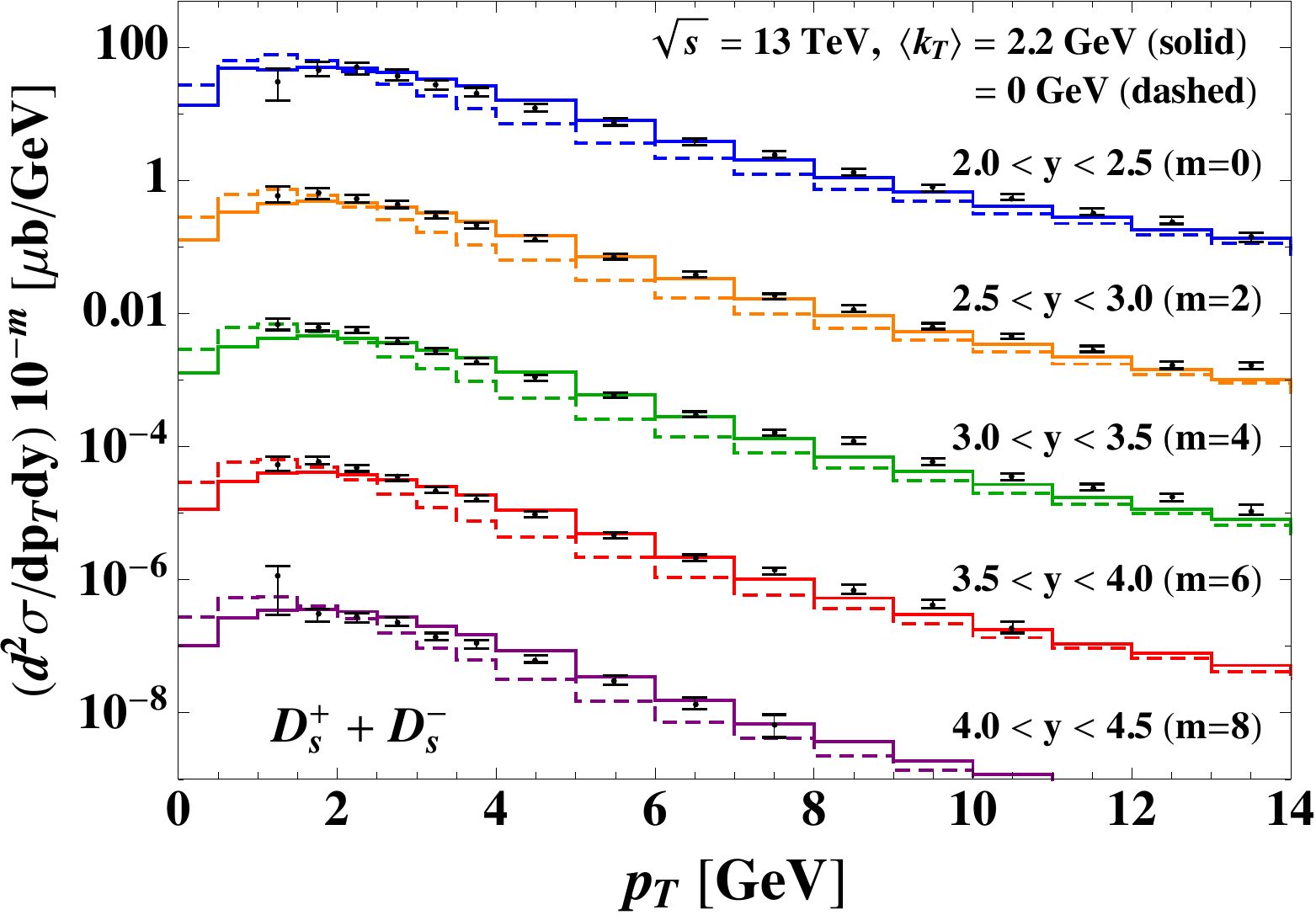} 
  \includegraphics[scale=0.48]{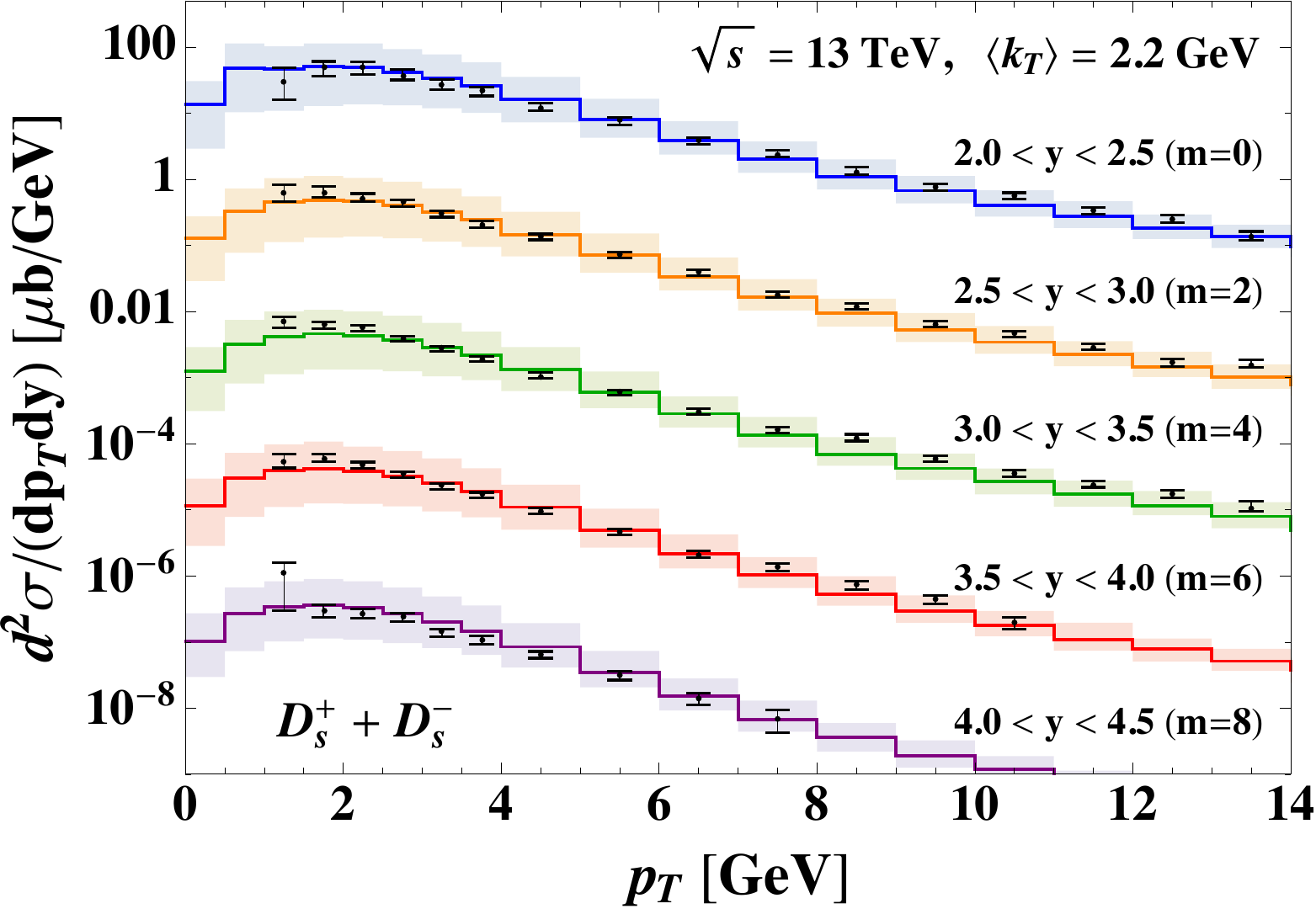} 
  \includegraphics[scale=0.48]{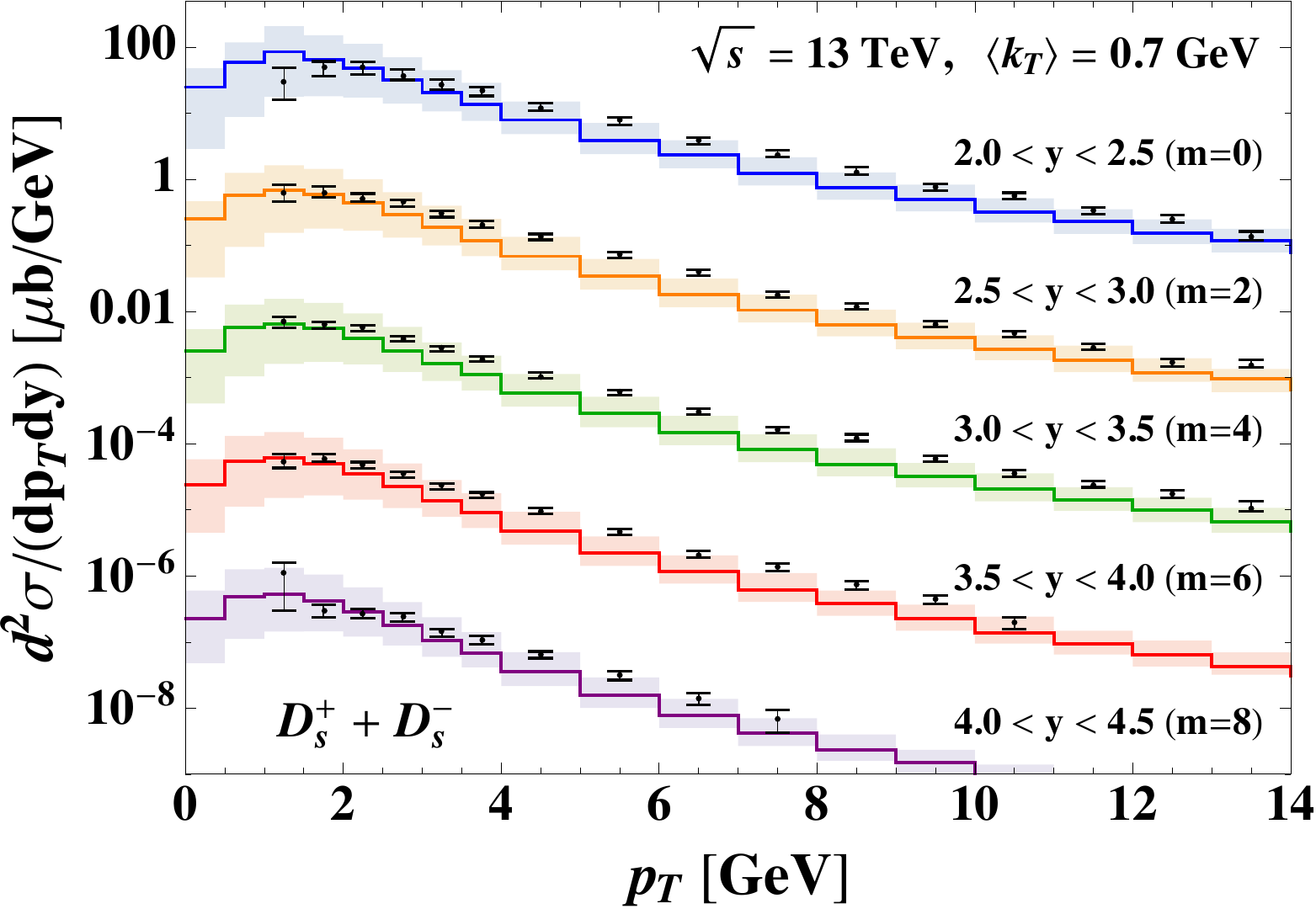}  
  \includegraphics[scale=0.48]{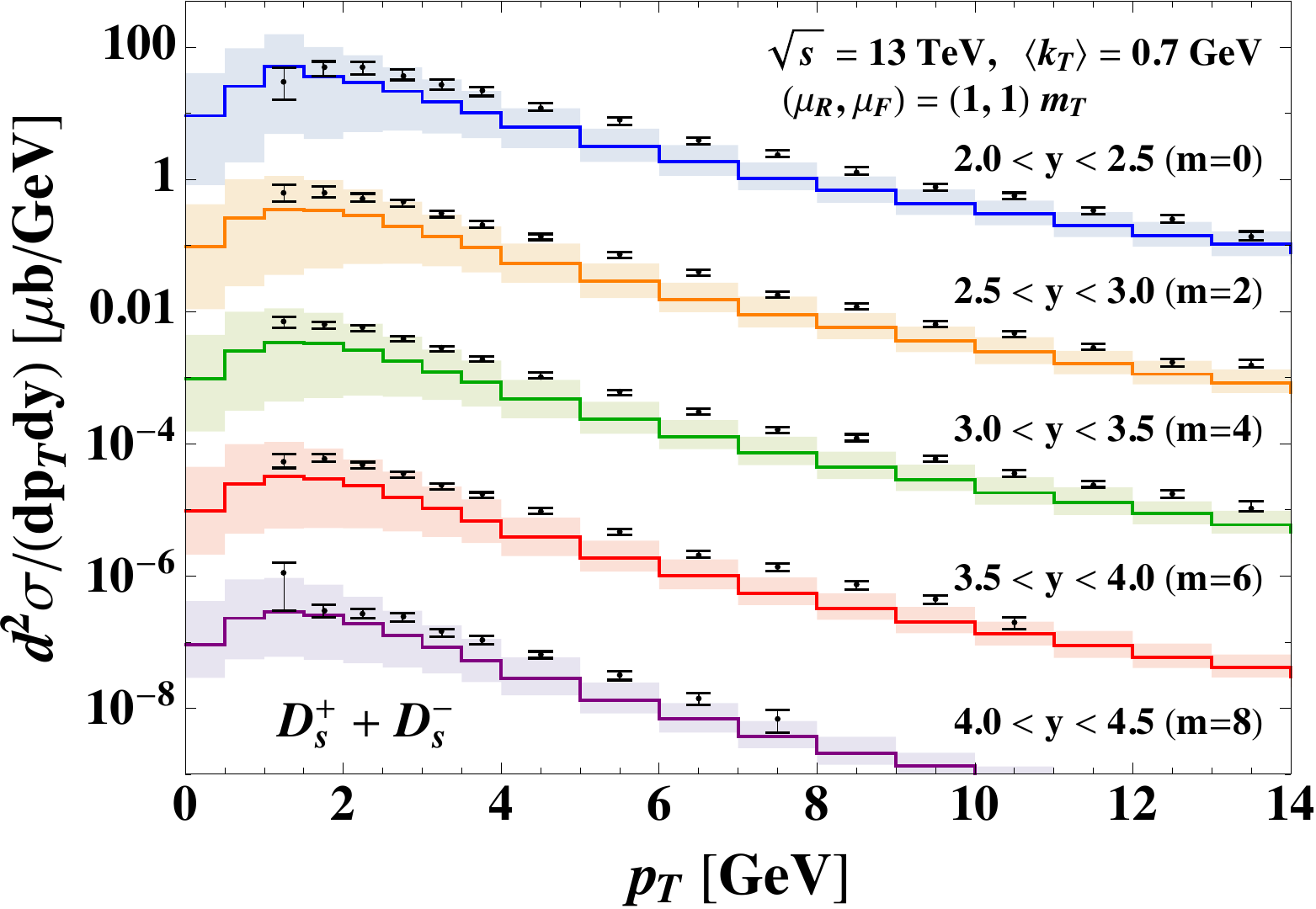}   

\par\end{centering}
\caption{Comparison between LHCb experimental data
on double-differential distributions in the meson $p_T$ and $y$ for $D_{s}^\pm$ production in $pp$ collisions \cite{Aaij:2015bpa} and our QCD predictions.
Data (and predictions) for different $\Delta y$ bins are shifted by $10^{-m}$ where values of $m=0$, 2, 4, 6 and 8.
The upper left panel refers to the case 
 $N_{R}=1.0 $ and $N_{F}=1.5$, 
where $(\mu_R, \mu_F) = (N_R, N_F) \, m_T$, both fixing
$\langle k_{T}\rangle = 2.2$ GeV and in the collinear approximation ($\langle k_{T}\rangle =$ 0 GeV).
The upper right and lower left panels refer to the same scale configuration with the shaded band showing the uncertainty
built by seven-point scale variation in the 
range  $N_{R,F}=0.5-2$ 
for $\langle k_{T}\rangle =2.2$ GeV (best fit to data) and $\langle k_{T} \rangle =0.7$ GeV (our default value), respectively.  
The lower right panel shows the same as lower left, but with $(N_R, N_F)$ = (1,1) (conventional scale choice).
\label{fig:fit-LHCb} }
\end{figure}

\begin{figure}[ht]
\begin{centering}
\includegraphics[width=0.48\textwidth]{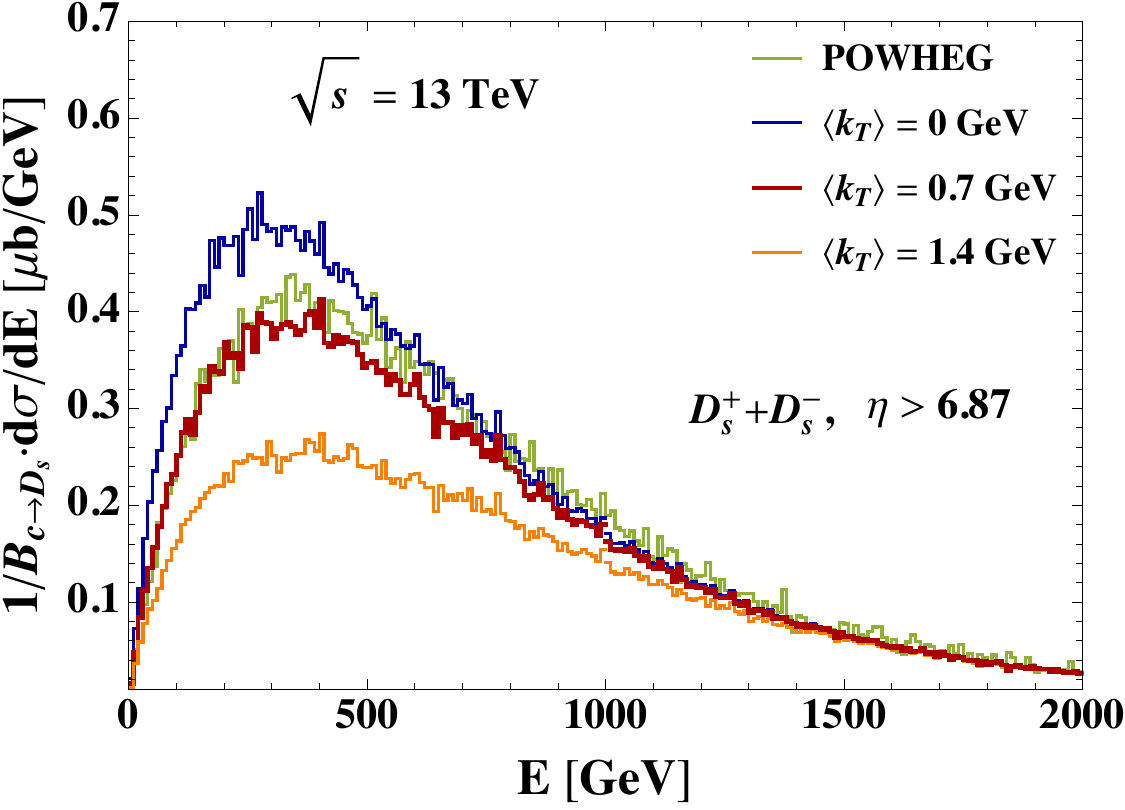} 
\includegraphics[width=0.48\textwidth]{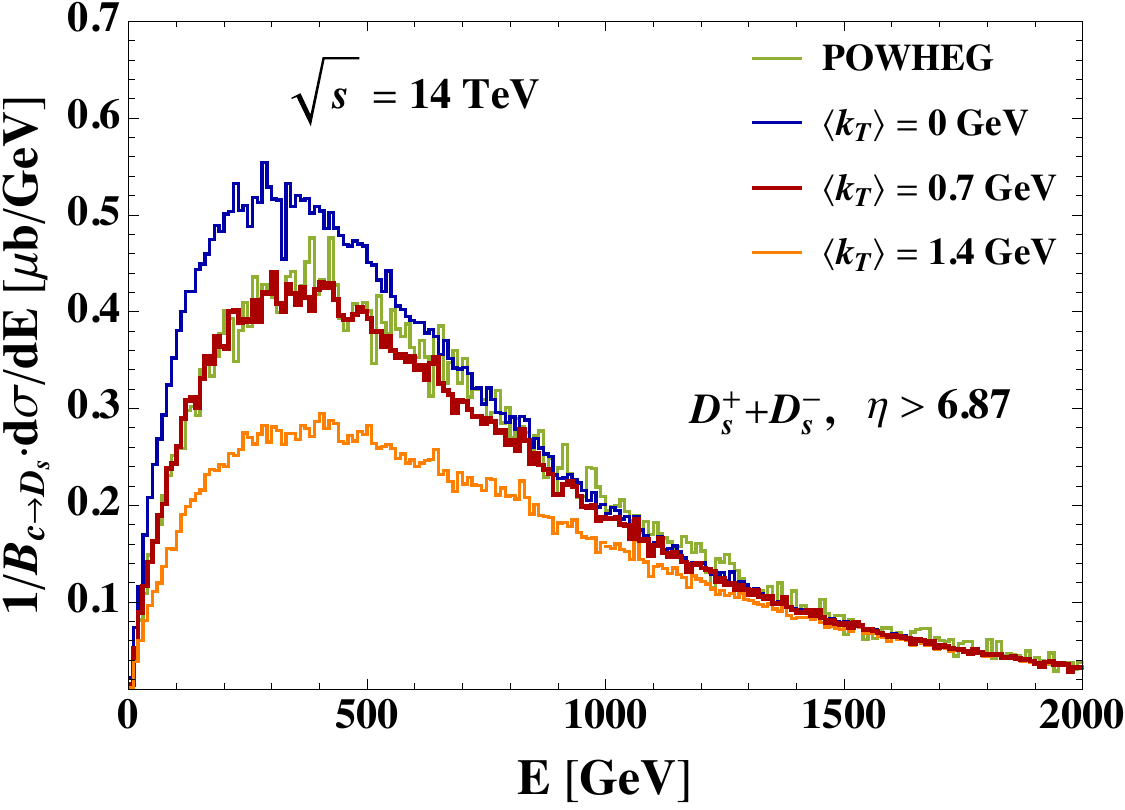} 
\end{centering}
\caption{Differential distributions of the energy $E$ of the $D_{s}$
produced in $pp$ collision at $\sqrt{s}=$ 13 (left) and 14 (right) TeV, for $D_s^\pm$ pseudorapidity $\eta>6.87$. Predictions obtained within our framework using $(\mu_R, \mu_F) = (1.0, 1.5) m_T$ and $\left\langle k_{T}\right\rangle = 0,\ 0.7, \ 1.4$ GeV, are compared to those of a computation based on NLO hard-scattering matrix-element matched to parton shower followed by hadronization, according to the \textsc{Powheg~+~Pythia} framework. See text for more detail. 
}
\label{fig:CDs-kT}
\end{figure}

The predictions with $\mu_R=m_T,\ \mu_F=1.5\,m_T$ and $\langle k_T\rangle=2.2$ GeV, together with those for a $\langle k_T\rangle=0$ evaluation using the sames scales and the LHCb data \cite{Aaij:2015bpa}, are shown in the upper left panel of
figure \ref{fig:fit-LHCb} with the solid and dashed histograms, respectively.
From top to bottom, the panel shows data and
theoretical evaluations in five different rapidity bins of $\Delta y=0.5$ width for $2.0 < y < 4.5$
and normalization  shifted by $10^{-m}$ where 
$m=0$, 2, 4, 6, 8. 
The upper right panel of
figure \ref{fig:fit-LHCb} shows, for $\kt=2.2$
GeV, the scale uncertainty band 
obtained as an envelope of seven
combinations of scales between factors of 0.5 and 2.0 of the central
scales $(\mu_R,\mu_F)=(1.0,1.5)\,m_T$, namely with the combinations of $(N_R,N_F)$ equal to factors of $m_T$ of $(0.5, 0.75)$, $(2, 3)$,
$(1.0, 0.75)$, $(0.5, 1.5)$, $(1, 3)$,  $(2.0, 1.5) $ and $(1.0, 1.5)$ \cite{Cacciari:2012ny}. The lower left panel of figure \ref{fig:fit-LHCb} shows the central results and uncertainty band for the same scale choices, now with $\kt=0.7$ GeV, the transverse momentum smearing that approximates the \textsc{Powheg~+~Pythia} results (see figure \ref{fig:CDs-kT}). Finally, the lower right panel shows results with $\kt=0.7$ GeV and the conventional central choice of QCD scales, namely, $N_{R,F}=1.0$.

The uncertainty bands in figure \ref{fig:fit-LHCb} are consistent across the rapidity bins. We expect that the large scale uncertainties here dominate the uncertainties at even higher rapidity. PDF uncertainties are typically smaller than the scale uncertainties for forward charm production at these energies \cite{Zenaiev:2019ktw}.
Results for the $p_T$ and $y$ distributions, in similar agreement with the LHCb 
data, are obtained for $D^\pm$ and
$D^0,\ \bar{D}^0$ production, used below as sources of 
$\nu_\mu+\bar{\nu}_\mu$ and $\nu_e+\bar{\nu}_e$. 

In general varying ($N_F$, $N_R$) changes both the rate and the shape of the distributions. The shape is particularly sensitive to $N_F$, whereas the normalization is particularly sensitive to $N_R$. At fixed $N_F$, the larger $N_R$ gives a smaller cross section. The variation of the shape with $N_F$ depends on the $\sqrt{s}$. At LHC energies, the $p_T$ distribution is steeper for larger $N_F$ than for smaller $N_F$.

Here, $B$-meson production follows from the same heavy-quark dynamics, with $c\bar{c}$ replaced by $b\bar{b}$, so it is generally expected that the renormalization and factorization scaling factors are the same for both heavy-quark flavors. To the
extent that $\langle k_T\rangle$ describes the intrinsic parton transverse momentum in the initial state protons, this parameter also
is process independent for $pp\to Q\bar{Q}X$ for $Q=c,\ b$.
For the same central scale and $\langle k_T\rangle$ choices,
the $p_T$ distribution of $B^\pm$ mesons at $\sqrt{s}=13$ TeV for
$y=2.0-4.5$ lies $\sim 10\%$ below the LHCb data
\cite{Aaij:2017qml}. 
The $B$-decay contribution to the total number of tau neutrino events amounts to less than 10\% , as shown in the following, so to approximate bottom production, we use the same scale and $\langle k_T\rangle$ choices as for charm production, with the replacement $m_c\to m_b$. Refinements in the parameters used for the description of the $B$-mesons will not change the conclusions of this paper.

Alternatively,
one could also consider varying $N_F$ and $\langle k_T\rangle$, with $N_R$ fixed, to find the two-parameter combination which provides the best fit to
the LHCb 13 TeV $D_s$  data on double-differential cross sections in $p_T$ and $y$ at $\sqrt{s}$ = 13 TeV. We find that $N_F=1.44$ and $\langle k_T\rangle=2.23$ GeV is the best fit in that case when $N_R=1$, with $\chi^2$/DOF=2.68 and with corresponding predicted $\sigma_{c\bar{c}}$ for 1 GeV$<p_T<8$ GeV and $2.0<y<4.5$ amounting to 87\% of the central value of the experimental result by the LHCb collaboration, which they extrapolate from $D_s$ data.
The 3$\sigma$ allowed interval 
of $\langle k_T\rangle\ (\langle k_T^2\rangle)$ values
turns out to be an ellipse that spans the range
$2.02-2.44$ GeV ($5.20-7.58$ GeV$^2$) for 
$N_F=1.26-1.62$, favoring a lower $N_F$ and higher $\langle k_T\rangle$ (and vice versa) for excursions from the best fit when $N_R=1.0$. An examination of the differences between the full charm meson production data from LHCb, their total charm-anticharm pair cross sections, and theoretical predictions varying simultaneously the three parameters $N_F$, $N_R$ and $\langle k_T\rangle$, would be interesting 
to better understand the roles of $\mu_R$, $\mu_F$ and $\langle k_T\rangle$ in theoretical predictions of
charm production \cite{underway}.\footnote{The parameter $\langle k_T \rangle$ is also tied to the choice of fragmentation function since both influence the 
heavy-meson rapidity and transverse momentum in the collider frame \cite{underway}.} Since our focus here is on $\nu_\tau+\bar{\nu}_\tau$ production, in the following sections we use $N_F=1.5$, $N_R=1.0$ and $\langle k_T\rangle=2.2$ GeV as representatives values of the range of parameter choices that lead to a reasonable description of the experimental LHCb data, together with more widely adopted choices of the same input parameters. As we will see, $\langle
k_T\rangle=2.2$ GeV leads to a suppression, relative to predictions with smaller $\langle k_T\rangle$, of neutrino production in the forward region.

The large value of $\langle k_T\rangle =2.2$ GeV is
difficult to reconcile with theoretical expectations for the strong
interaction effects that we approximately model with the Gaussian
factor. Large $\langle k_T\rangle$ values have been used in some analyses \cite{Apanasevich:1998ki,Miu:1998ju,Balazs:2000sz}. For example, the NLO evaluation of direct photon production in $p\bar{p}$ collisions at the Tevatron, without resummation effects but including $k_T$-smearing with a Gaussian function, shows good agreement with CDF and D0 data when
$\langle k_T\rangle =2.5$ GeV ($\langle k_T^2\rangle = 8.0$ GeV$^2$) for the photon \cite{Apanasevich:1998ki}.
A smaller value of $\langle k_T^2\rangle=1$ GeV$^2$ for the gluon PDF
in the unpolarized proton is used to
describe polarized proton-unpolarized proton scattering to 
$ J/\psi X$ and $DX$ in ref. \cite{DAlesio:2017rzj}. On the other
hand, in an
analysis of di-$J/\psi$ production at LHCb \cite{Aaij:2016bqq}, the unpolarized transverse
momentum gluon distribution, factorized in terms of the usual
collinear PDF and a Gaussian as in eq. (\ref{eq:gaussian}), yields
$\langle k_T^2\rangle = 3.3\pm 0.8$ GeV$^2$
\cite{Lansberg:2017dzg}. Non-perturbative Sudakov factors are 
introduced on top of resummation procedures to describe the transverse momentum
distributions of
Drell Yan and $W,Z$ production. For example, non-perturbative
parameters in impact parameter space that translate to values of $\langle
k_T^2\rangle$ in the range of $\sim 1-2$ GeV$^2$ are obtained in
ref. \cite{Bacchetta:2018lna} to augment the
corrections from resummations to better fit the experimental data.
One interpretation of the need for a large value of
$\langle k_T\rangle$ is that it compensates for missing higher-order perturbative QCD corrections.

The effect of a more conservative value of $\langle k_T\rangle=0.7$ GeV ($\langle k_T^2\rangle=0.6$ GeV$^2$) is shown in
the lower left panel of figure \ref{fig:fit-LHCb} along with the scale
uncertainty band, again with our default central scale choices $(\mu_R,\mu_F)=(1.0,1.5) m_T$.
Figure \ref{fig:CDs-kT} shows predictions for energy distributions for $D_s$ production in
the very forward region, with $\eta > 6.87$,
using different approaches, at two different center-of-mass energies.  
Predictions of a computation relying on NLO QCD matrix-elements matched to the parton shower and hadronization algorithms implemented in \textsc{Pythia},
with matching performed according to the \textsc{Powheg} \cite{Frixione:2007vw,Frixione:2007nw} method, 
are compared to those obtained by combining NLO QCD results with the Gaussian transverse momentum smearing model using $\langle k_T \rangle = 0,\ 0.7$ and 1.4 GeV, followed by fragmentation. 
All the distributions shown use the same
default central scales and charm quark mass. 
In the NLO + shower Monte Carlo computation, the parameters related to fragmentation and intrinsic
transverse momenta are tuned to pre-existing
experimental data.
The parton shower algorithm accounts for the effect of multiple soft and collinear partonic emissions on top of the hard-scattering process, affecting the kinematics and the dynamics of the event, inducing  modifications of the distributions at the fixed-order level.   
The energy distributions of the \textsc{Powheg~+~Pythia} computation for $D_s$ production in the forward region agree well with our Gaussian
smeared
NLO predictions when $\langle k_T\rangle=0.7$
GeV, as shown in both panels of figure~2. 
The lower left panel of figure~\ref{fig:fit-LHCb} shows that the lower
value of $\langle k_T\rangle$ agrees less well with the data at higher
transverse momentum but agrees well in case of lower $p_T$ of the meson, the kinematic region that dominates in the total cross section and in the production of a forward neutrino beam.

\begin{figure}
\begin{centering}
\includegraphics[width=0.48\textwidth]{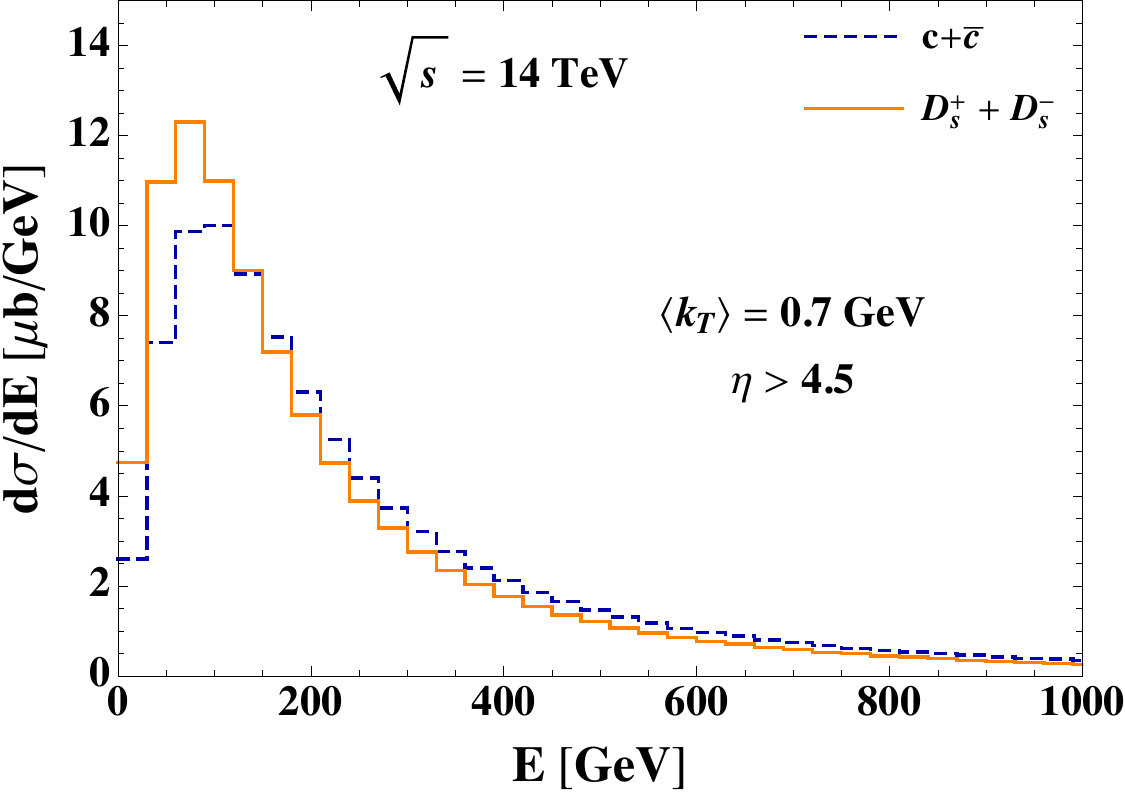}
\includegraphics[width=0.48\textwidth]{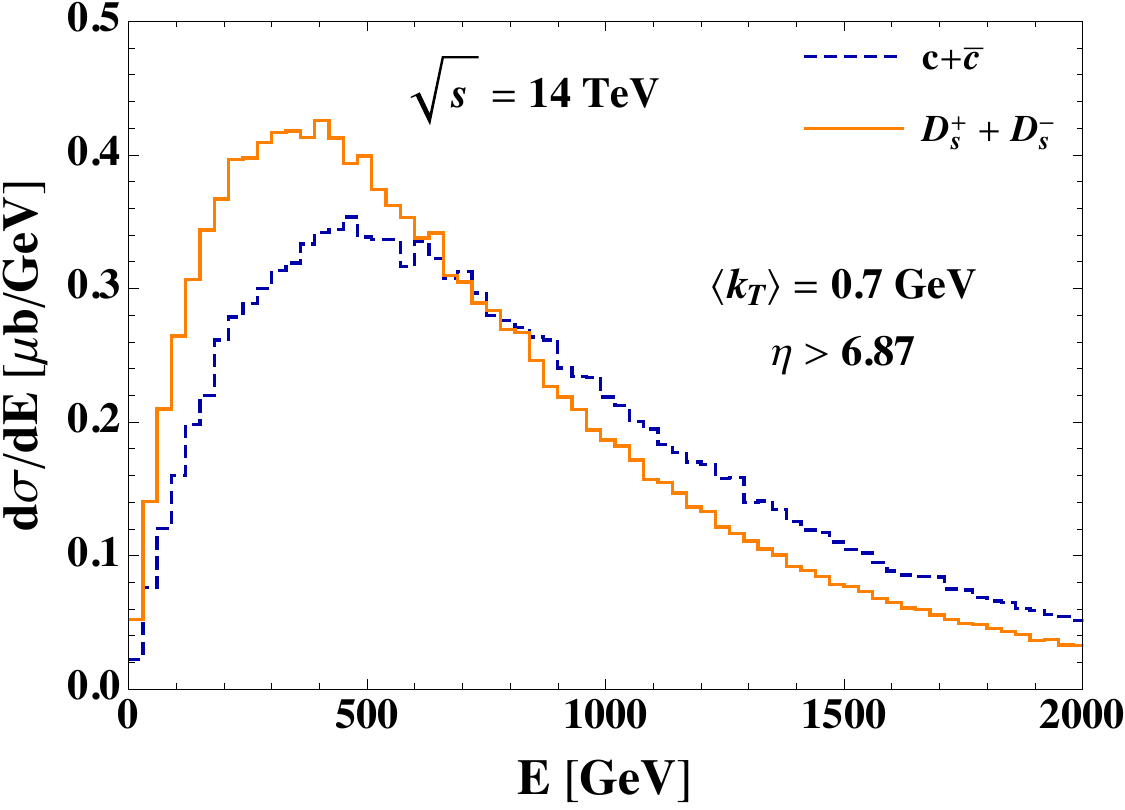} 
\end{centering}
\caption{{The effect of fragmentation of charm quark to $D_{s}$ which shifts
the energy distribution to lower energy for both $\eta>4.5$ (left)
and $\eta>6.87$ (right).  Shown are the differential cross sections} of charm quarks and $D_{s}$ mesons produced in $pp$ collision at $\sqrt{s}=14$ TeV for $(\mu_R, \mu_F) = (1.0, 1.5) m_T$ and $\left\langle k_{T}\right\rangle =0.7$ GeV, as a function
of the energy $E$. Note that the $D_s$ fragmentation fraction has been set
to unity to demonstrate this effect clearly. 
\label{fig:CDsEeta}}
\end{figure}

As already mentioned, it is more conventional to use scales equal to $(\mu_R,\mu_F) = (1.0,1.0) m_T$. We show in the lower right panel of figure
\ref{fig:fit-LHCb}
the double differential distributions $d^2 \sigma / dp_T dy$ for $D_s$ production in different rapidity bins, rescaled as above, obtained using as input of our computation these scales and the same value of $\langle k_T\rangle=0.7$ GeV as in
the lower left panel. The conventional scale combination $(\mu_R,\mu_F)=(1.0,1.0)m_T$
typically gives rise to predicted cross sections lower than the LHCb data, at least considering the present limited accuracy of the theoretical calculations, but the data are still included within the large theoretical scale uncertainty bands.

To allow for comparisons with what could be considered more conventional $\kt$, we show results for both the $\kt =0.7$ 
GeV, a value consistent with a fully non-perturbative interpretation of $\kt$ effects
that most closely follows the \textsc{Powheg~+~Pythia} result, and for  $\kt=2.2$ GeV, the value corresponding to our best fit of LHCb experimental data. 
We also show some results with $\kt=0-1.4$ GeV. Also for comparisons with other QCD evaluations of heavy-flavor production, we show results for both central
scales $(\mu_R,\mu_F) = (1.0,1.5) m_T$ and for conventional central scales $(\mu_R,\mu_F) = (1.0,1.0) m_T$, with their respective uncertainty bands that cover the seven point scale variation of 0.5-2 times the central scale.
With an interpretation of $\kt=2.2$ GeV, rather than with a value $\sim m_p$, as compensating for missing higher order QCD corrections, the seven point scale variations encompass $\kt$ uncertainties, at least to a degree.

Finally, we illustrate the effect of fragmentation in
figure \ref{fig:CDsEeta}. In the left panel, the charm and $D_s$ energy
distributions are shown for $\eta>4.5$ at $\sqrt{s}=14$ TeV, while on
the right, the energy distribution is shown when $\eta>6.87$. For both
panels, the fragmentation fraction for $c\to D_s$ is set to unity to allow
a direct comparison of the charm and meson distributions. The
right panel of figure \ref{fig:CDsEeta} shows that the impact of the fragmentation function extends to very high energy in
the very forward region.

%% file: results-14TeV-comb.tex
\section{Neutrinos from the heavy-flavor hadrons}
\label{sec:results}

\subsection{Tau neutrinos}
Tau neutrinos and antineutrinos arise primarily from the prompt decays of
$D_s$ mesons into $\tau + \nu_\tau$, where $B(D_s\to\tau\nu_\tau)=(5.48\pm 0.23)\times
10^{-2}$ \cite{pdg:2019}.
The decay of the $\tau$ itself is also
prompt and produces a tau neutrino, which is the dominant source of high-energy tau neutrinos in the forward region since the tau carries most
of the energy of the $D_s$, given  $m_{D_s} = 1.97 \ {\rm GeV}$ and
$m_{\tau} = 1.78 \ {\rm GeV}$.
The energy distribution of the $\nu_\tau$ that comes directly from the $D_s$ leptonic decay (called the ``direct'' neutrino) is straightforward to calculate from the isotropic decay of the $D_s$ in its rest frame, followed by a boost to the collider frame where the $D_s$ has a four-momentum $p_D$. Keeping the
polarization of the tau \cite{Barr:1988rb}, the energy
distribution of the $\nu_\tau$ from the tau decay  (here
called the ``chain'' decay neutrino) can also be obtained. The latter distribution, after
integrating over angles relative to the tau momentum
direction, is discussed in refs.
\cite{Pasquali:1998xf,Bhattacharya:2016jce}
in the context of the atmospheric tau neutrino flux. Here, we use the full energy and angular distribution in the collider frame to apply the requirement that the
neutrino pseudorapidity fulfills the $\eta>6.87$ constraint. Details of the tau neutrino
energy and angular distribution from $D_s\to \tau\to \nu_\tau$ appear in
ref. \cite{Bai:2018xum}. The distributions of tau neutrinos and
antineutrinos from the $D_s$ decays are identical because of the zero-spin of the meson (direct neutrinos) and the compensation of
particle/antiparticle and left-handed/right-handed effects in the
chain decays of the tau \cite{Barr:1988rb}. 

$B$-meson production and decay also contributes to the number of tau
neutrinos, but at a level of less than 10\% of
the contribution from $D_s$ production and decay. The branching fractions to taus from charged and neutral $B$'s are 
$  B(B^+\to \bar{D}^0\tau^+\nu_\tau) =  (7.7\pm 2.5)\times 10^{-3}$, 
 $B(B^+\to \bar{D}^*(2007)^0\tau^+\nu_\tau)  =  (1.88\pm 0.20)\times
 10^{-2}$, 
$ B(B^0\to {D}^-\tau^+\nu_\tau =  (1.08\pm 0.23)\times 10^{-2}$,  and
 $B(B^0\to {D}^*(2010)^-\tau^+\nu_\tau)  =  (1.57\pm 0.10)\times
                                           10^{-2}$  \cite{pdg:2019}.
We use the central values of all the branching fractions. 
The energy and angular distributions of tau
neutrinos from $B$ meson decays are discussed in appendix A,
as are the energy and angular distributions of muon and electron neutrinos from 
heavy-meson decays.

\begin{figure}
\begin{centering}
\includegraphics[width=0.48\textwidth]{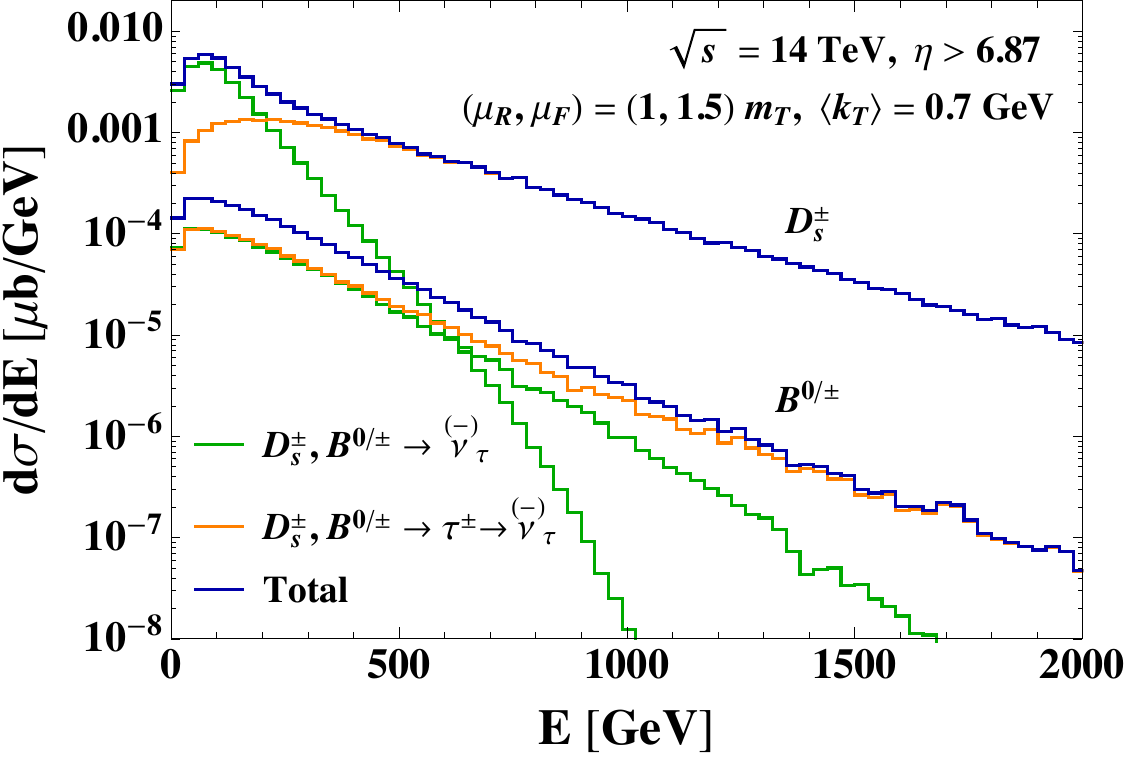} 
\includegraphics[width=0.48\textwidth]{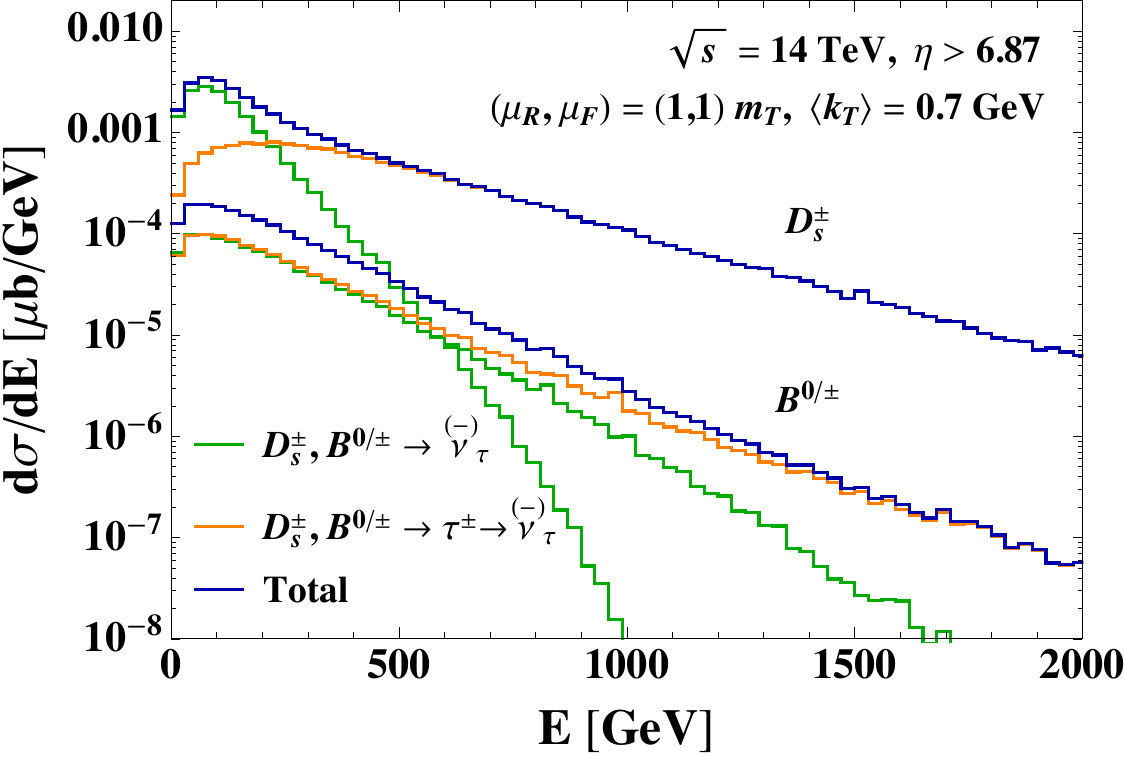} 
\par\end{centering}
\caption{
The neutrino energy distributions for tau neutrinos and antineutrinos from the direct decay $D_{s}^\pm / B^{0,\pm}\rightarrow\overset{{\scriptscriptstyle (-)}}{\nu}_{\tau}$   (green)  and the chain decay 
  $D_{s}^\pm/  B^{0,\pm}\rightarrow\tau^\pm\rightarrow
  \overset{{\scriptscriptstyle(-)}}{\nu}_{\tau}
  $ 
  (orange) and their
  sum (blue) for neutrinos with pseudorapidity $\eta>6.87$, produced in
  $pp$ collisions at
$\sqrt{s}=14$ TeV. Predictions are obtained using as input
$\langle k_T\rangle =0.7$ GeV  with
our default scale combination $(\mu_R, \mu_F) = (1.0,1.5) \, m_T$ (left) and with
the conventional scale combination $(\mu_R, \mu_F) = (1.0,1.0) \, m_T$ (right).
The contributions from both $D_s$ and $B$ mesons are shown separately. 
}
\label{fig:NutauCS}
\end{figure}

Figure \ref{fig:NutauCS} shows the energy distributions for tau neutrinos and antineutrinos from both the $D_s$ and $B$ meson decays with neutrino pseudorapidity $\eta>6.87$. 
The contribution of the direct and the chain decays are shown separately, as well as their sum.  
The left panel
shows the distributions with our default scale combination
$(\mu_R,\mu_F)=(1.0, 1.5) \, m_T$, while the right panel shows the same
distributions, but using as input
$(\mu_R,\mu_F)=(1.0, 1.0) \, m_T$. Qualitatively, the
distributions are similar, although, as expected from the discussion in section 3, the differential
distributions using the conventional scale are lower than with our default scale choice.

The number of neutrino events per unit energy can be written as 
\begin{equation}
\frac{dN}{dE} =\frac{d \sigma (pp \rightarrow \nu X)}{d E} \times {\cal P}_{\rm int} \times  {\mathcal L}\ ,
\end{equation}
where the interaction probability in the detector is
\begin{equation}
{\cal P}_{\rm int} = \bigl( \rho_{\rm Pb} \times l_{\rm det} \times N_{\rm avo} \bigr)\,  \frac{ \sigma_{\nu_\tau Pb}}{A_{\rm Pb}} \ .
\label{eq:Pint}
\end{equation}
Here, we use an integrated luminosity ${\mathcal L} = 3000 {\ \rm fb}^{-1}$ and a lead density $\rho_{Pb} =
11.34 {\ \rm g/cm^3}$. The nucleon number of lead is $A_{Pb}= 208$
and
$l_{\rm det}$ is the length of the lead neutrino target in the
detector which is also characterized by a cross sectional area of radius 1 m
(thus $\eta>6.87$) for our discussion here.
For reference, a detector of lead with radius of 1 m and length $l_{\rm det}$~=~1~m 
has a mass of $\sim$ 35.6 ton. For the number of events, we quote
the number of events per ton of lead (per 2.8 cm depth of a lead
disk with radius 1 m). As discussed in more detail in ref. \cite{Bai:2018xum}, we evaluate
the neutrino and antineutrino charged-current cross sections at NLO, including mass effects \cite{Kretzer:2002fr,Kretzer:2003iu,Jeong:2010za,Jeong:2010nt,Reno:2006hj}, using the nCTEQ15 PDFs for lead
\cite{Kovarik:2015cma}. 
For neutrino energies above $\sim 10$ GeV, deep inelastic scattering (DIS) dominates quasi-elastic scattering and few-pion production \cite{Lipari:1994pz,Kretzer:2004wk}.
At the energies of interest, the neutrino DIS cross section is roughly a factor of $\sim 2$ larger than the antineutrino cross section.
Kinematic corrections due to the tau mass reduce the 
charged-current cross section by $\sim 25\%$ for a neutrino energy $E_{\nu}$ =  100 GeV, and by $\sim 5\%$ for $E_{\nu}$ = 1000 GeV \cite{Jeong:2010nt}. 
NLO corrections are small, less than $\sim 5\%$ for neutrino energies of 100 GeV, as shown e.g. in ref. \cite{Jeong:2010za}.

\begin{figure}
\begin{centering}
\includegraphics[width=0.48\textwidth]{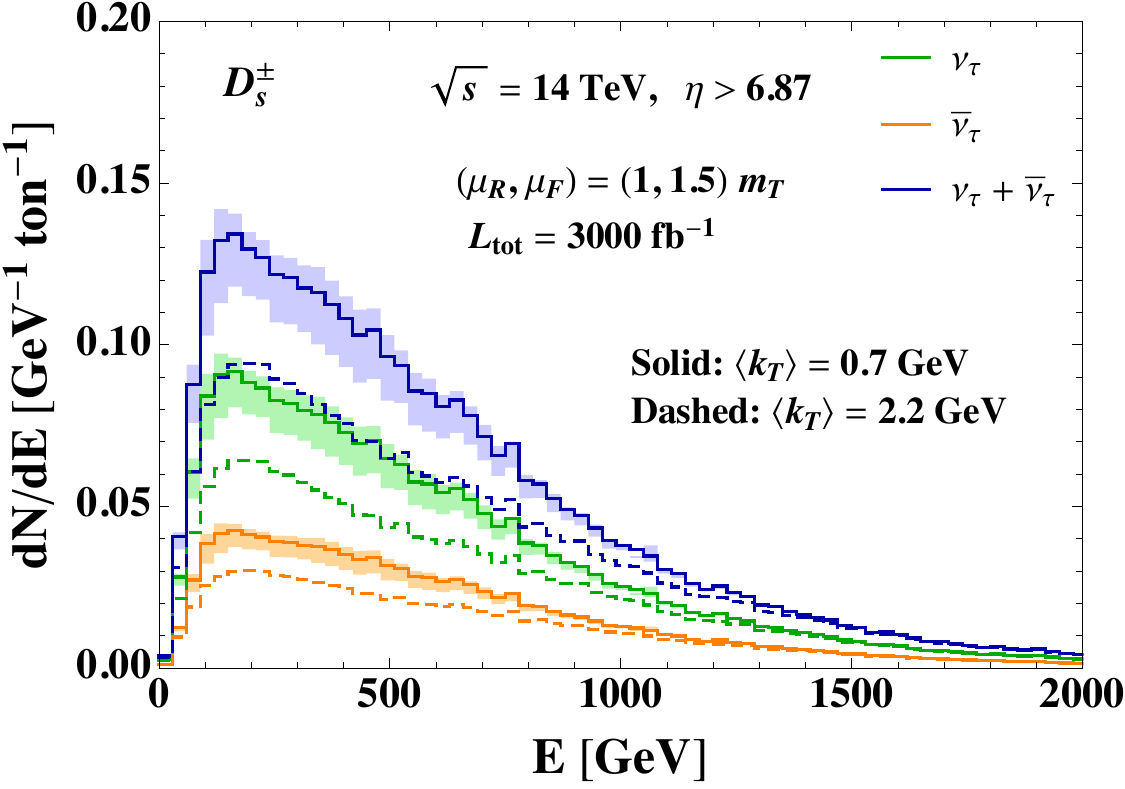}
\includegraphics[width=0.48\textwidth]{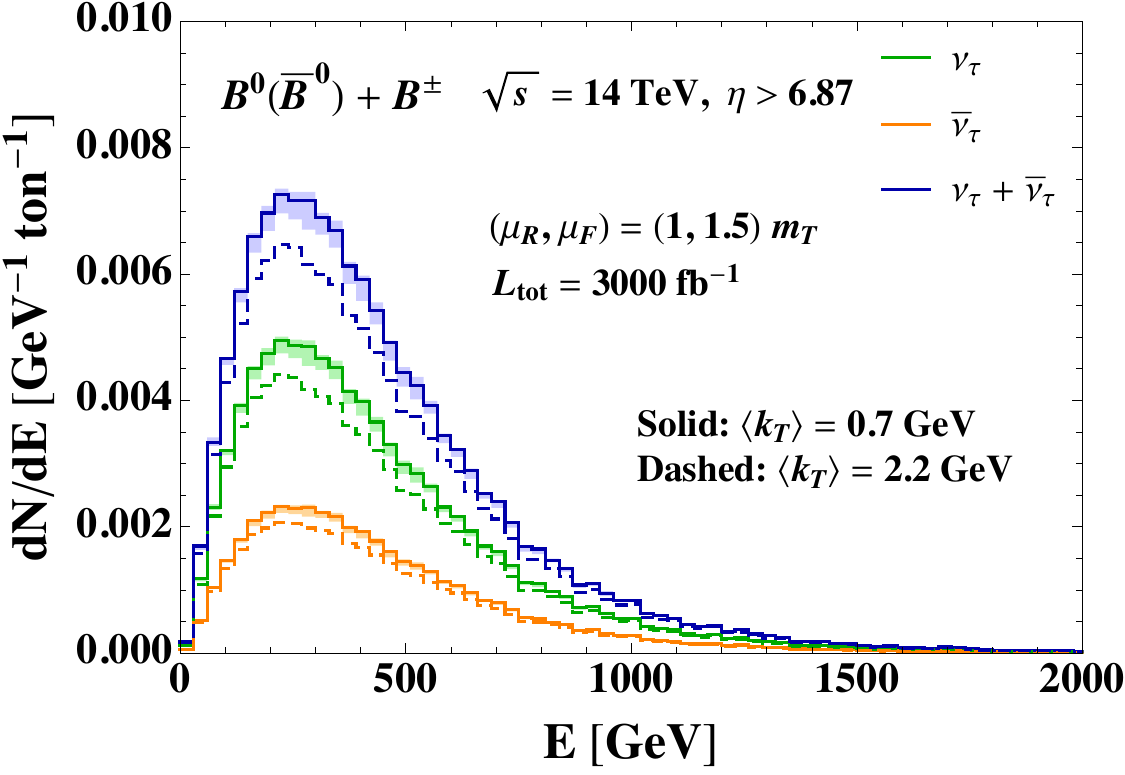}  
\includegraphics[width=0.48\textwidth]{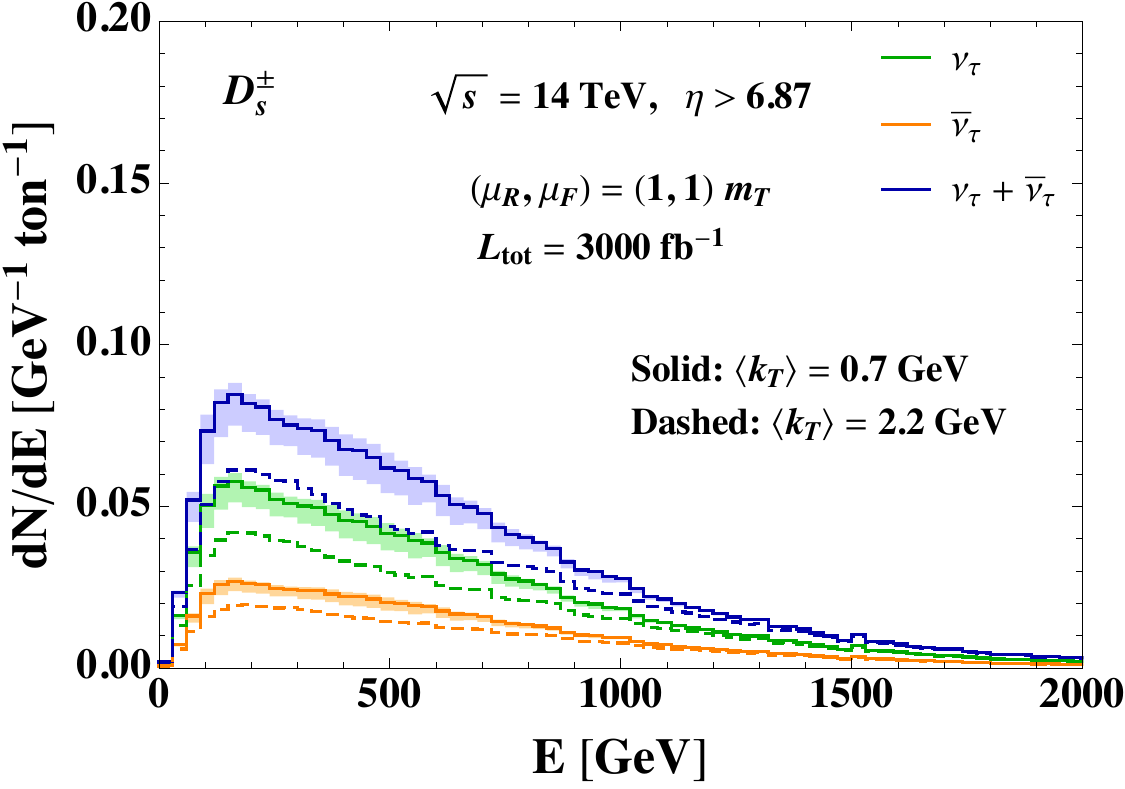}
\includegraphics[width=0.48\textwidth]{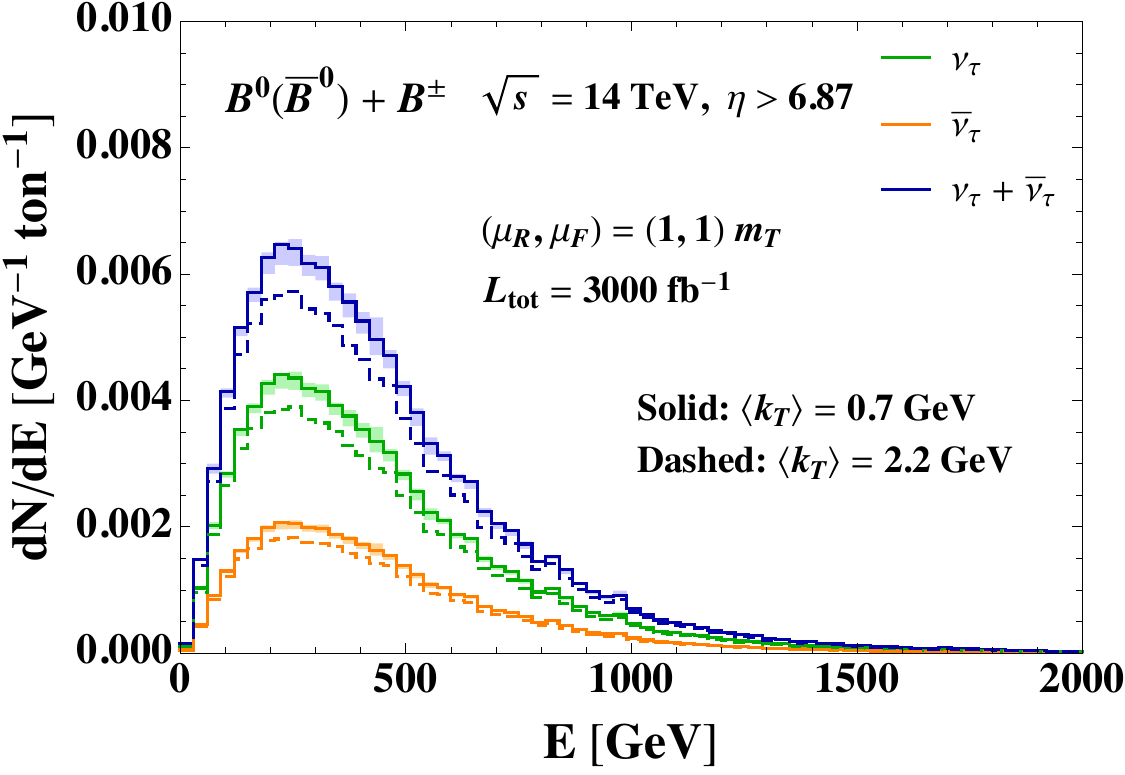}   
\par\end{centering}
\caption{
Our predictions for the tau neutrino and/or antineutrino number of charged-current events per GeV per ton as a function of the incident neutrino energy for neutrino 
pseudorapidity $\eta>6.87$ for $pp$ collisions with $\sqrt{s}=14$ TeV and integrated luminosity ${\cal L}=3000$ fb$^{-1}$.  
The central predictions (solid histogram) refer to the  $\left\langle k_{T}\right\rangle$ = 0.7 GeV value and the bands arise from the $\left\langle k_{T} \right\rangle$ variation in the range $0 <  \left\langle k_{T}\right\rangle  < 1.4$ GeV. 
The results obtained with $\left\langle k_{T}\right\rangle$ = 2.2 GeV are also shown (dashed histogram).
    The upper plots are obtained by setting the QCD scales to $(\mu_R, \mu_F) = (1.0, 1.5) m_T$, while the lower plots refer to the conventional scale combination $(\mu_R, \mu_F) = (1.0, 1.0) m_T$. NLO  corrections are included in the DIS cross sections.
\label{fig:NutauEventLHC-kT}}
\end{figure}

In figure \ref{fig:NutauEventLHC-kT}, we show the number of events per unit neutrino energy per ton of detector lead for tau neutrinos and antineutrinos from the $D_s^\pm$ (left panels) and $B$ meson (right panels) decays, evaluated for a total integrated luminosity ${\cal L} = 3000 \, {\rm  fb^{-1}}$.
The upper panels include results with our default scales $(\mu_R, \mu_F) = (1.0, 1.5)\, m_T$, 
while the lower panels show the same
quantities using as input the conventional scale combination $(\mu_R, \mu_F) = (1.0, 1.0)\, m_T$.
The central solid histograms are obtained for $\left\langle k_{T}\right\rangle$ = 0.7 GeV,
and the bands reflect the uncertainty range due to the 
variation of $\left\langle k_{T}\right\rangle$ 
in the range $0 -1.4$ GeV. 
We also present predictions for $\left\langle k_{T}\right\rangle$
= 2.2 GeV, that are shown with the dashed histograms.

\begin{figure}[ht]
\begin{centering}
\includegraphics[width=0.48\textwidth]{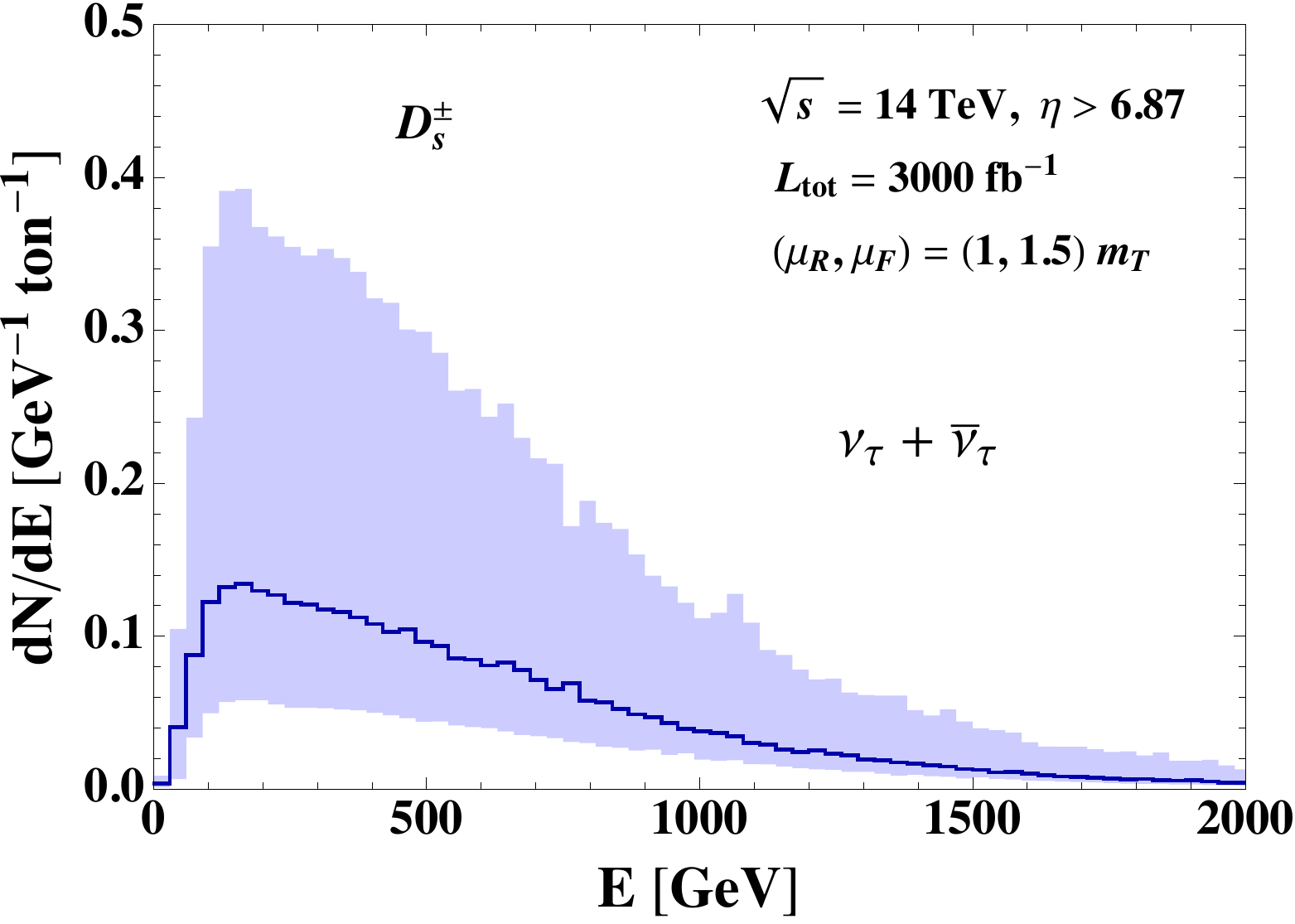}
\includegraphics[width=0.48\textwidth]{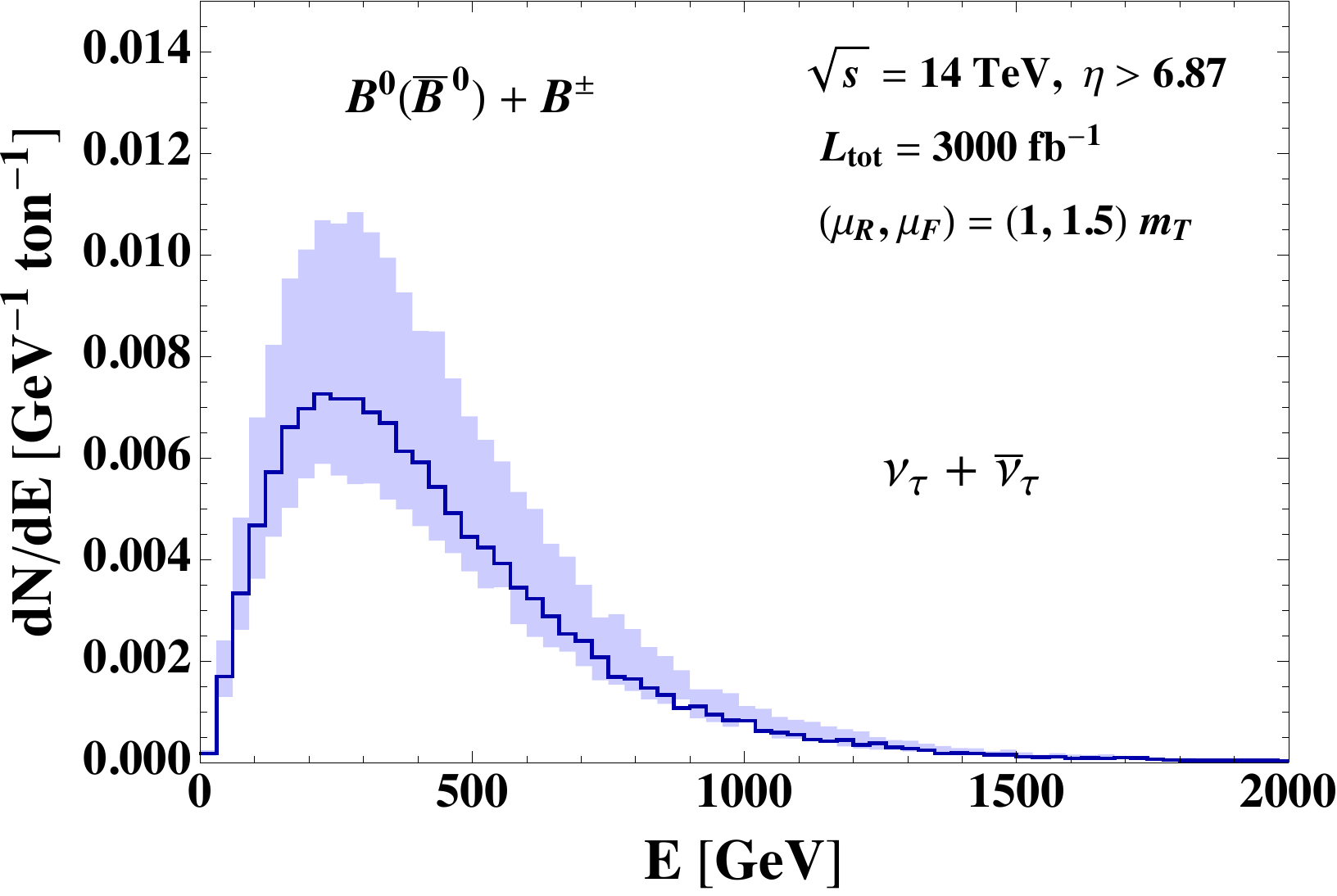}
\includegraphics[width=0.48\textwidth]{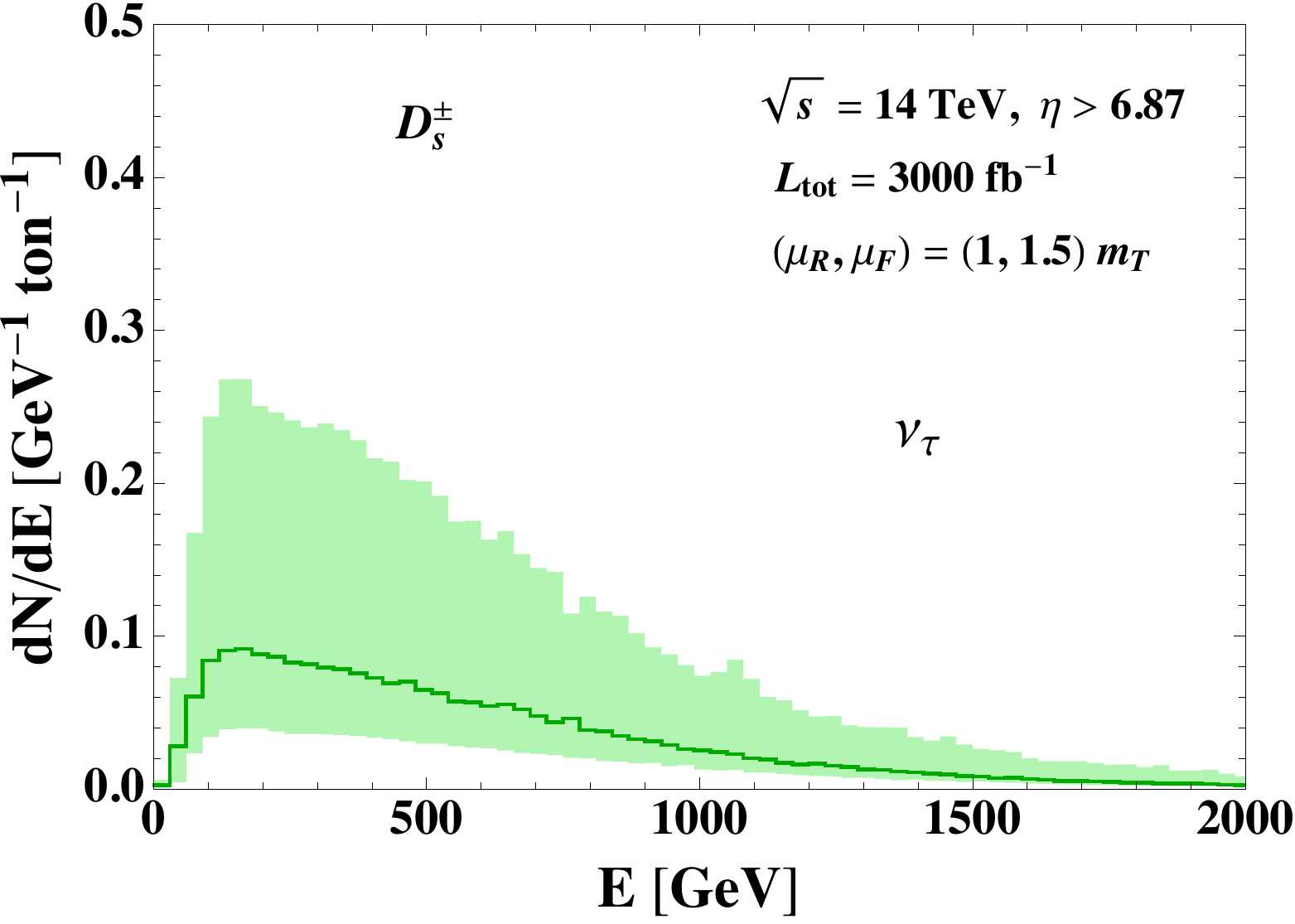} 
\includegraphics[width=0.48\textwidth]{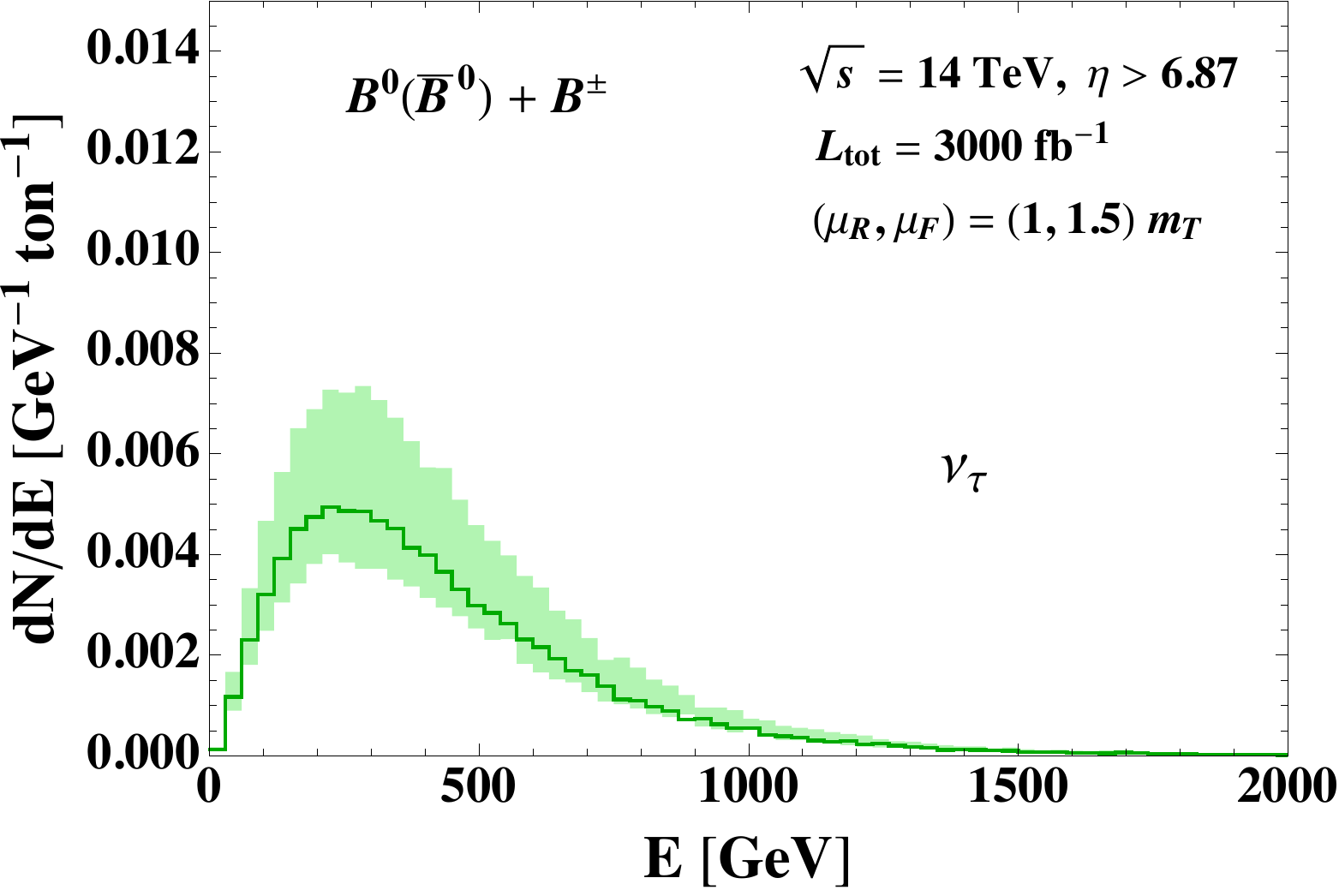} 
\includegraphics[width=0.48\textwidth]{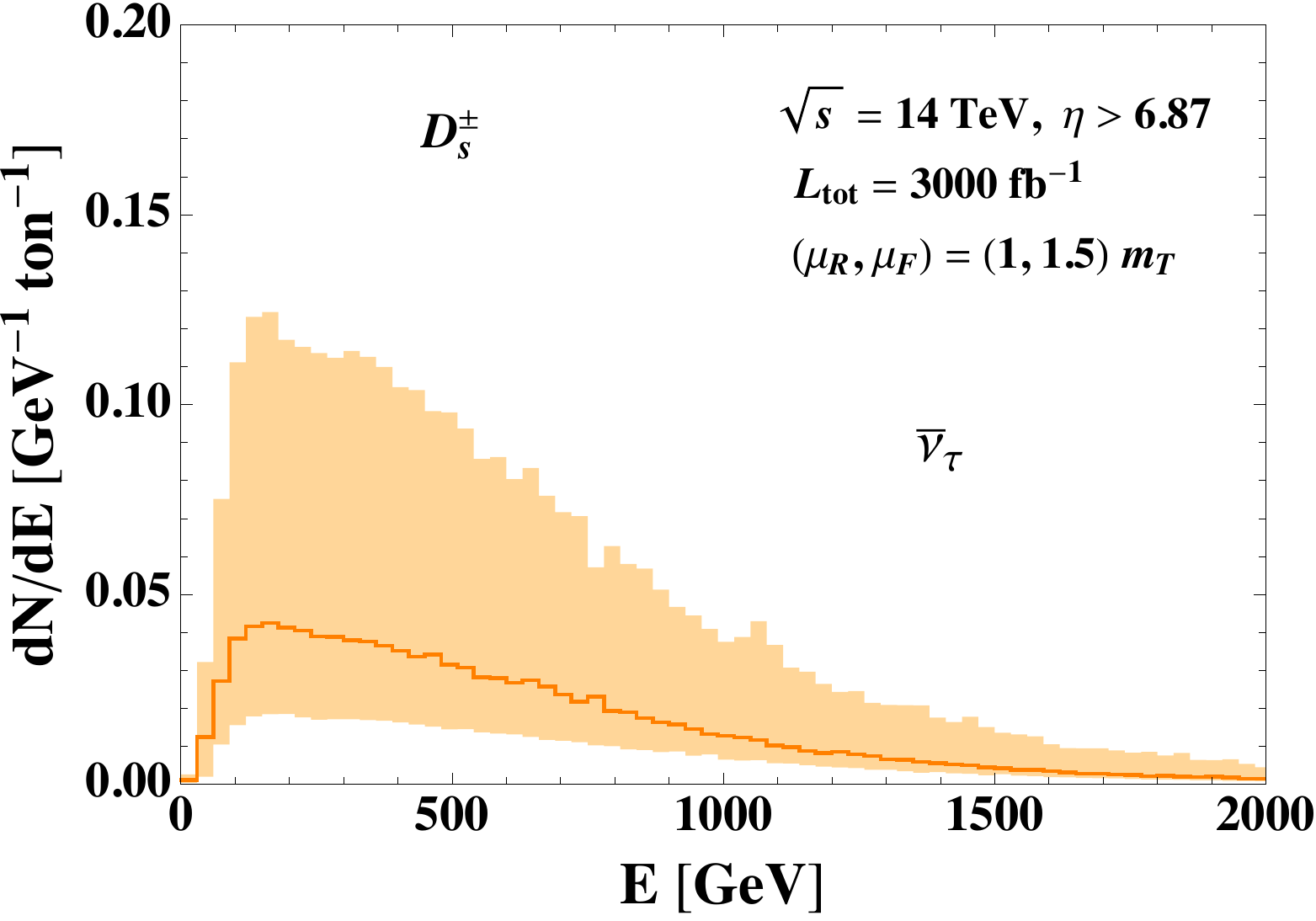}
\includegraphics[width=0.48\textwidth]{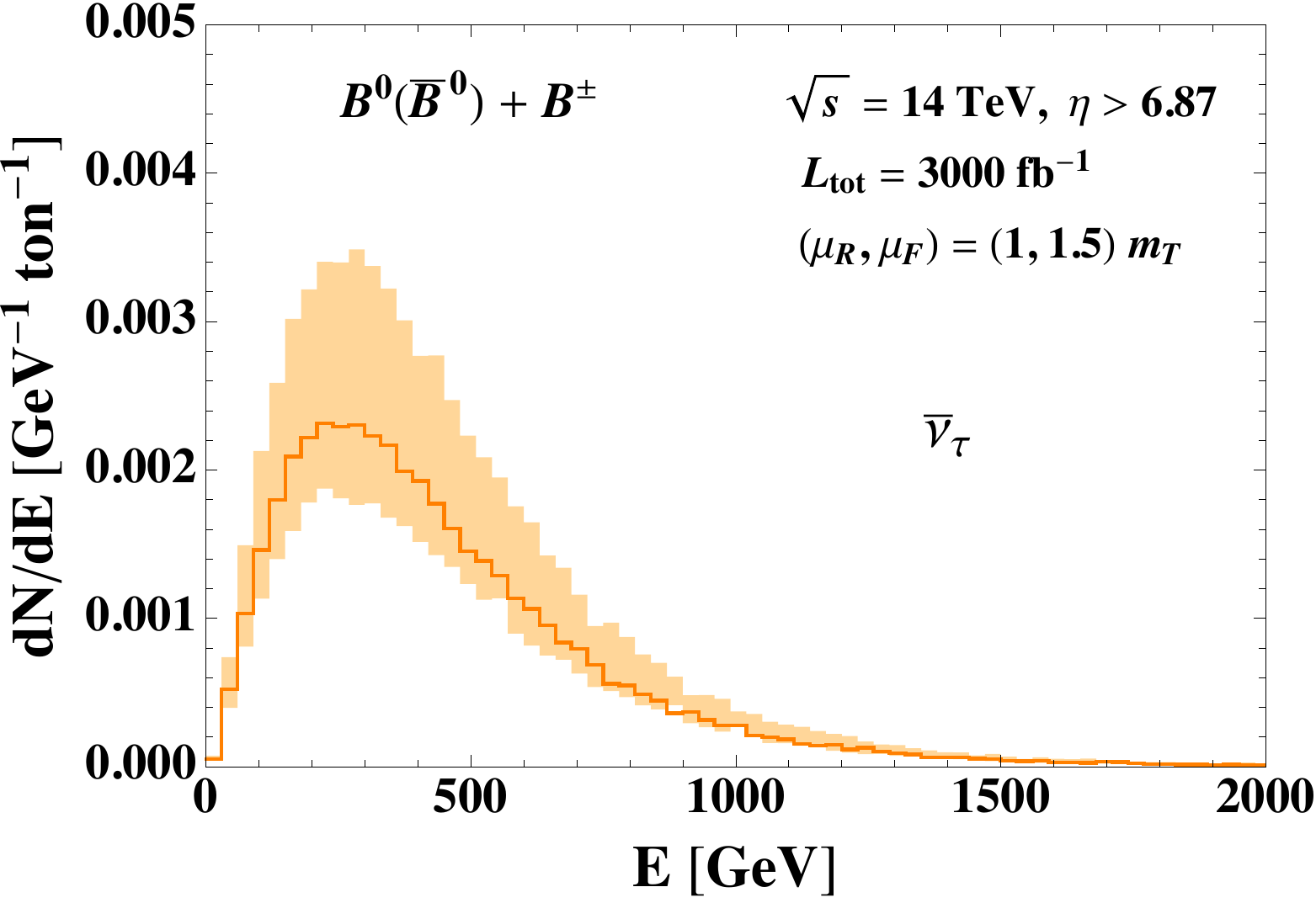}\par\end{centering}
\caption{Uncertainty range due to the QCD scale variation in the tau neutrino and antineutrino number of charged-current events per GeV per ton as a function of the incident neutrino energy for neutrino pseudorapidity $\eta>6.87$ and for the $pp$ collision with $\sqrt{s}=14$ TeV and integrated luminosity ${\cal L}=3000$ fb$^{-1}$.
The central predictions are obtained using as input $(\mu_R, \mu_F) = (1, 1.5) m_T$ and $\left\langle k_{T}\right\rangle$~=~0.7~GeV. The upper and lower limits arise from the QCD 7-point scale variation range in figure \ref{fig:fit-LHCb}. NLO QCD corrections are accounted for in the DIS pQCD cross section.
\label{fig:NutauEventLHC-scale}}
\end{figure}

Incorporating $\langle k_T \rangle$ effects has a large impact on the predictions at low energies as expected from figure \ref{fig:CDs-kT}. 
In particular, computing charm production with $\langle k_T \rangle = 0$ GeV, enhances the number of events per unit energy by
up to $\sim 8\%$ at $E_\nu \sim$ 100 GeV with respect to the case with the default $\langle k_T \rangle = 0.7$ GeV, whereas the differences are less than 1\% for $E_\nu \gsim$ 1000 GeV.
The lower limit on the number of events per unit energy, corresponding to the case $\langle k_T \rangle = $ 1.4 GeV, is lower by $\sim$ 16 (5)\% at  $E_\nu \sim$ 100 (1000) GeV with respect to the case with $\langle k_T\rangle=0.7$ GeV, whereas the difference reduces to 1\% or even less for $E_\nu \gsim$ 1500 GeV. 
On the other hand, if $\left\langle k_{T} \right\rangle$ = 2.2 GeV, the number of events from $D_s^\pm$ in the peak of the distribution is a factor of $\sim 2/3$  lower than in the peak for $\left\langle k_{T}\right\rangle$ = 0.7 GeV. The $\left\langle k_{T} \right\rangle$ sensitivity of the predictions of the number of ($\nu_\tau$ + $\bar{\nu}_\tau$) neutrinos from $B$ meson decays is much smaller than for $D_s$ meson decays. A comparison of
upper and lower panels shows the stronger impact of the factorization scale dependence on the predictions for $\nu_\tau+\bar{\nu}_\tau$ from charm mesons than from $B$ mesons.

\begin{figure}[ht]
\begin{centering}
\includegraphics[width=0.48\textwidth]{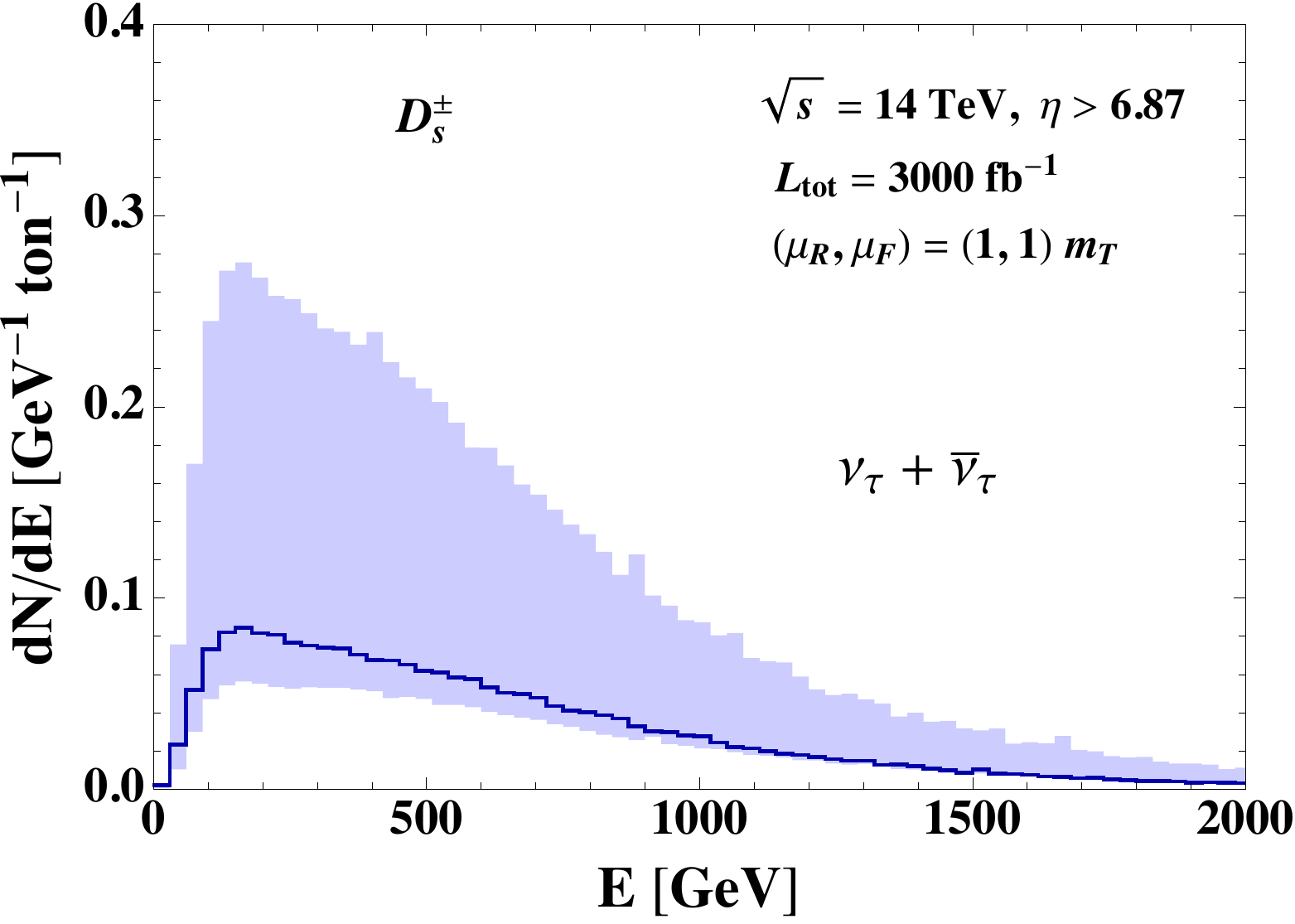}
\includegraphics[width=0.48\textwidth]{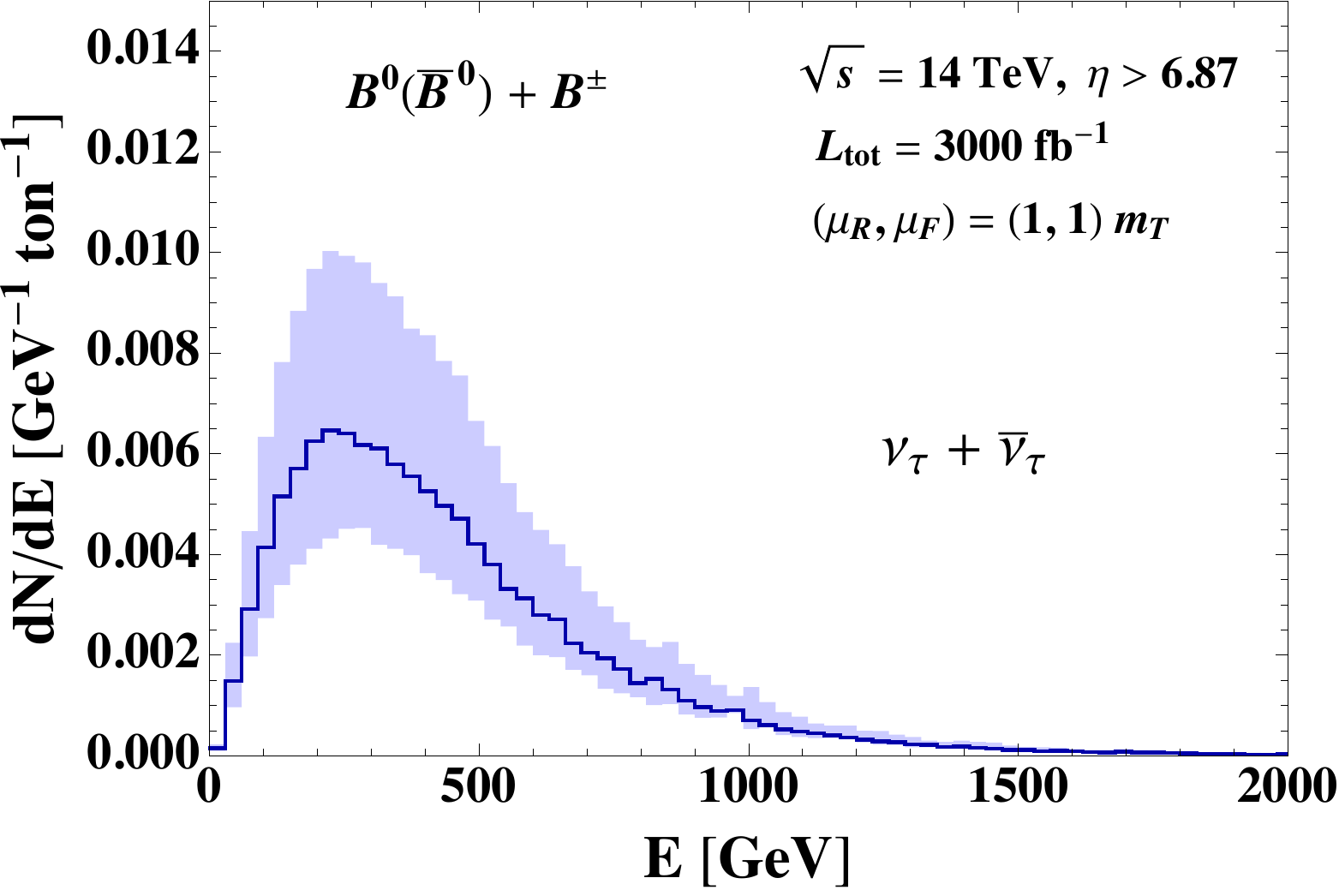}
\includegraphics[width=0.48\textwidth]{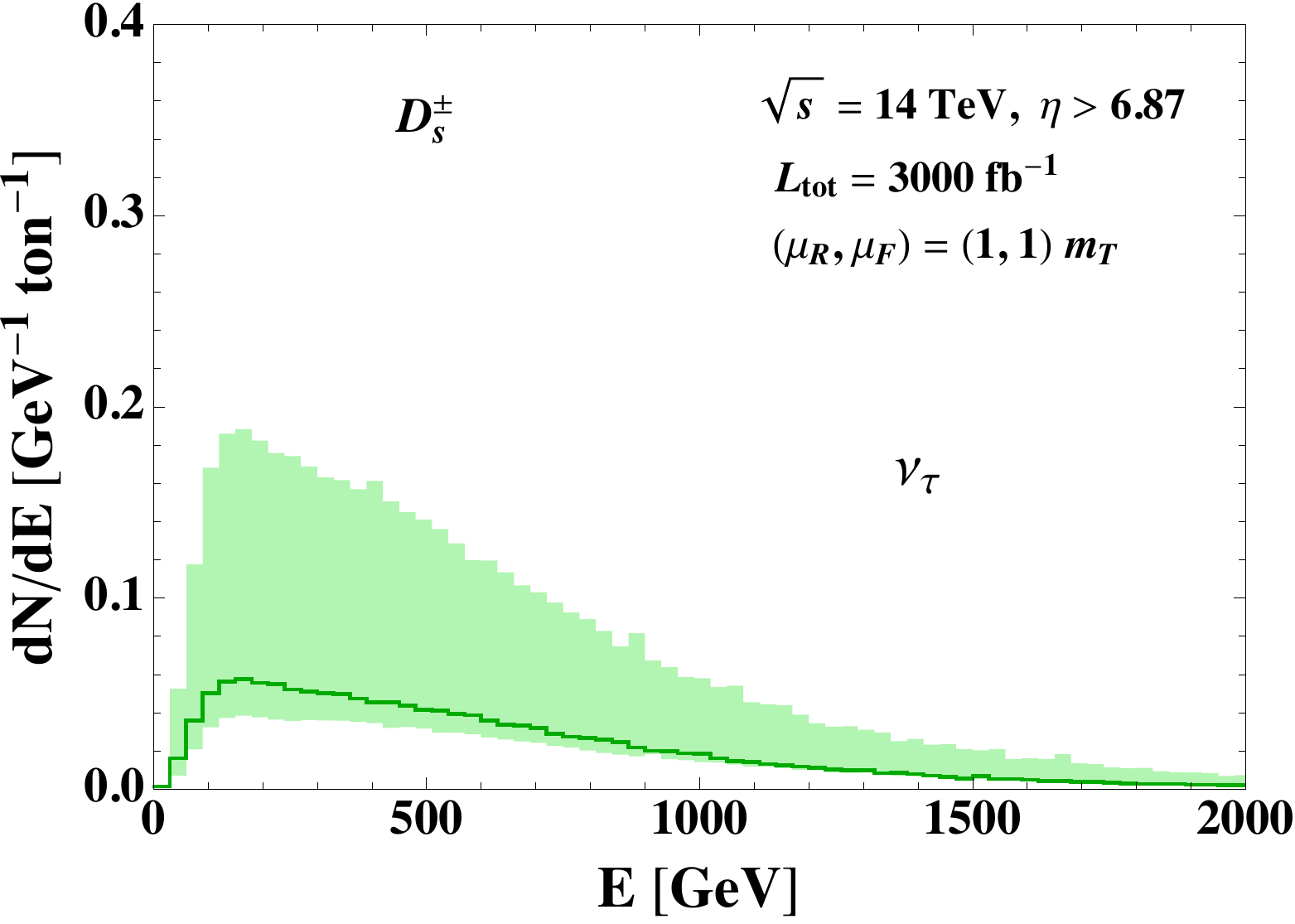} 
\includegraphics[width=0.48\textwidth]{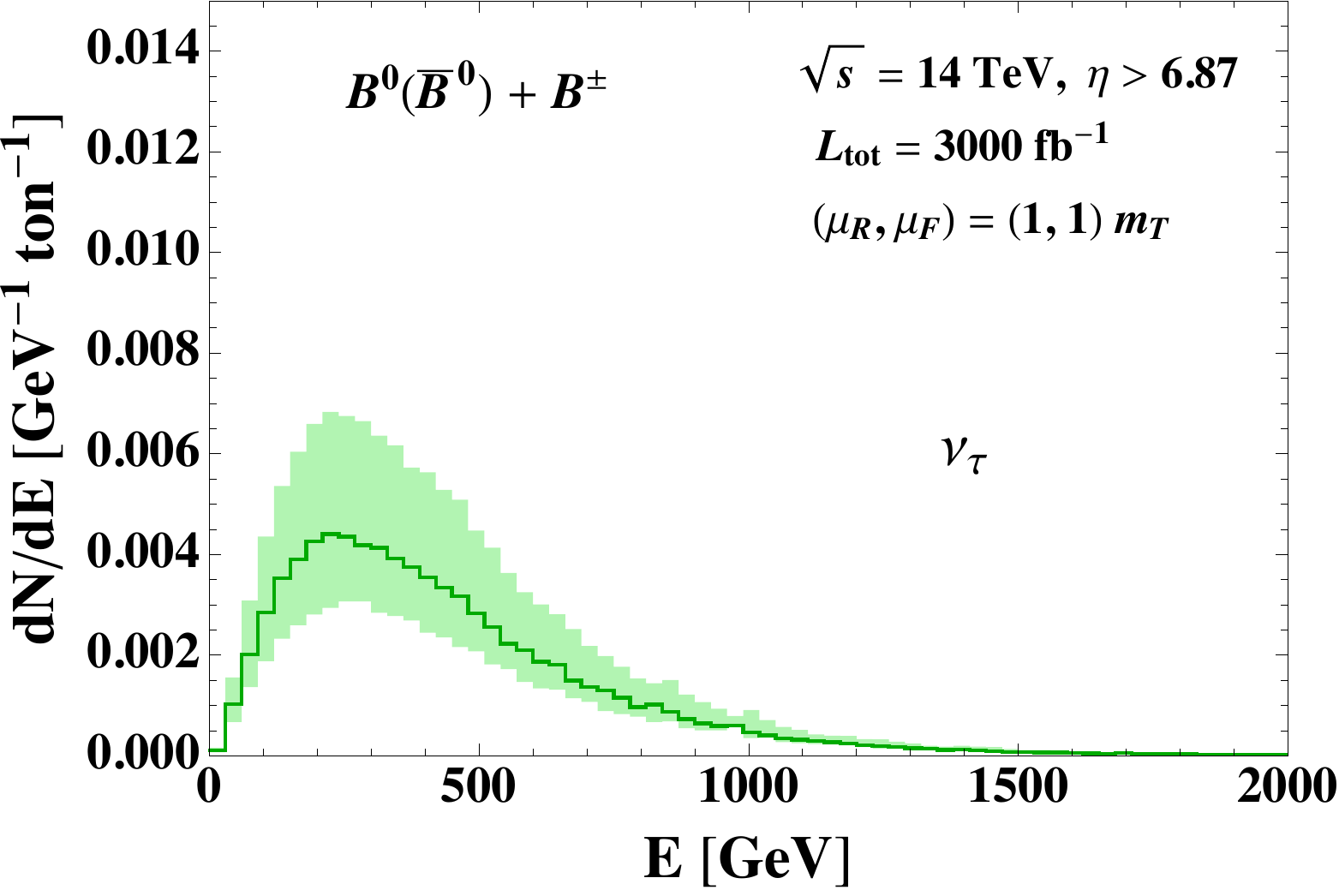} 
\includegraphics[width=0.48\textwidth]{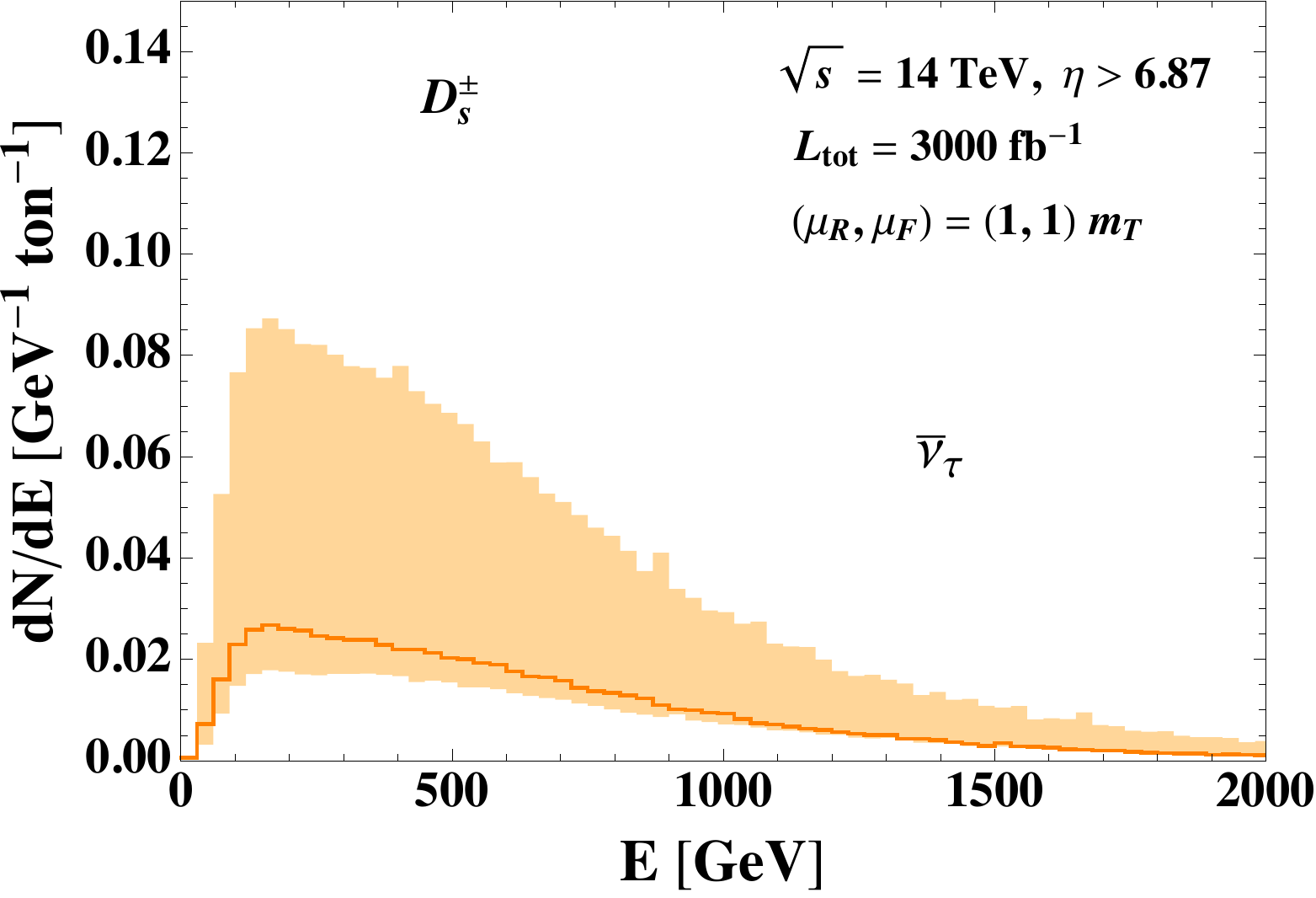}
\includegraphics[width=0.48\textwidth]{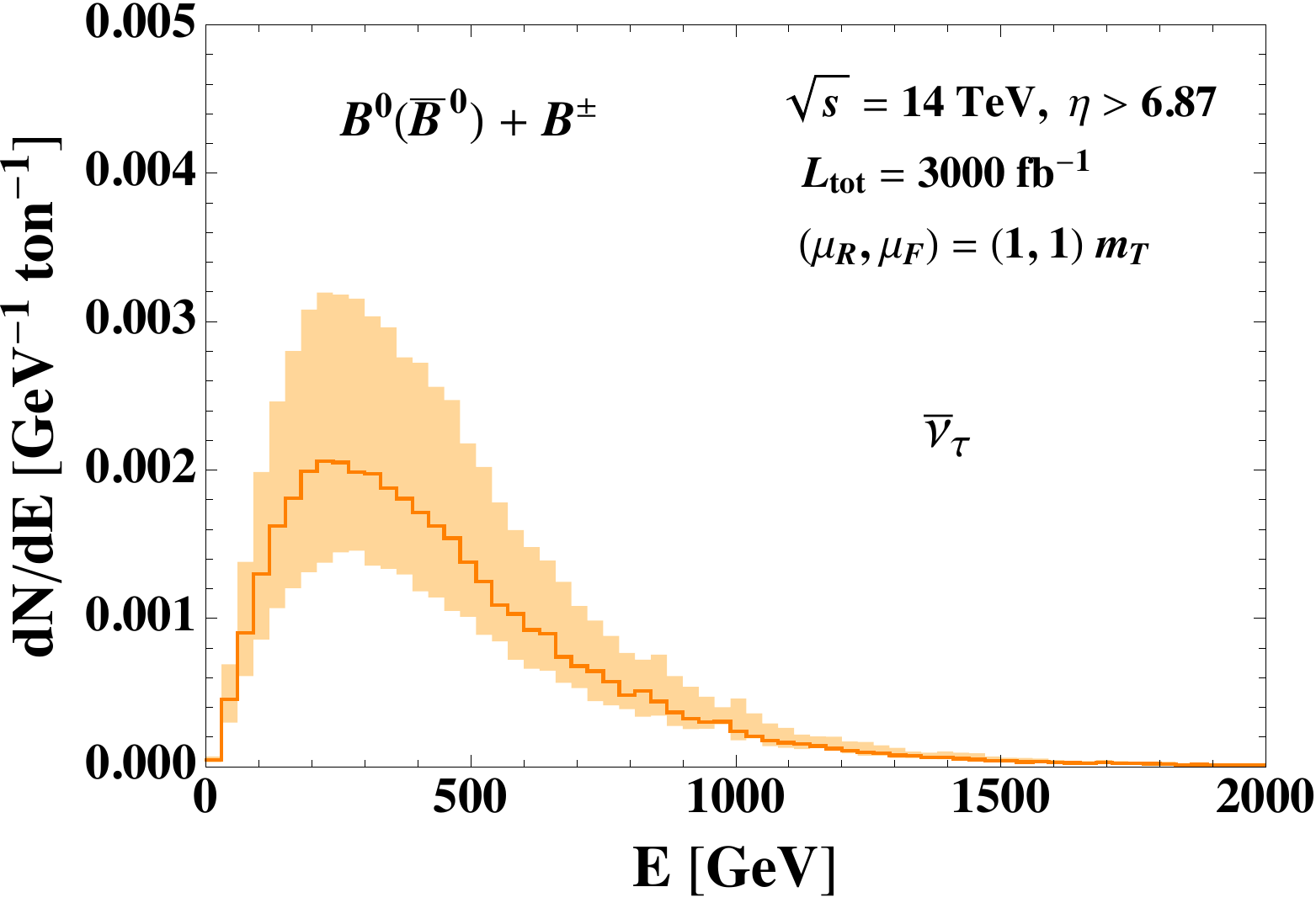}\par\end{centering}
\caption{The same as figure \ref{fig:NutauEventLHC-scale}, but adopting the central scale conventional choice  $(\mu_R, \mu_F) = (1.0, 1.0) \, m_T$ and seven-point scale variation around it, again with $\left\langle k_{T}\right\rangle$ = 0.7 GeV. 
\label{fig:NutauEventLHC1}}
\end{figure}

Figure \ref{fig:NutauEventLHC-scale} shows the uncertainty bands associated with the QCD scale variation
in the range considered in figure \ref{fig:fit-LHCb}, for a central scale choice $(\mu_R, \mu_F) = (1, 1.5)\,m_T$. These bands are computed as envelopes of
seven combinations of $(N_R, N_F)$, equal to factors of $m_T$ of $(0.5, 0.75)$, $(2.0, 3.0)$,
$(1.0, 0.75)$, $(0.5, 1.5)$, $(1.0, 3.0)$,  $(2.0, 1.5)$ and $(1.0, 1.5)$.
For neutrinos from the $D_s$ decay, the upper boundary of the band is larger than the central prediction by a factor of $\sim 3-4$, while the lower edge of the band is 40 $-$ 60\% smaller than the central prediction for $E_\nu \lesssim 1500$ GeV. Thus, the QCD scale uncertainty band in our evaluation with 
$\left\langle k_{T}\right\rangle$ = 0.7
GeV has overlap with the prediction using $\left\langle k_{T}\right\rangle$ = 2.2 GeV, as follows from comparing figure \ref{fig:NutauEventLHC-kT} and \ref{fig:NutauEventLHC-scale}. 
For neutrinos from $B$ meson decays, the scale uncertainty bands are smaller than for neutrinos from $D$ mesons. In particular, the edge of the upper uncertainty band is a factor $1.5 - 2$ larger than the central prediction, whereas
the lower uncertainty band extends to about 20\% below the central prediction for $E_\nu \lesssim 1000$ GeV.
The scale uncertainty bands around central predictions with the conventional scale combination $(\mu_R, \mu_F) = (1.0, 1.0) m_T$, as shown in figure \ref{fig:NutauEventLHC1}, have
similar sizes to those in figure
\ref{fig:NutauEventLHC-scale}. 
However, the conventional scale choice leads to an overall lower central prediction for the number of events from both $D_s$ and $B \to \nu_\tau$ and $\bar{\nu}_\tau$ than the case $(\mu_R, \mu_F) = (1.0, 1.5) m_T$, 
as follows by comparing figures \ref{fig:NutauEventLHC-scale} and \ref{fig:NutauEventLHC1}.

\begin{table}[h]
	\vskip 0.35in
	\begin{center}		    
		\begin{tabular}{|c||c|c|c||c|c|c||c|c|}
			\hline
                         &  $\nu_\tau$ &  $\bar{\nu}_\tau $  &   $\nu_\tau + \bar{\nu}_\tau $ &
                         \multicolumn{5} {c|}{ $\nu_\tau + \bar{\nu}_\tau $}  \\	
                       \hline  
		      $ (\mu_R, \ \mu_F) $  & \multicolumn{3}{c||} {(1, 1.5)  $m_T$} & \multicolumn{3}{c||} {(1, 1.5)  $m_T$} &  {(0.5, 1.5)  $m_T$} &  {(1, 0.75)  $m_T$} \\	                       
		      \hline	                              
		      $\langle k_T \rangle $  & \multicolumn{3}{c||} {0.7 GeV} & 0 GeV  & 1.4 GeV & 2.2 GeV & \multicolumn{2}{c|} {0.7 GeV}  \\		                 
			\hline
		 	$D_s$ & 2451 & 1191 &3642& 3799 & 3261  &  2735 & 11008 & 1716\\ 
			\hline
			$B^{\pm,0}$& 96 & 46 & 142  & 144 & 137 & 127 & 214 & 115 \\
			\hline
			Total & 2547 & 1237 & 3784 & 3943 & 3398 & 2862 & 11222 & 1831\\
			\hline
		\end{tabular}
	\end{center}
	\caption{The charged-current event numbers for tau neutrinos and antineutrinos in 1 m length of the lead detector (equivalent to  $M_{\rm pb} \simeq$ 35.6 ton) assuming central scales 
	 ($\mu_R, \mu_F) = (1.0, 1.5) \, m_T$ in the computation of heavy-meson production in $pp$ collisions at $\sqrt{s}$ = 14 TeV and an integrated luminosity ${\cal L}=3000$ fb$^{-1}$. The scale combinations $(0.5,1.5)m_T$ and $(1,0.75)m_T$ give the upper and lower limits of the scale variation envelope obtained from the seven combinations of $(N_R,N_F)$ shown in figure \ref{fig:NutauEventLHC-scale}. 
	}
	\label{table:events1.5}
\end{table}

\begin{table}[h]
	\vskip 0.35in
	\begin{center}		    
		\begin{tabular}{|c||c|c|c||c|c|c||c|c|}
			\hline
                         &  $\nu_\tau$ &  $\bar{\nu}_\tau $  &   $\nu_\tau + \bar{\nu}_\tau $ &
                         \multicolumn{5} {c|}{ $\nu_\tau + \bar{\nu}_\tau $}  \\	
                       \hline  
		      $ (\mu_R, \ \mu_F) $  & \multicolumn{3}{c||} {(1, 1)  $m_T$} & \multicolumn{3}{c||} {(1, 1)  $m_T$} &  {(0.5, 1)  $m_T$} &  {(1, 0.5)  $m_T$} \\	                       
		      \hline	                              
		      $\langle k_T \rangle $  & \multicolumn{3}{c||} {0.7 GeV} & 0 GeV  & 1.4 GeV & 2.2 GeV & \multicolumn{2}{c|} {0.7 GeV}  \\		                 
			\hline
		 	$D_s$ & 1591 & 774  & 2365 & 2455 & 2143  & 1822 & 7834 &1779 \\ 
			\hline
			$B^{\pm,0}$& 87 & 42 & 129  & 131 & 124 & 115 & 202 & 91 \\	
			\hline
			Total & 1678 & 816 & 2494 & 2586 & 2267 & 1937 & 8036 & 1870 \\ 
			\hline
		\end{tabular}
	\end{center}
	\caption{Same as Table 1, but adopting ($\mu_R, \mu_F) = (1.0, 1.0) m_T$ as central scales in the computation of heavy-meson hadroproduction (conventional scale choice). The scale
	combinations $(0.5,1)m_T$ and $(1,0.5)m_T$ give the upper and lower limits of the scale variation envelope from the seven combinations of $(N_R,N_F)$ shown in figure \ref{fig:NutauEventLHC1}.
	}
	\label{table:events1}
\end{table}

In tables \ref{table:events1.5} and \ref{table:events1}, we present the total event numbers 
for our preferred scale choice $(\mu_R, \mu_F) = (1.0, 1.5)\,m_T$ and 
the conventional QCD scale choice $(\mu_R, \mu_F) = (1.0, 1.0)\,m_T$,
respectively, for 1 m of lead (35.6 ton).
Assuming an integrated luminosity of 3,000 fb$^{-1}$,
each table shows separately the number of $\nu_\tau$ and $\bar{\nu}_\tau$ events from $D_s^\pm$ and $B$
meson decays, respectively, and the total number, each for selected ($\mu_R$, $\mu_F$) scale and $\langle k_T\rangle$ choices. 
Table \ref{table:events1.5} shows that when using $(\mu_R, \mu_F) = (1.0, 1.5) m_T$ and the $\langle k_T\rangle$ value that yields the best match to the 
\textsc{Powheg + PYTHIA} $D_s^\pm$ energy distribution in the far forward $\eta>6.87$ region ($\langle k_T\rangle=0.7$ GeV), the total number of tau neutrino plus antineutrino events is
predicted to be $\sim 3,800$. 
Factorization and renormalization scale variations around the QCD central scales yield a very broad uncertainty band on the number of events, varying in the interval $\sim 1,800 - 11,200$. 
The large scale uncertainties indicate that contributions from missing higher orders in the perturbative calculation in collinear factorization are relatively large. On the other hand, the central scale choice, with $\langle k_T \rangle$ variation in the range $\langle k_T\rangle=0-2.2$ GeV, produces a smaller uncertainty in the number of events, which span the range $\sim 3,900-2,900$, 
with the lower end of the range of number of events corresponding to the ($\mu_R$, $\mu_F$ and $\kt$)-combination that is favored by LHCb data. Given our expectation that, for $\kt=2.2$ GeV, some effects of missing higher-order perturbative corrections are compensated by the large value of this parameter, an uncertainty in $\kt$ cannot be added to the scale uncertainty to get a total uncertainty for the number of events. In any case, the scale uncertainty yields a range of number of events that is significantly larger than the range found by varying $\kt$ from 0
to 2.2 GeV.

We set $(\mu_R, \mu_F)
= (1.0, 1.0)\,m_T$, the conventional scales used in most QCD theory evaluations of heavy-flavor production, for the predictions in Table \ref{table:events1}. This table shows as well a broad range of predicted number of events, with a central value of $\sim 2,500$ events for $\langle k_T\rangle=0.7$ GeV. This number amounts to a factor of $\sim 2/3$ of the one obtained in the evaluation with $(\mu_R, \mu_F) = (1, 1.5)\,m_T$ for $\kt=0.7$ GeV, 
however, it is comparable to the $2,900$ events for $\kt=2.2$ GeV with $(\mu_R, \mu_F) = (1, 1.5)\,m_T$.

\subsection{Muon neutrinos}

\begin{figure}[ht]
\begin{centering}
\includegraphics[width=0.48\textwidth]{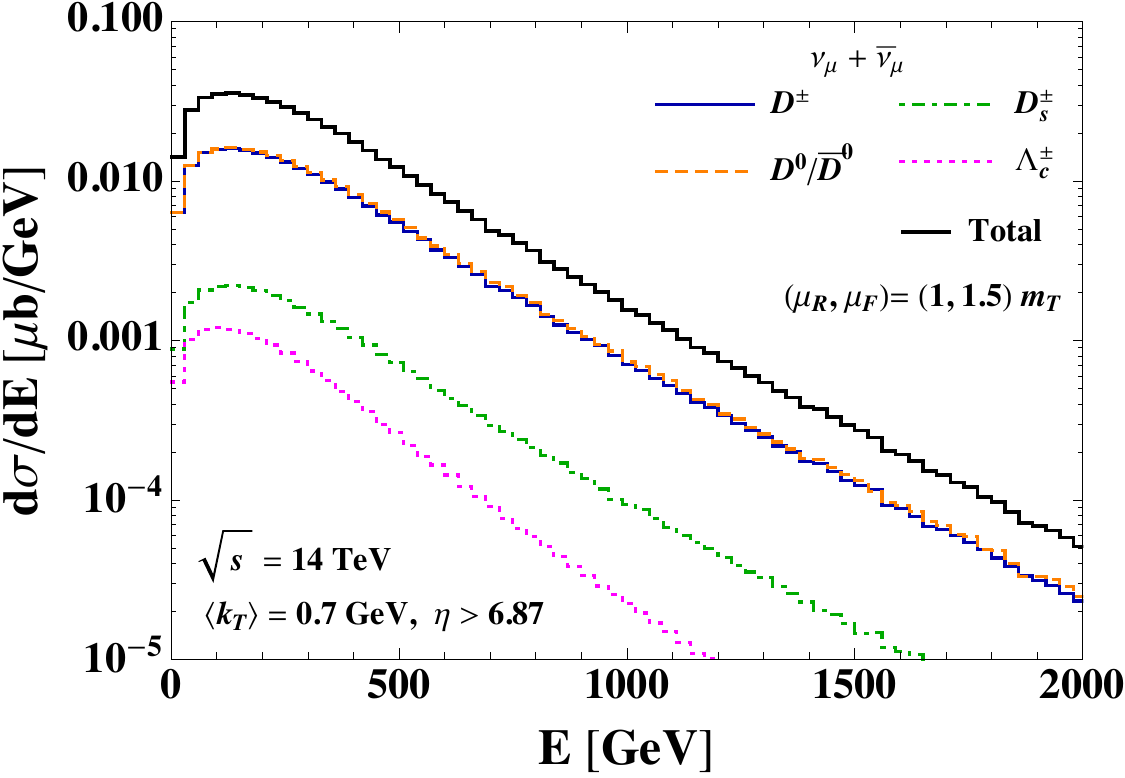} 
\includegraphics[width=0.48\textwidth]{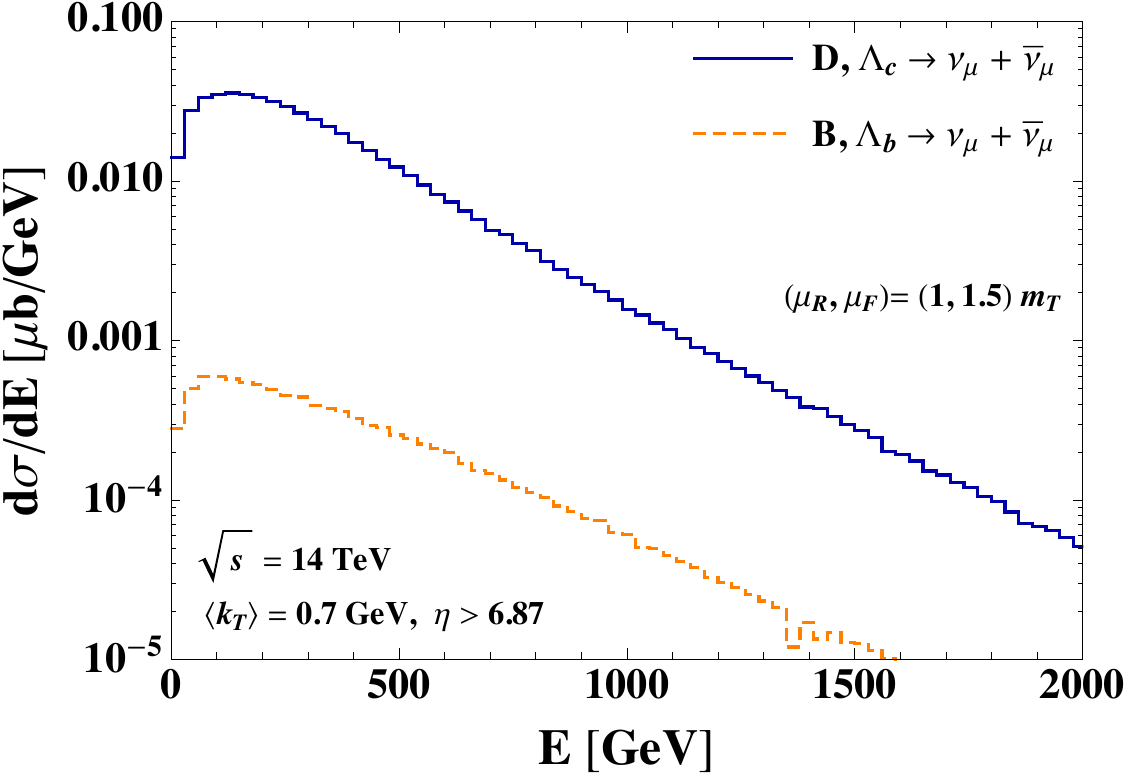} 
\includegraphics[width=0.48\textwidth]{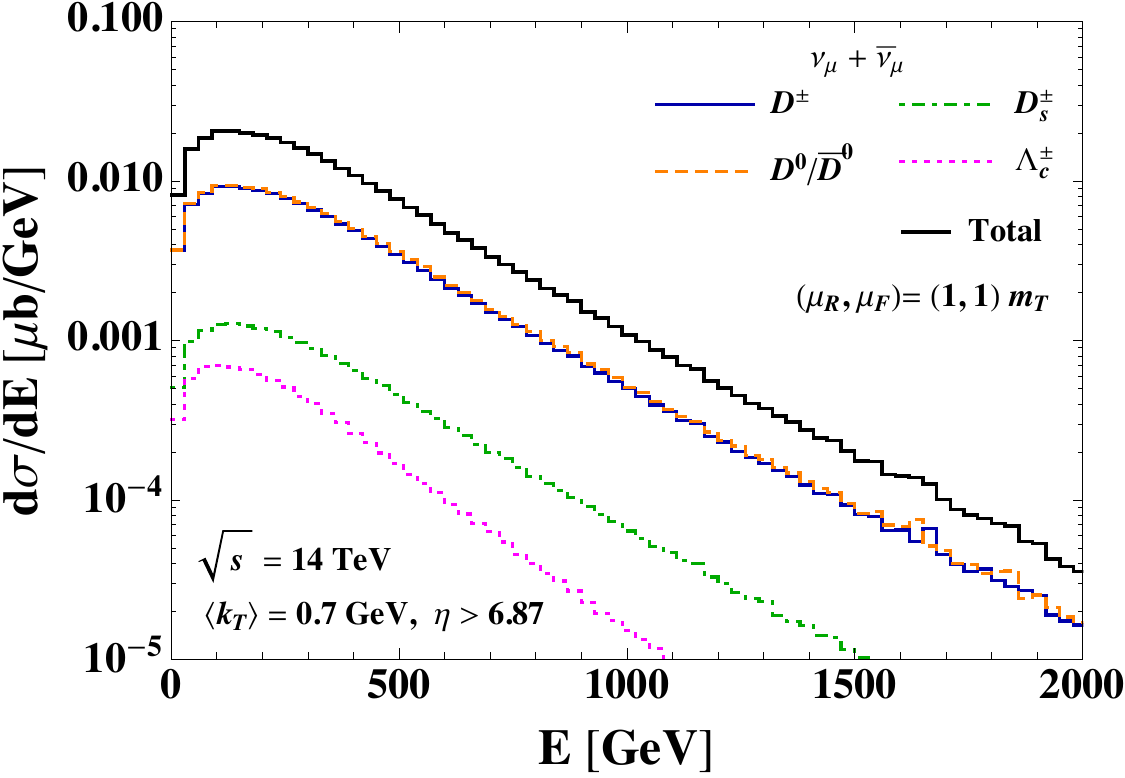} 
\includegraphics[width=0.48\textwidth]{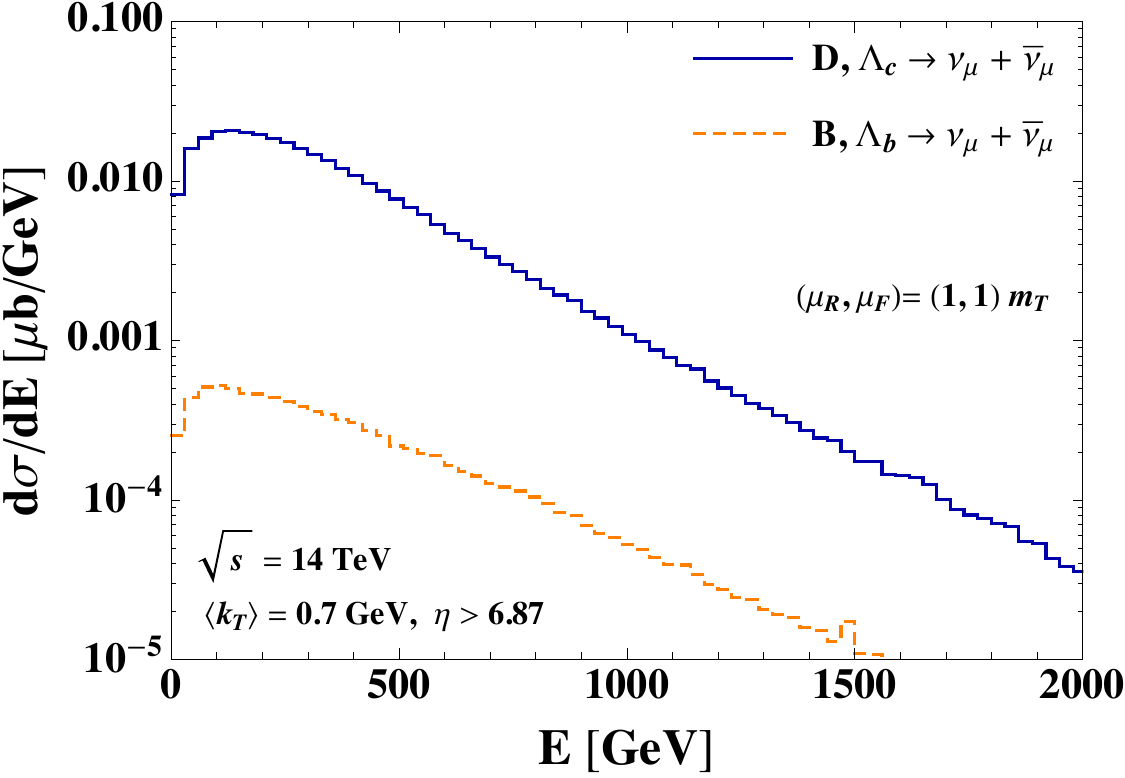} 
\par\end{centering}
\caption{Our predictions for the  
muon neutrinos plus antineutrinos energy distribution for $pp$ collisions at $\sqrt{s}=14$ TeV,  with neutrino pseudorapidity  $\eta > 6.87$. Left: Shown are the contributions from each charm hadron $D^+$, $D^{0}$, $D_{s}^+$ and
$\Lambda_{c}$ and their antiparticles, together with their sum.  
Right: Total contributions from the charm hadrons and bottom hadrons are presented, respectively. The total for $B$-hadrons accounts for the contributions from  $B^+$, $B^{0}$, $B_{s}^+$ and $\Lambda_{b}$ and their antiparticles.
 The upper plots refer to the scale choice  $(\mu_R, \mu_F) = (1.0, 1.5)\,m_T$, while the lower plots correspond to the conventional choice $(\mu_R, \mu_F) = (1.0, 1.0)\,m_T$. The value $\left\langle k_{T}\right\rangle = 0.7$ GeV is used as input in all cases.
\label{fig:NumuCS}}
\end{figure}

At the interaction region, the muon neutrino and electron neutrino
fluxes from heavy-flavor production and decay will be nearly the same, coming primarily from 
the neutral and charged $D$ semileptonic decays.
Figure \ref{fig:NumuCS} shows the energy distributions of the sum of muon neutrinos and antineutrinos. The
upper two panels are for $(\mu_R, \mu_F)
= (1.0, 1.5)\,m_T$ and the lower two panels are for $(\mu_R, \mu_F)
= (1.0, 1.0)\,m_T$.
The left plots show the contributions from the 
charm hadrons, $D^+$, $D^{0}$, $D_{s}^+$ and
$\Lambda_{c}$ and their sum.
In the right plots, the total contributions from all charm hadrons
and the bottom hadrons are shown. 
As can be seen in the left plots, the decays of $D^{\pm}$, $D^{0}$ and $\bar{D}^{0}$ dominate the muon neutrino and
antineutrino distributions. 
This is due to larger fragmentation fractions and decay branching fractions to muon neutrinos compared to those of $D_s^\pm$ and $\Lambda_c$.
Similarly, the muon neutrinos and antineutrinos from $B$ meson decays are mainly from $B^{\pm}$, $B^{0}$ and $\bar{B}^{0}$. The bottom hadron contributions, compared to the charm hadron contributions, to the inclusive muon neutrino plus antineutrino energy distribution at $\sqrt{s} = 14$ TeV with $\eta>6.87$ is about a factor $\sim 1/60$ smaller than the distributions from charm at low energy, and a factor of
$\sim 1/20$ at high energy. 
As in figure \ref{fig:NutauCS} for $\nu_\tau + \bar{\nu}_\tau$, figure \ref{fig:NumuCS} 
shows that the predicted energy distribution of $\nu_\mu +\bar{\nu}_\mu$ from charm is $\sim 1.5$ larger for $(\mu_R, \mu_F)
= (1.0, 1.5)\,m_T$ than for $(\mu_R, \mu_F)
= (1.0, 1.0)\,m_T$, while the smaller contributions from $B$ hadrons
are much less sensitive to the scale choice.

\begin{figure}
\begin{centering}
\includegraphics[width=0.48\textwidth]{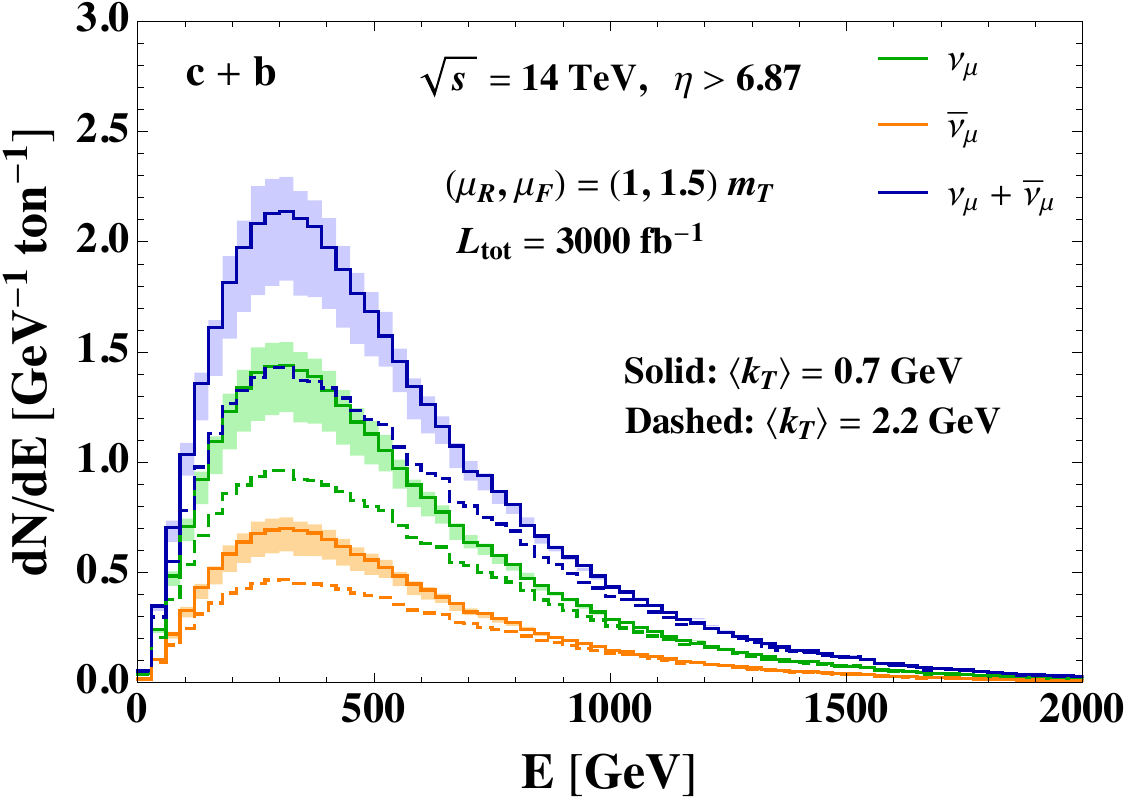} 
\includegraphics[width=0.48\textwidth]{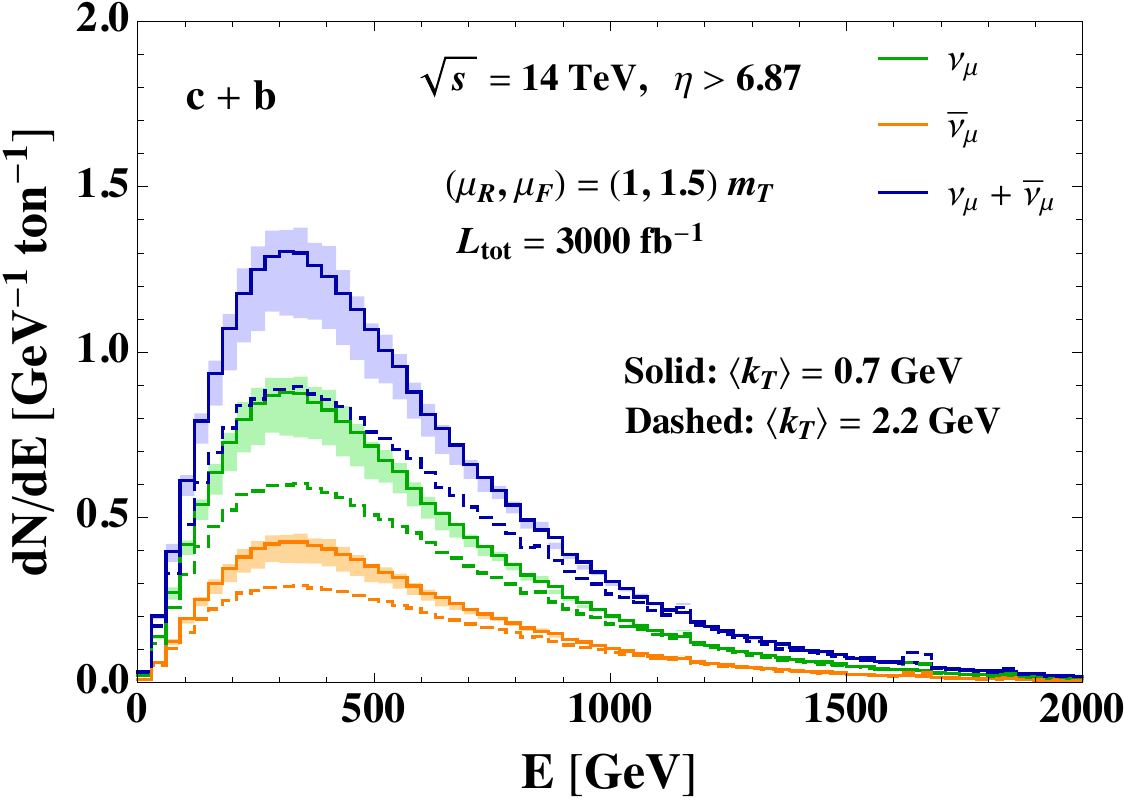} 
\par\end{centering}
\caption{Our predictions for the muon neutrino and antineutrino number of charged current events per GeV for 1 ton of lead target as a function of the incident neutrino energy for neutrinos with a pseudorapidity $\eta>6.87$ generated by heavy-flavor decays in $pp$ collisions with $\sqrt{s}=14$ TeV. 
A value of $\left\langle k_{T}\right\rangle$ = 0.7 GeV is adopted for the central predictions, while the uncertainty bands accounts for the effect of $\left\langle k_{T}\right\rangle$ variation in the interval $0 -1.4$ GeV. NLO QCD corrections are accounted for in the DIS cross section. The integrated luminosity amounts to ${\cal L}=3000$ fb$^{-1}$.
 \label{fig:NumuEventLHC}}
\end{figure}

\begin{figure}
\begin{centering}
\includegraphics[width=0.48\textwidth]{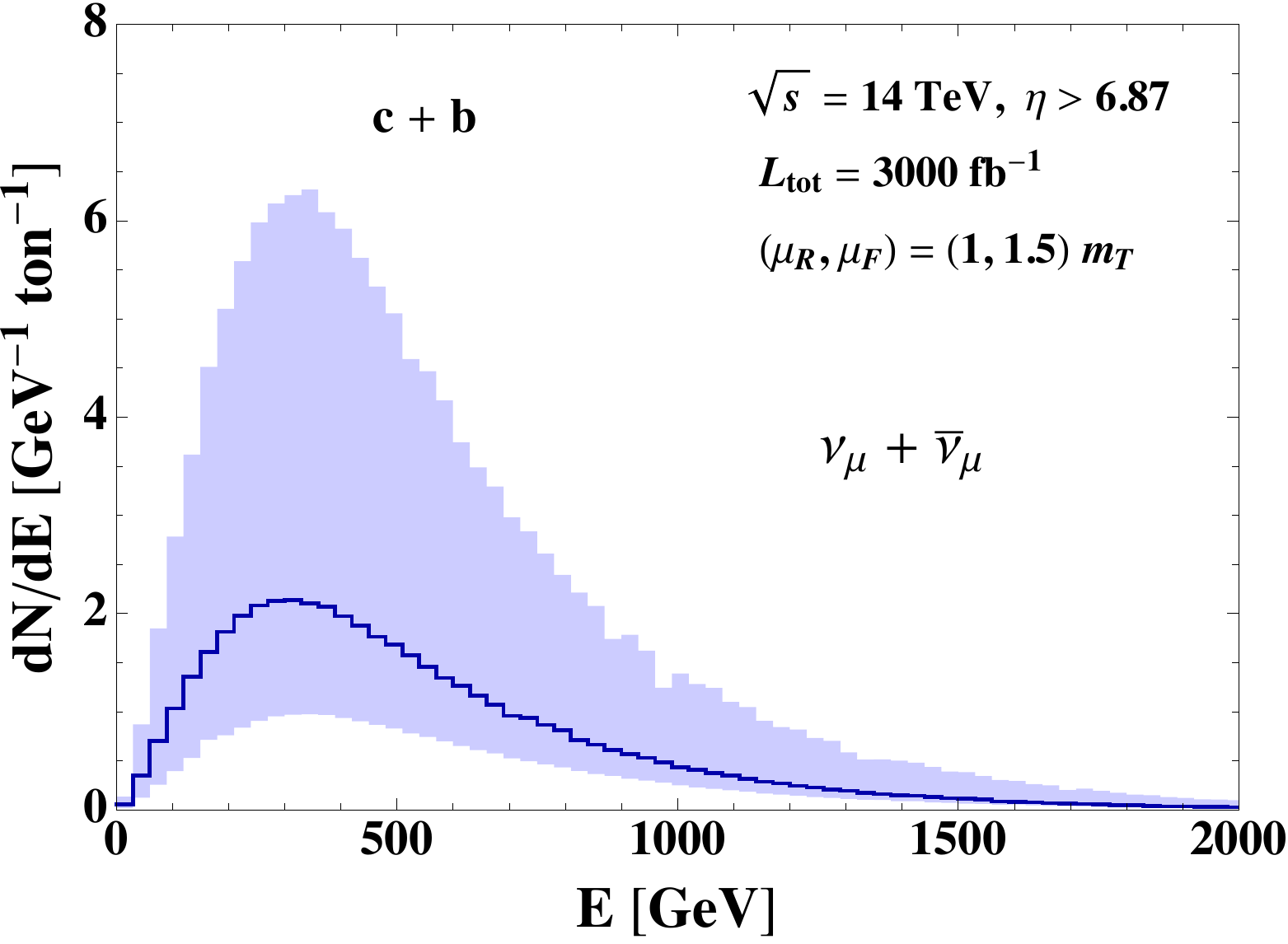}
\includegraphics[width=0.48\textwidth]{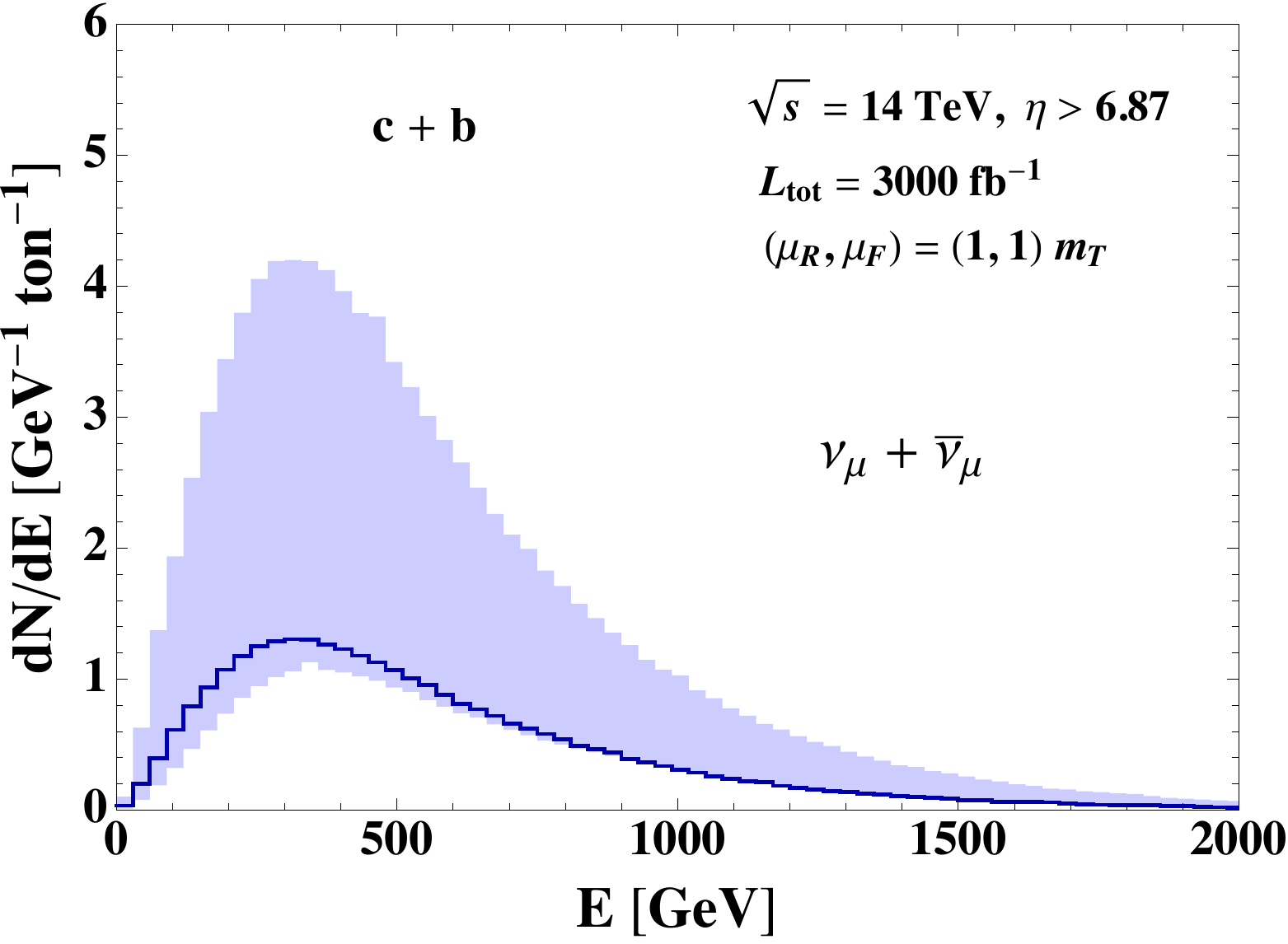}
\includegraphics[width=0.48\textwidth]{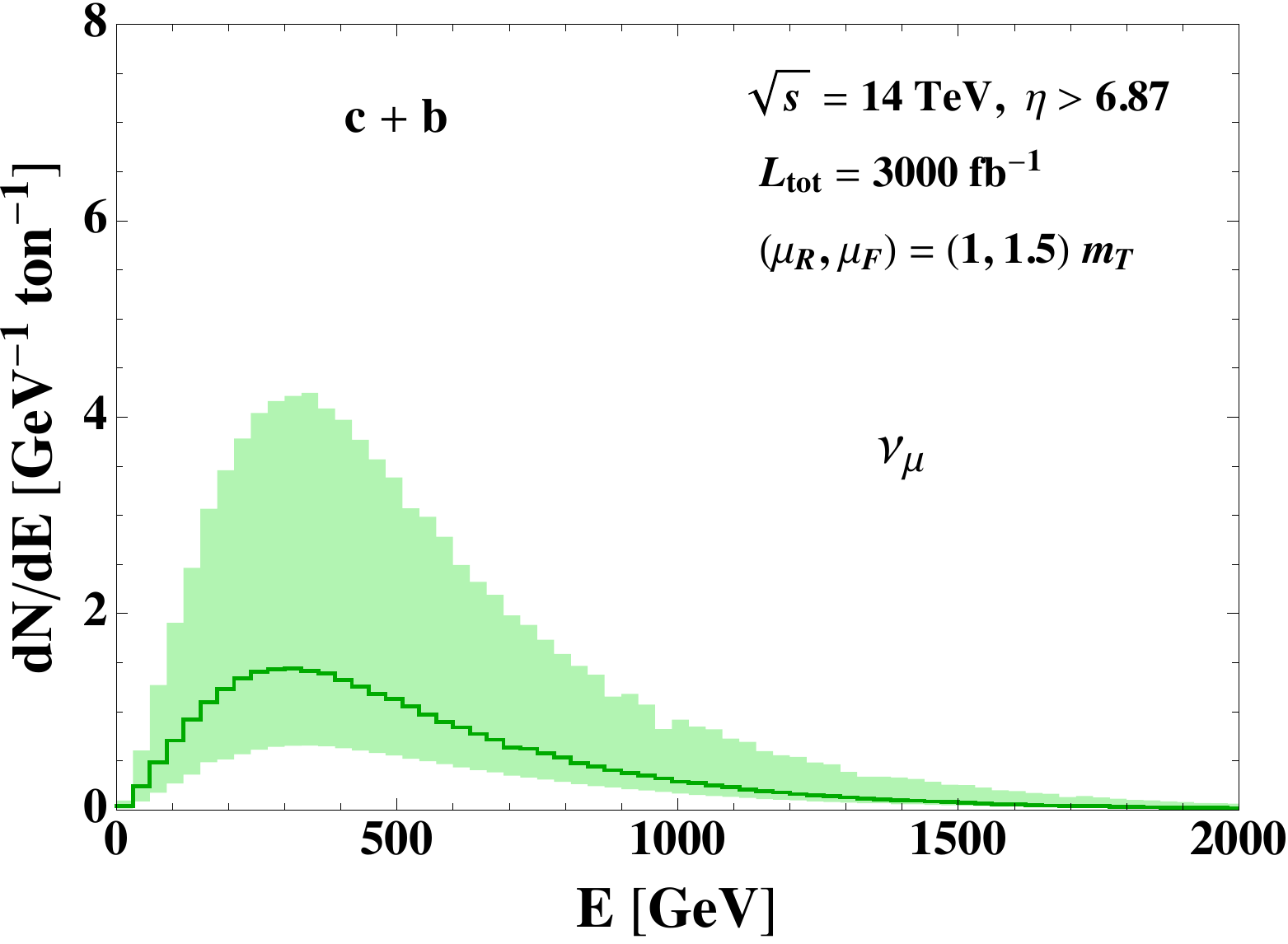} 
\includegraphics[width=0.48\textwidth]{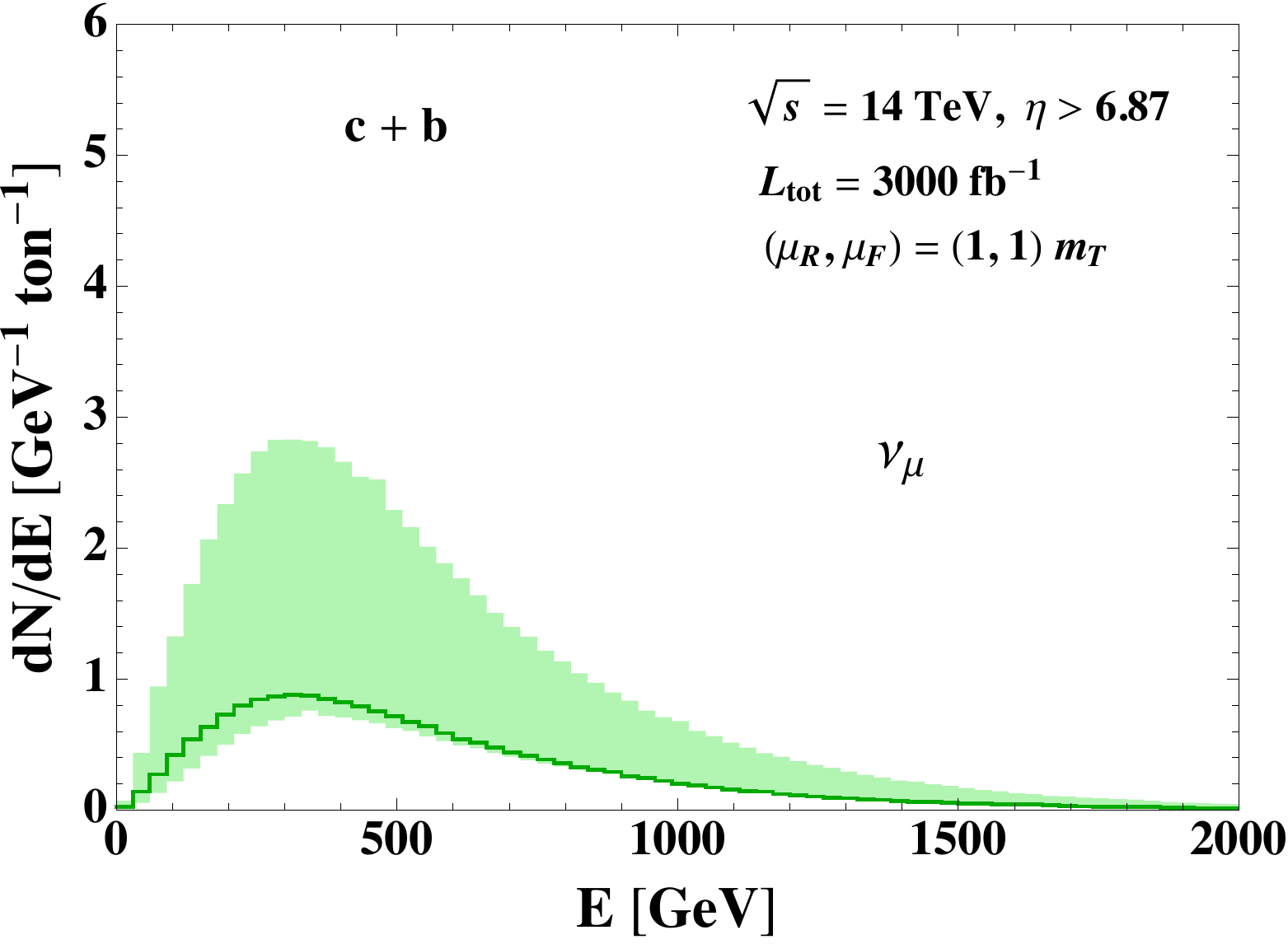} 
\includegraphics[width=0.48\textwidth]{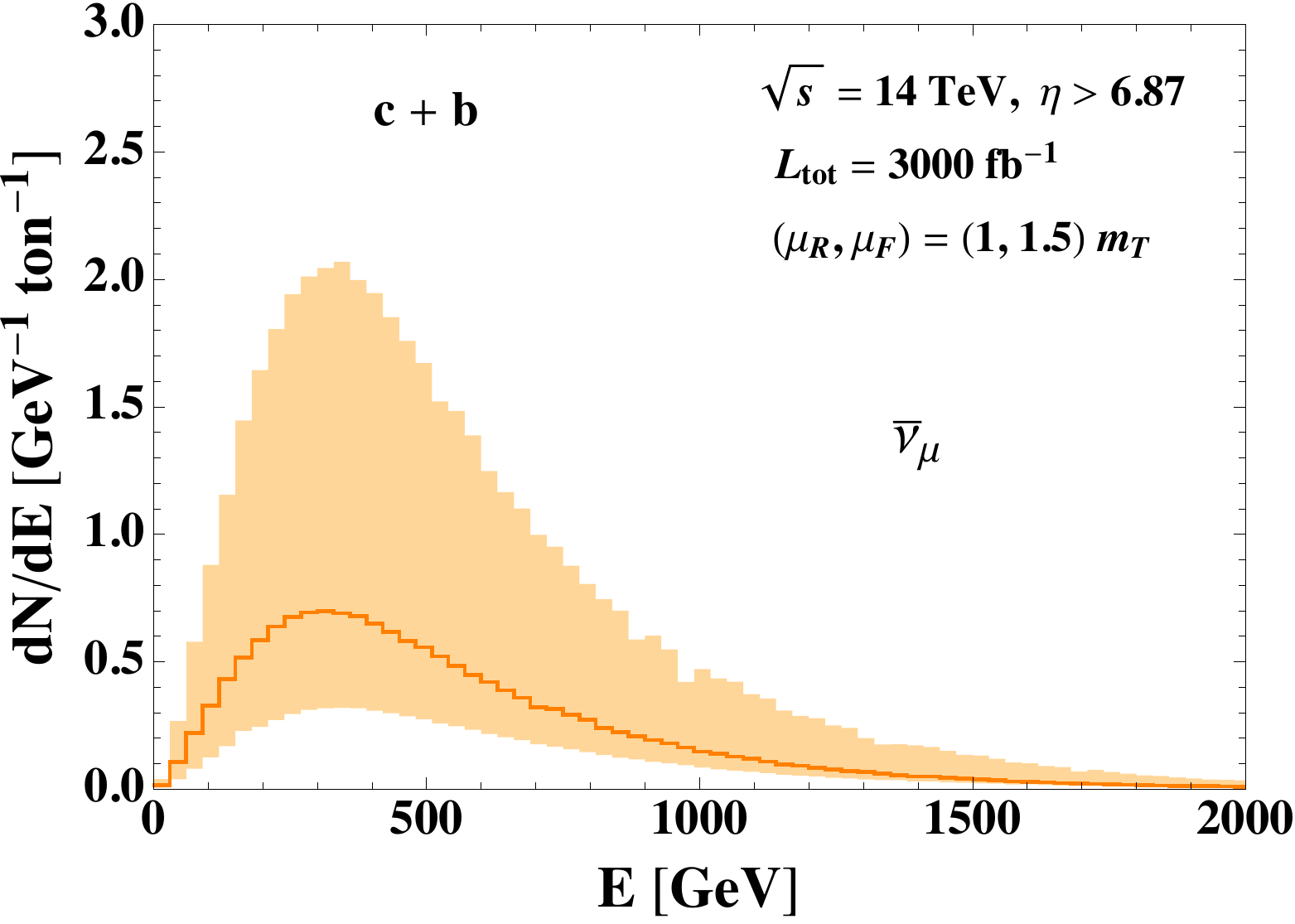}
\includegraphics[width=0.48\textwidth]{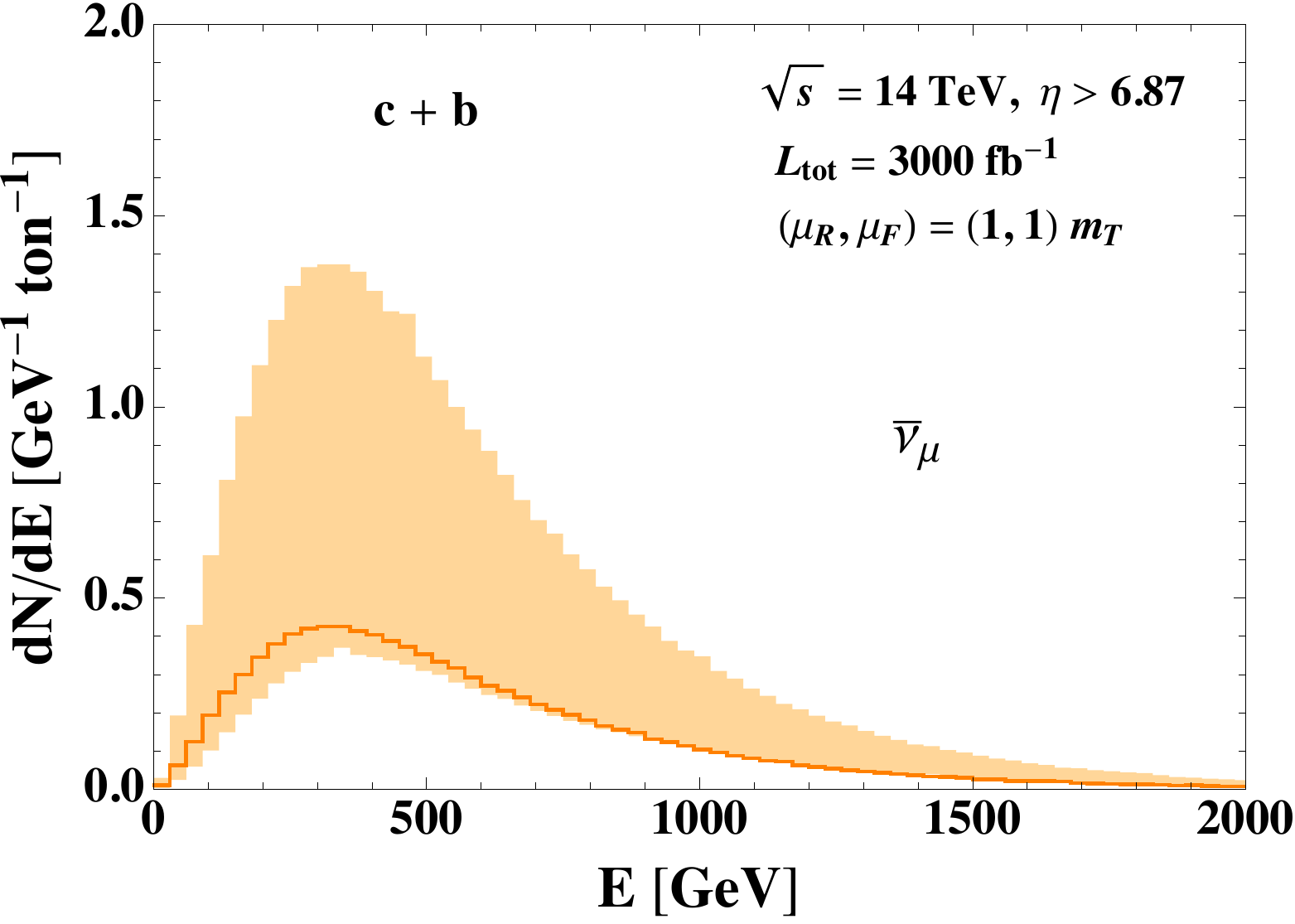}\par\end{centering}
\caption{Uncertainty range due to the QCD scale variation in the muon neutrino and antineutrino number of charged current events per GeV per ton as a function of the incident neutrino energy for neutrinos with pseudorapidity $\eta>6.87$ generated by heavy-flavor decays in $pp$ collisions with $\sqrt{s} = 14$ TeV and integrated luminosity ${\cal L}=3000$ fb$^{-1}$.
The central predictions are obtained using as input $(\mu_R, \mu_F) = (1, 1.5)\,m_T$ (left) and $(\mu_R, \mu_F) = (1.0, 1.0)\,m_T$ (right), respectively. The upper and lower limits in each panel arise from 7-point scale variation in the same range as in figure \ref{fig:fit-LHCb}.
\label{fig:NumuEventLHC1.5}}
\end{figure}

The corresponding predictions for the number of
muon neutrino and antineutrino charged-current events per unit energy for a 1 ton lead target for
$\eta>6.87$ are shown in figures \ref{fig:NumuEventLHC} and \ref{fig:NumuEventLHC1.5}. 
Figure \ref{fig:NumuEventLHC} shows the number of muon neutrino and muon antineutrino  events per unit energy from heavy flavor, including uncertainty bands from $\langle k_T\rangle$ variation in the range [0, 1.4] GeV for $(\mu_R, \mu_F) = (1.0, 1.5)\, m_T$.
Again, the dashed histograms correspond to results with $\langle k_T\rangle=2.2$ GeV. The left panels
in figure \ref{fig:NumuEventLHC1.5} show, from top to bottom, the sum
of $\nu_\mu+\bar{\nu}_\mu$ charged current events per ton of lead, for $\nu_\mu$ and
for $\bar{\nu}_\mu$, all for $(\mu_R, \mu_F) = (1.0, 1.5)\, m_T$. The right
panels show the same, but with $(\mu_R, \mu_F) = (1.0, 1.0)\, m_T$. Each
panel includes a wide uncertainty band reflecting the theoretical
uncertainties associated with scale variation. Our evaluation using NLO perturbative QCD gives a factor of
$\sim 13$ in the ratio of $\nu_\mu+\bar{\nu}_\mu$ to
$\nu_\tau+\bar{\nu}_\tau$ events based on heavy flavor alone, with our default input parameters.

Heavy-flavor hadron decays, however, are not the only sources of $\nu_\mu+\bar{\nu}_\mu$. In principle, pion, kaon and weak gauge boson decays can also contribute.
In ref. \cite{Beni:2019gxv}, the contributions from $W$
and $Z$ production and decay to neutrinos are studied. They show that neutrinos from weak gauge boson decays populate mostly in the pseudorapidity
range of $|\eta|<4.5$, but are negligible in the region $\eta>6.5$
\cite{Beni:2019gxv}.

At first glance, one might expect that 
pion and kaon decays will not be important sources of neutrinos. Pions with $E_\pi>9$ GeV have $\gamma c \tau>500$ m. Charged kaons have $\gamma c \tau>500$ m for $E_K>67$ GeV.
Pion and kaon contributions to the number of $\nu_\mu+\bar{\nu}_\mu
$ at high energy were neglected in the earlier work of
ref. \cite{DeRujula:1992sn}. Park, in ref. \cite{Park:2011gh}, used \textsc{Pythia} to
account for pions and kaons, requiring the decay to occur within 50 m
of the interaction point to guarantee that particles decay inside the
beam pipe, leading to a factor of $\sim 100$ times
more $\nu_\mu+\bar{\nu}_\mu$ events than $\nu_\tau+\bar{\nu}_\tau$
events. 

In ref. \cite{Abreu:2019yak}, a more detailed evaluation of $\nu_\mu+\bar{\nu}_\mu$ events
is performed for a detector with $25\times 25$ cm$^2$ cross-sectional area at 480 meters from the \textsc{Atlas} interaction point.
They find a factor of $\sim 1000$ more interactions by $\nu_\mu+\bar{\nu}_\mu$ than by
$\nu_\tau+\bar{\nu}_\tau$  when pions and kaons are included, although with a different energy distribution.  Their evaluation of heavy-flavor contributions is done using \textsc{Pythia} \cite{Sjostrand:2019zhc}, while their light hadron production is estimated using the \textsc{Crmc} \cite{crmc} simulation package with \textsc{Epos-LHC} \cite{Pierog:2013ria}, \textsc{Qgsjet-II-04 } \cite{Ostapchenko:2010vb} and \textsc{Sibyll} 2.3c
\cite{Engel:2019dsg,Fedynitch:2018cbl,Riehn:2017mfm}. 
They
find that most of the neutrinos with energies above 1 TeV from charged light hadron decays come from pion and kaon decays in a region within $\sim 55$ m from the \textsc{Atlas}
interaction point.
They find that magnetic fields sweep lower-energy (below $\sim 100$ GeV) charged particles away
if the particles have travelled 20 m downstream from the interaction point and have passed through
the front quadrupole absorber of inner radius 17 mm \cite{Abreu:2019yak}. 
They also find that two-body decays of charged pions and charged kaons are the dominant
sources of $\nu_\mu+\bar{\nu}_\mu$ production.

Our primary focus here is on the heavy-flavor contributions to the number of events from
$\nu_\mu+\bar{\nu}_\mu$ and $\nu_\tau+\bar{\nu}_\tau$.
Instead of simulating light charged hadron trajectories in magnetic fields, guided
by the results of ref. \cite{Abreu:2019yak}, we approximately
evaluate the number of $\nu_\mu+\bar{\nu}_\mu$ that come from pions and kaons as follows. We
evaluate the
$\pi^\pm  $ and $K^\pm $ two-body decay contributions to the flux of
$\nu_\mu+\bar{\nu}_\mu$
with the requirement that the mesons decay within 55 m of the
interaction point and the decaying hadron's momentum
lies within an angle 
of $\theta< 1$ mrad relative to the beam axis to stay within the opening of the quadrupole absorber. While the 2-body light meson decays dominate, a more complete calculation of the light meson contributions to $\nu_\mu+\bar{\nu}_\mu$ would include $K_L$ semileptonic decays.

\begin{figure}
\begin{centering}
\includegraphics[width=0.48\textwidth]{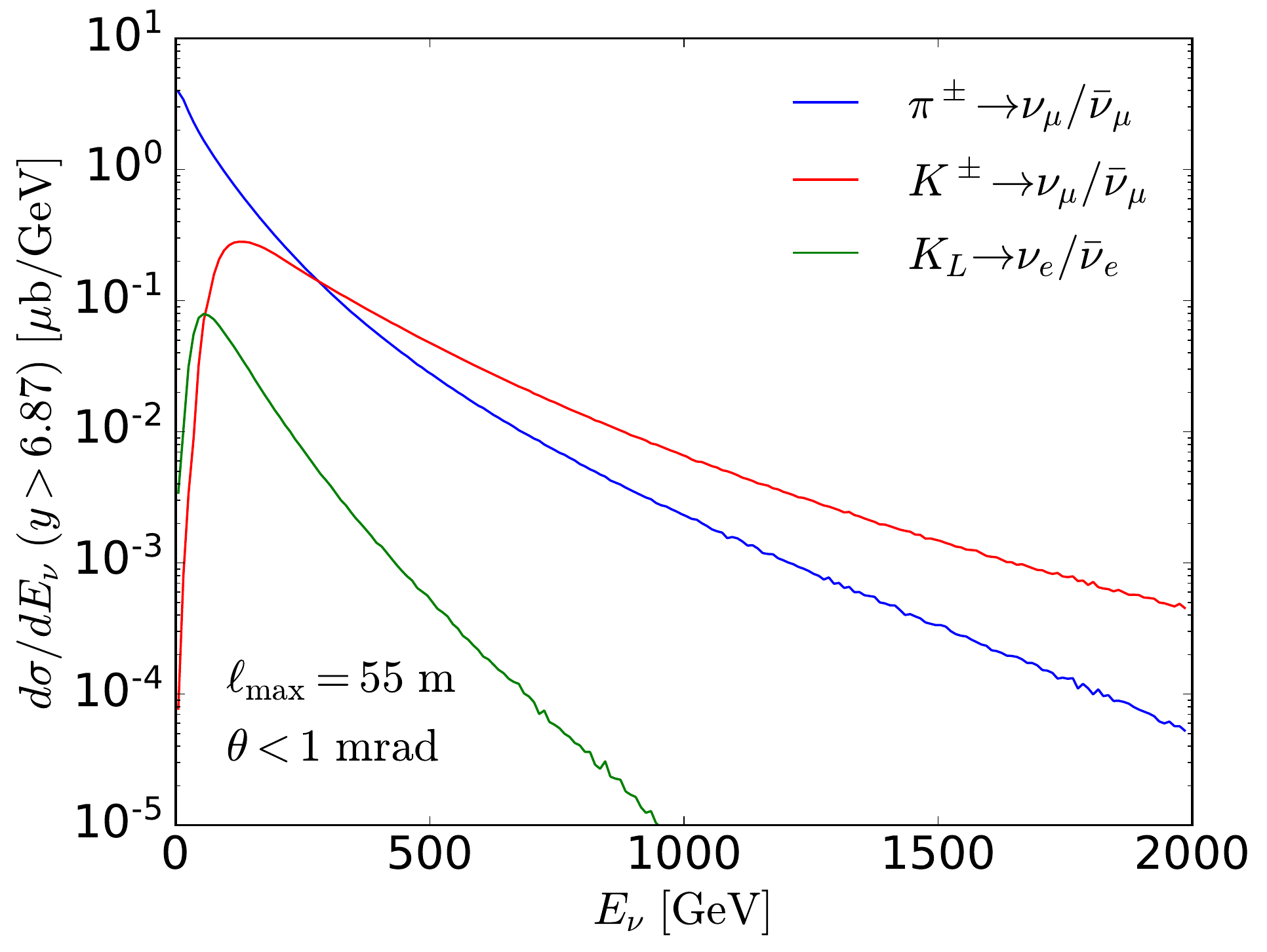} 
\par\end{centering}
\caption{The $\nu_\mu+\bar{\nu}_\mu$ energy distributions from charged pions (blue) and kaons (red) and the $\nu_e+\bar{\nu}_e$ energy distributions from $K_L$ semileptonic decays (green) for forward neutrinos ($\eta>6.87$) from the decay of mesons produced in $pp$ interactions at $\sqrt{s}=14$ TeV, considering those mesons whose decay occurs within 55 m of the interaction point and whose momentum lies within an angle of $\theta<1$ mrad from the beam axis. The meson energy and rapidity distributions are evaluated using the parametrization of Koers et al. \cite{Koers:2006dd}. 
 \label{fig:pionkaon}}
\end{figure}

We use
the parametrization of Koers et al. \cite{Koers:2006dd}, based on fits
to \textsc{Pythia} distributions for charged pions and kaons as a function of
energy and rapidity, to generate the pion and kaon distributions.
For reference, pion and kaon 
charged-particle multiplicities per interaction in $pp$ collisions at $\sqrt{s}=14$ TeV are $\sim
50$ and $\sim 6$, respectively \cite{Koers:2006dd}.
Two-body decays are implemented, as for the $D_s^\pm \to \nu_\tau \tau$ case, with
the requisite changes to the initial and final particle masses.  Compared to figure \ref{fig:NumuCS},
charged pions and kaons contribute a factor of $\sim 100$ more $\nu_\mu+\bar{\nu}_\mu$ than
heavy flavors do, as shown by the blue and red curves in figure \ref{fig:pionkaon}.
We will turn to the issue of the oscillations of neutrinos of different flavors to tau neutrinos.  To estimate the number of
$\nu_e+\bar{\nu}_e$, we consider the dominant contribution which is from $K_L\to \pi e\nu_e$. With the same geometry requirements, we show with the green curve in figure \ref{fig:pionkaon} the contribution from $K_L$ production and decay into $\nu_e+\bar{\nu}_e$.  We use the Koers distributions for $K^+ +K^-$, then divide by two for $K_L$. The three-body semileptonic kaon decay is evaluated following ref. \cite{Cirigliano:2001mk}. The peak of the electron neutrino distribution from $K_L$ decays is about a factor of $\sim 2$ larger than
the peak of the electron neutrino distribution 
from heavy-flavor decays. While $K_L$ semileptonic decays to $\nu_e$ are dominant, a more complete calculation would include the $K^+\to \nu_e$ semileptonic decays as well.

Three-flavor neutrino oscillations of the much larger number of $\nu_\mu+\bar{\nu}_\mu$ from charged pions and kaons to $\nu_\tau+\bar{\nu}_\tau$ could, in principle, overwhelm the number of $\nu_\tau+\bar{\nu}_\tau$ from heavy-flavor decays. However, the baseline of 480 m is not at all optimal for $\nu_\mu\to\nu_\tau$ oscillations in the standard scenario with 3 active flavors. 
In two flavor approximation, for a mass squared difference of $\Delta m_{32}^2\simeq 2.5\times 10^{-3}$ eV$^2$ and a mixing angle $\theta_{23}\simeq \pi/4$, the oscillation probability is 
$P(\nu_\mu\to \nu_\tau)\simeq 2.3\times 10^{-6}/(E^2/{\rm GeV^2}$), so for energies above $E_{\nu_\mu}=10$ GeV, $\nu_\mu\to \nu_\tau$ oscillations give a negligible
contribution to the number of $\nu_\tau+\bar{\nu}_\tau$, 
even given the large theoretical uncertainties in the number of $\nu_\tau+\bar{\nu}_\tau$ events per unit energy. The $\nu_e\to \nu_\tau$ oscillation probability is even smaller.

In the next section, we use the approximate numbers of 
electron and muon neutrinos from light mesons, together with our heavy flavor results for all three neutrino flavors and turn to tau neutrino oscillations in a 3+1 neutrino mixing framework. This  illustrates an example of a signal of new physics that could be probed by a forward neutrino detector at the LHC when heavy-flavor uncertainties are under better theoretical control.
While not necessary for an analysis of $\nu_\tau+\bar{\nu}_\tau$ because tau neutrinos come from $D_s$ and $B$ decays, for an analysis of all three neutrino flavors, a full accounting of light meson production and decay to neutrinos, using all available data on forward production (e.g., from LHCf \cite{Adriani:2015iwv}, TOTEM
\cite{Antchev:2014lez} and CMS \cite{Sirunyan:2017nsj}) along with magnet and detector configurations in the interaction region, will be necessary. A forward tune of \textsc{Pythia} that is underway \cite{Abreu:2020ddv} would also guide future work.

%% file: newphysics.tex
\section{New Physics}
\label{sec:newphysics}

The detection of a large number of identifiable $\nu_\tau+\bar{\nu}_\tau$ events would offer opportunities
to explore a new corner of parameter space for neutrino oscillations into a fourth ``sterile'' neutrino. The baseline of 480 m is most sensitive to a fourth neutrino mass $m_4$ of the order of tens of eV.

\begin{figure}
\begin{centering}
\includegraphics[width=0.7\textwidth]{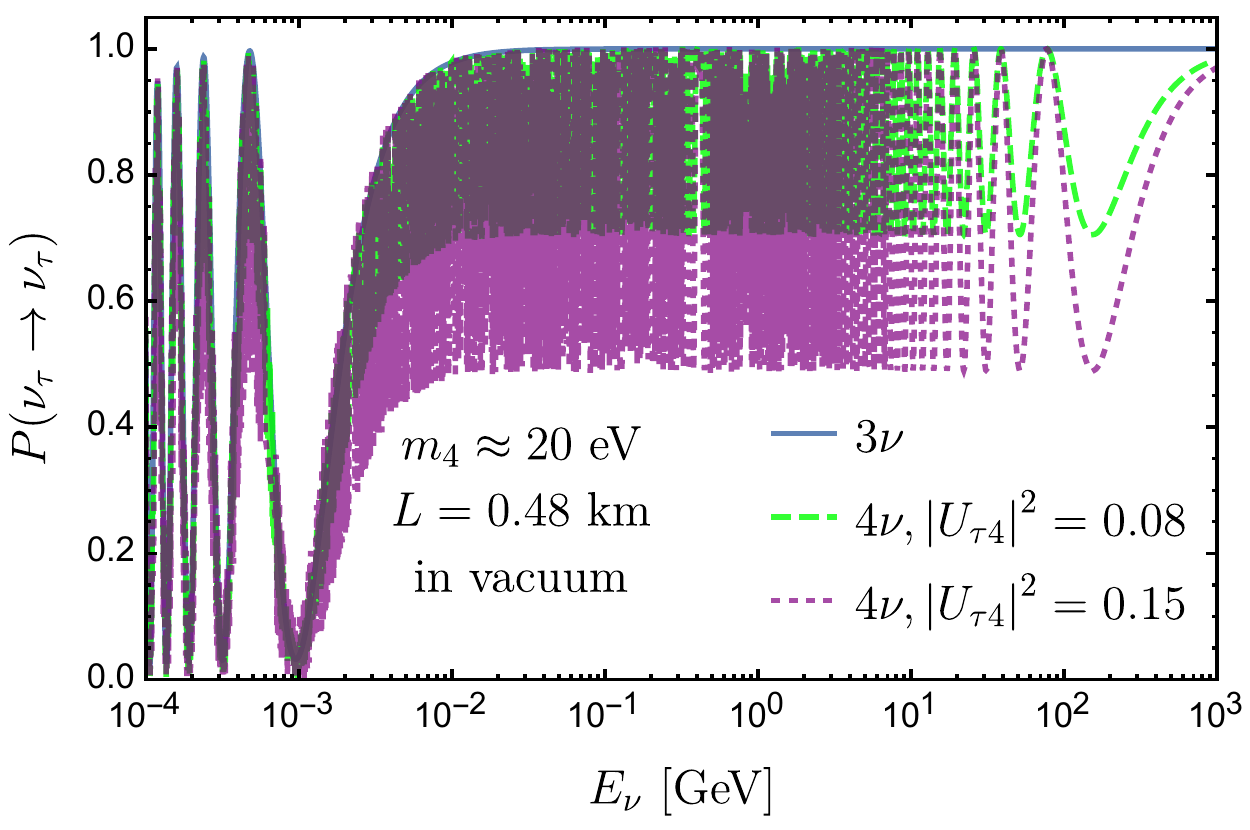}
\par\end{centering}
\begin{centering}
\includegraphics[width=0.7\textwidth]{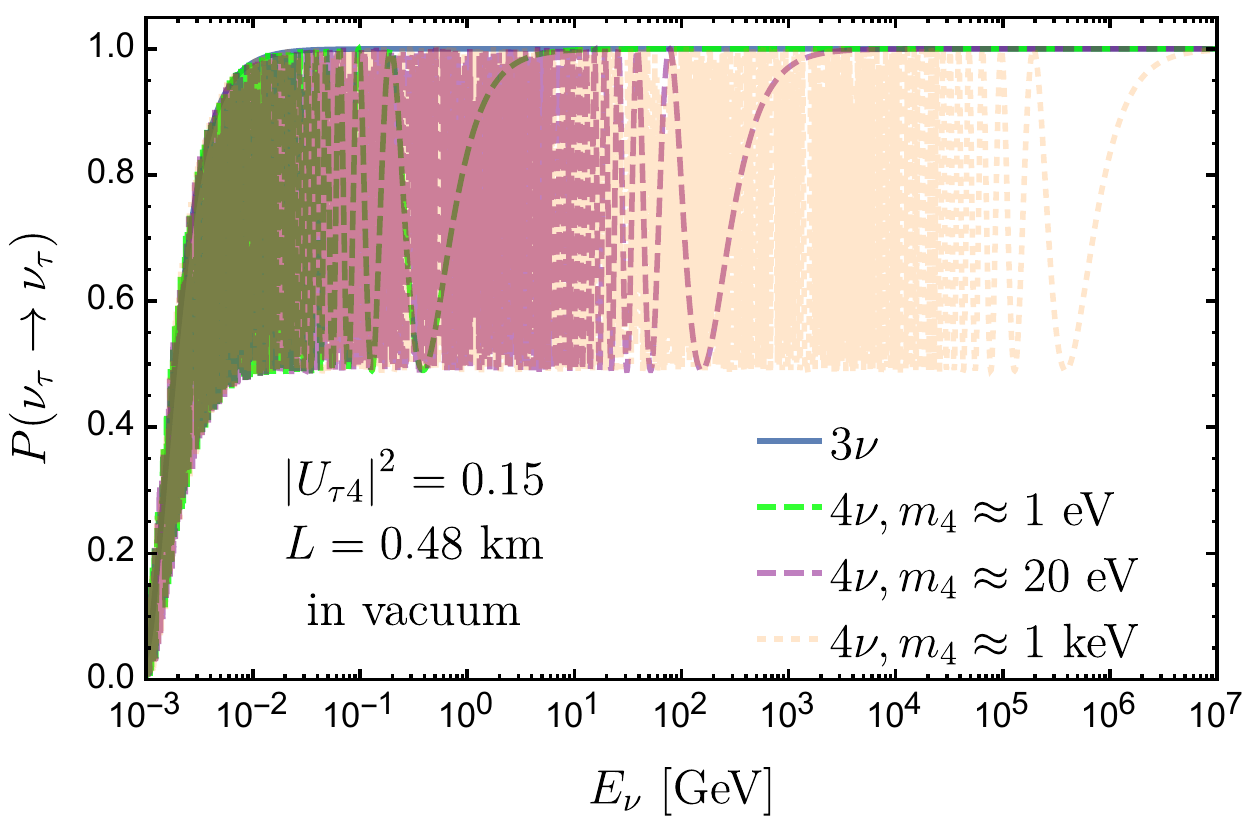}
\par\end{centering}
\caption{The survival probability
$P(\nu_\tau\to \nu_\tau)$ as a function of the tau neutrino energy $E_\nu$, for the standard 3-active-flavor oscillation framework \cite{pdg:2016} and in a $3+1$ oscillation scenario, considering a baseline of $L=480$ m. Besides results in the standard framework, the upper panel shows results for the 3+1 scenario with $m_4 = 20$ eV, for $|U_{\tau 4}|^2=0.08$ and 0.15, whereas
the lower panel shows results for the 3+1 scenario with $|U_{\tau 4}|^2=0.15$ and $m_4=1$ eV, 20 eV and 1 keV. \label{fig:dips-nuOsci}}
\end{figure}

The flavor eigenstates
$\nu_{l}$ can be expressed as a superposition of the mass eigenstates $\nu_{j}$ according to the formula
\begin{equation}
\nu_{l}=\sum_{j = 1}^{n_{\nu}}U_{lj}\nu_{j}\ ,
\end{equation}
where $n_{\nu}$ is the total number of neutrinos subject to oscillations: for the standard oscillation scenario with three active neutrinos, $n_{\nu}=3$, while for an oscillation scenario with three active neutrinos plus one sterile neutrino, $n_{\nu}=4$. 
The full transition probability appears in, for example, ref. \cite{pdg:2016}. 
The blue lines in the two panels of figure
\ref{fig:dips-nuOsci} show that tau neutrino survival probability
$P(\nu_\tau\to \nu_\tau)$ in the three-flavor scenario approaches unity in the energy range $E_{\nu}\in[1,1000]$
GeV. The values of the parameters used here for three-flavor mixing are 
$\sin^{2}(\theta_{12})=0.310$,
$\sin^{2}(\theta_{13})=0.02241$, $\sin^{2}(\theta_{23})=0.580$, $\Delta m_{21}^{2}=7.5\times10^{-5}$
eV$^{2}$ and $\Delta m_{31}^{2}=2.457\times10^{-3}$ eV$^{2}$ \cite{Esteban:2018azc}. The values of all CP phases are taken to be zero.
For a baseline of 480 m, the tau neutrino survival probability has significant features in the few MeV energy range and below, but not in the energy range of interest (see the position of the peaks of the energy distributions in section 4).

The transition
probability in the scenario with three active and one sterile neutrino flavor, in the case when the mass eigenstates fulfill the hierarchy $m_4\gg  m_{1,2,3}$, can be simplified to \cite{Giunti:2019aiy}
\begin{equation}
P(\nu_{\alpha}\rightarrow\nu_{\beta})  \simeq  \delta_{\alpha\beta}-4(\delta_{\alpha\beta}-\mid U_{\beta n_\nu}\mid^2)\mid U_{\alpha n_\nu}\mid^2 \sin^2\Biggl(\frac{\Delta m^{2}L}{4E_{\nu}}\Biggr)\ ,
\label{eq:probability}
\end{equation}
where $\Delta m^2=m_4^2-(m_1^2+m_2^2+m_3^2)/3\simeq m_4^2$.
For $m_4\sim 20$ eV, oscillation effects in the tau neutrino survival probability can be pronounced. The highest energy oscillation node with this hierarchy occurs at 
\begin{equation}
E_{\nu\textrm{-max [GeV]}}=  \frac{\Delta m^{2}L}{2\pi} = 0.807\,\Delta m_{\textrm{[eV]}}^{2}L_{\textrm{[km]}}\ .\label{eq:Enu-max}
\end{equation}
Introducing a heavy sterile neutrino extends the region of pronounced oscillation dips in 
the tau neutrino survival probability
to a higher energy 
with examples shown in figure \ref{fig:dips-nuOsci}. When $m_4 = 20$ eV, a survival probability dip occurs at $E_{\nu_\tau} = 155$ GeV, an energy near that of the peak of the unoscillated tau neutrino number of events per unit energy. Using eq. 
(\ref{eq:probability}) and $|U_{\tau 4}|^2 = 0.08$ and 0.15, the $\nu_\tau\to\nu_\tau$ survival probability is shown for $m_4 = 20$ eV (upper panel) and $m_4 = 1$ eV, 20 eV and 1 keV (lower panel), all for a baseline of $L = 480$ m.
The values of $|U_{\tau 4}|^2$ are acceptable according to current IceCube constraints
\cite{Jones:2019nix,Blennow:2018hto}.
Matter effects can be neglected with the density of the Earth's
crust $\rho\approx2.6$ g/cm$^{3}$ and the electron fraction $Y_{e}\approx0.5$
\cite{Ohlsson:1999xb,Li:2018ezt}.

As the lower panel of figure \ref{fig:dips-nuOsci} shows, 
for increasing $m_4$ masses,
oscillations become rapid over the full neutrino energy range, even at low energies, so 
their average determines the $\nu_\tau$ survival probability.
Given the uncertainties in the absolute scale of the $\nu_\tau + \bar{\nu}_\tau$ flux in the very forward direction, an average decrease in the number of events due to oscillations into sterile neutrinos
would be difficult to extract with measurements of forward LHC neutrinos. For this reason, we focus on our example of the case $m_4 = 20$ eV.

\begin{figure}
\begin{centering}
\includegraphics[width=0.49\textwidth]{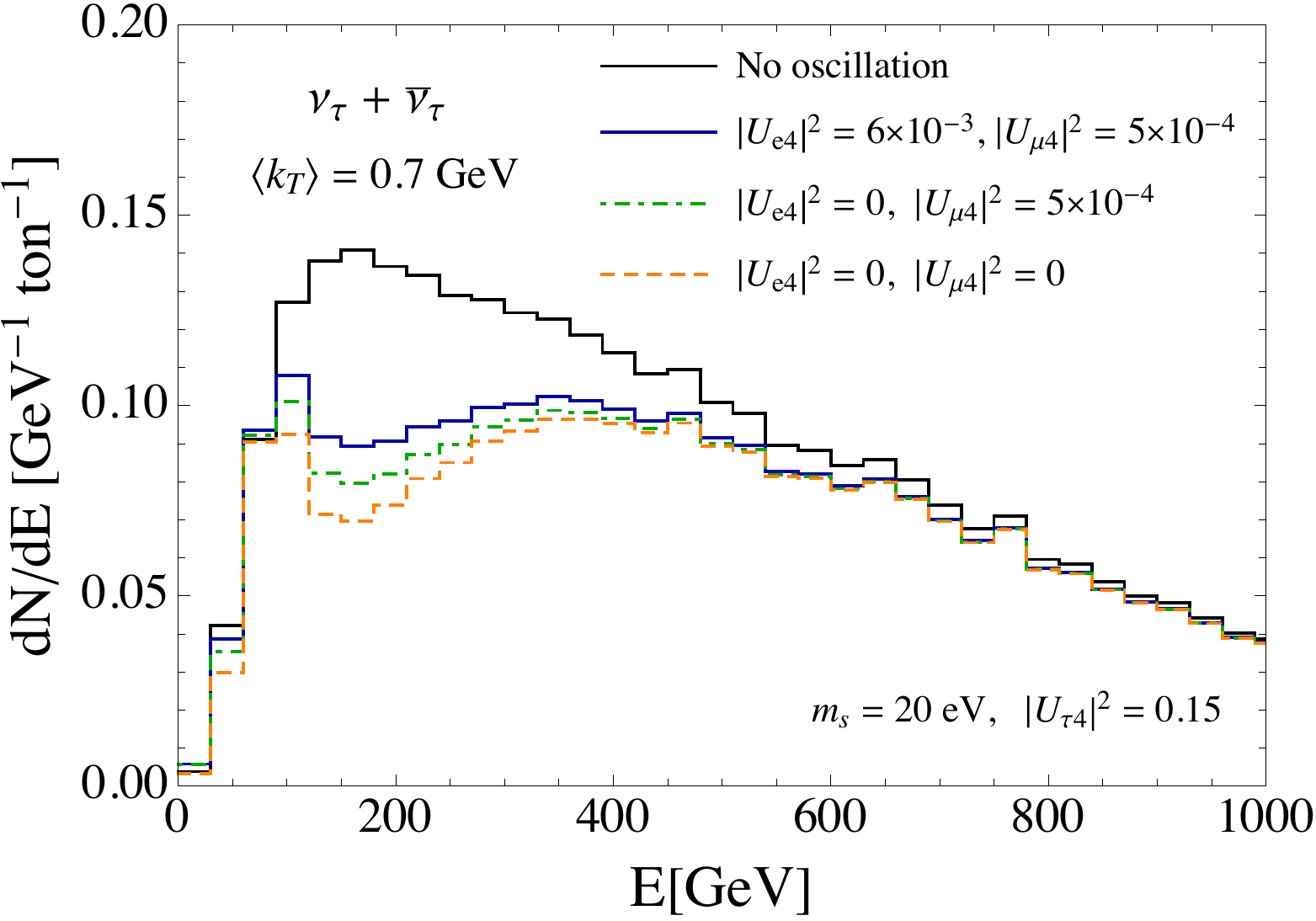}
\includegraphics[width=0.49\textwidth]{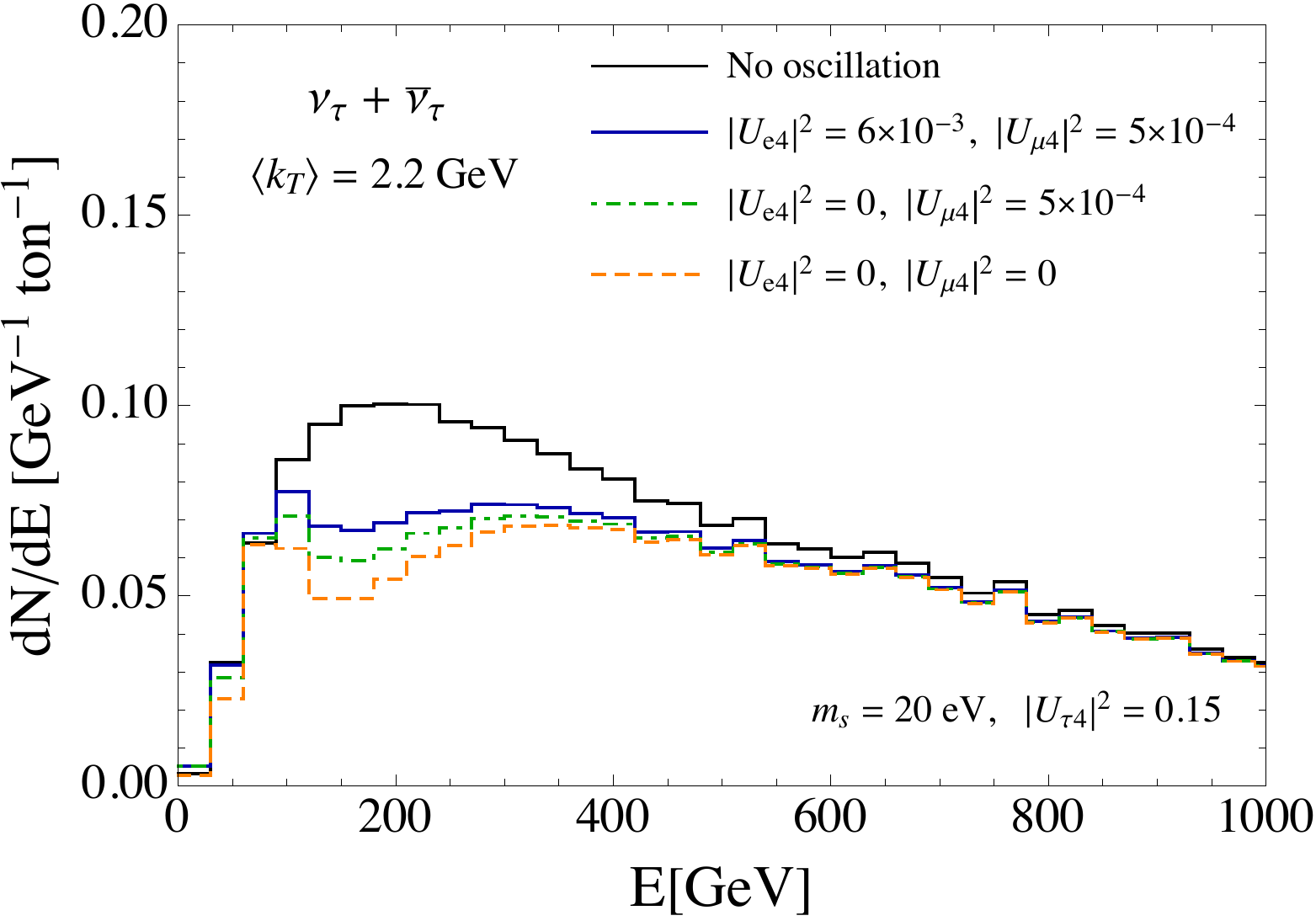}
\par\end{centering}
\caption{Predictions of the number of $\nu_\tau + \overline{\nu}_\tau$ charged-current events as a function of neutrino energy in absence of oscillations and in presence of oscillations in a 3+1 mixing framework, for various choices of the oscillation parameters in the same experimental setup already used in section 4. Numbers of events are reported for a ton of lead detector, for $\langle k_T \rangle$ = 0.7 GeV (left) and 2.2 GeV (right).}   \label{fig:nu_tau-nuOsci}
\end{figure}

Figure \ref{fig:nu_tau-nuOsci} shows the number of events as a function of energy, in the standard three-active-flavor oscillation framework (black histogram) and in the 3+1 oscillation framework with $m_4 = 20$ eV and $|U_{\tau 4}|^2=0.15$, considering our default heavy-flavor QCD input parameter set and $\langle k_T\rangle = 0.7$ GeV (left) and 2.2 GeV (right). The orange-dashed histogram shows the effect of $\nu_\tau$ disappearance due to oscillations 
in a 3+1 scenario 
with $|U_{e4}|^2=|U_{\mu 4}|^2=0$, where a dip is visible at neutrino energies $\sim$ 150 GeV. 

\begin{figure}
\begin{centering}
\includegraphics[width=0.49\textwidth]{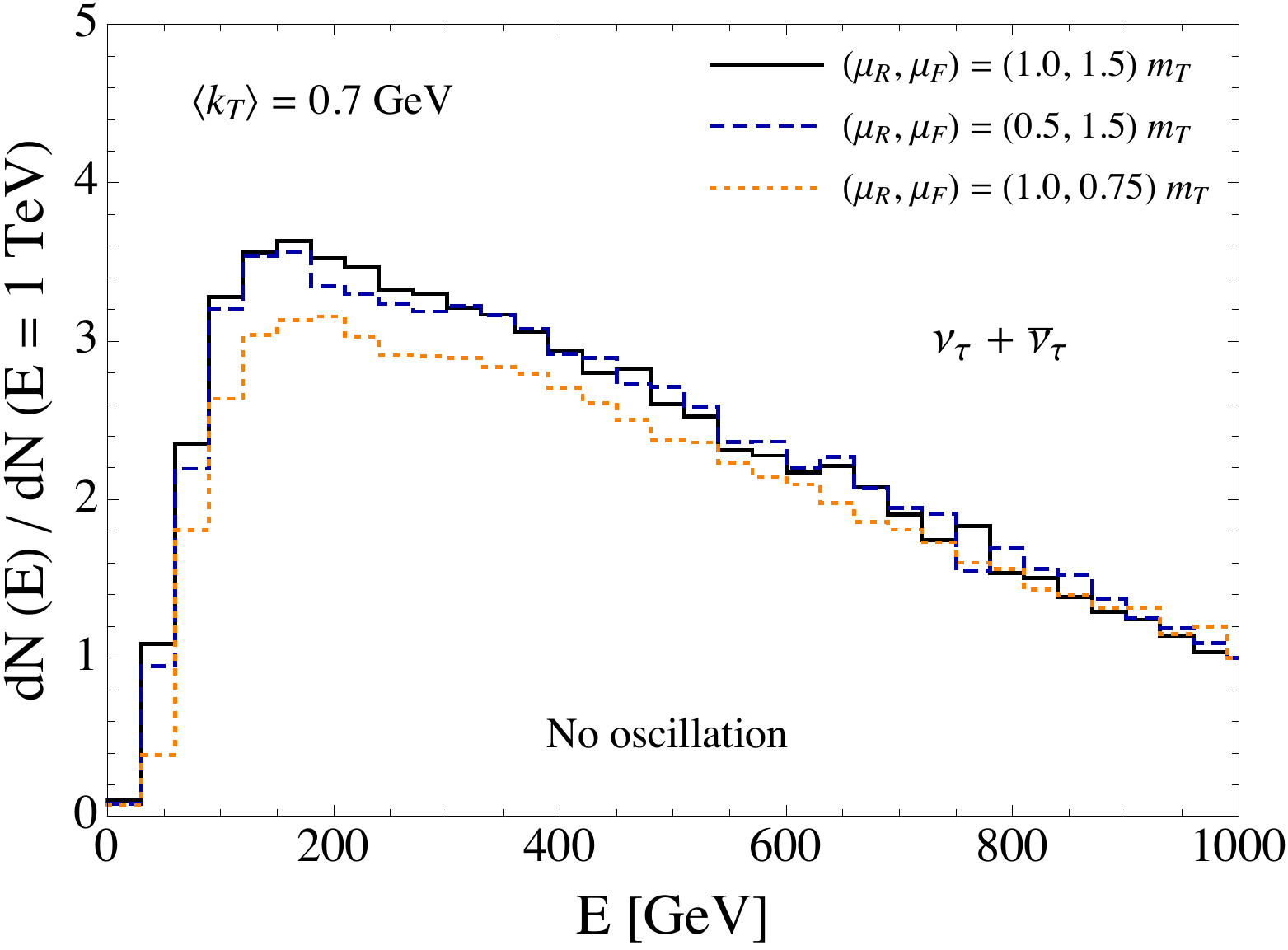}
\includegraphics[width=0.49\textwidth]{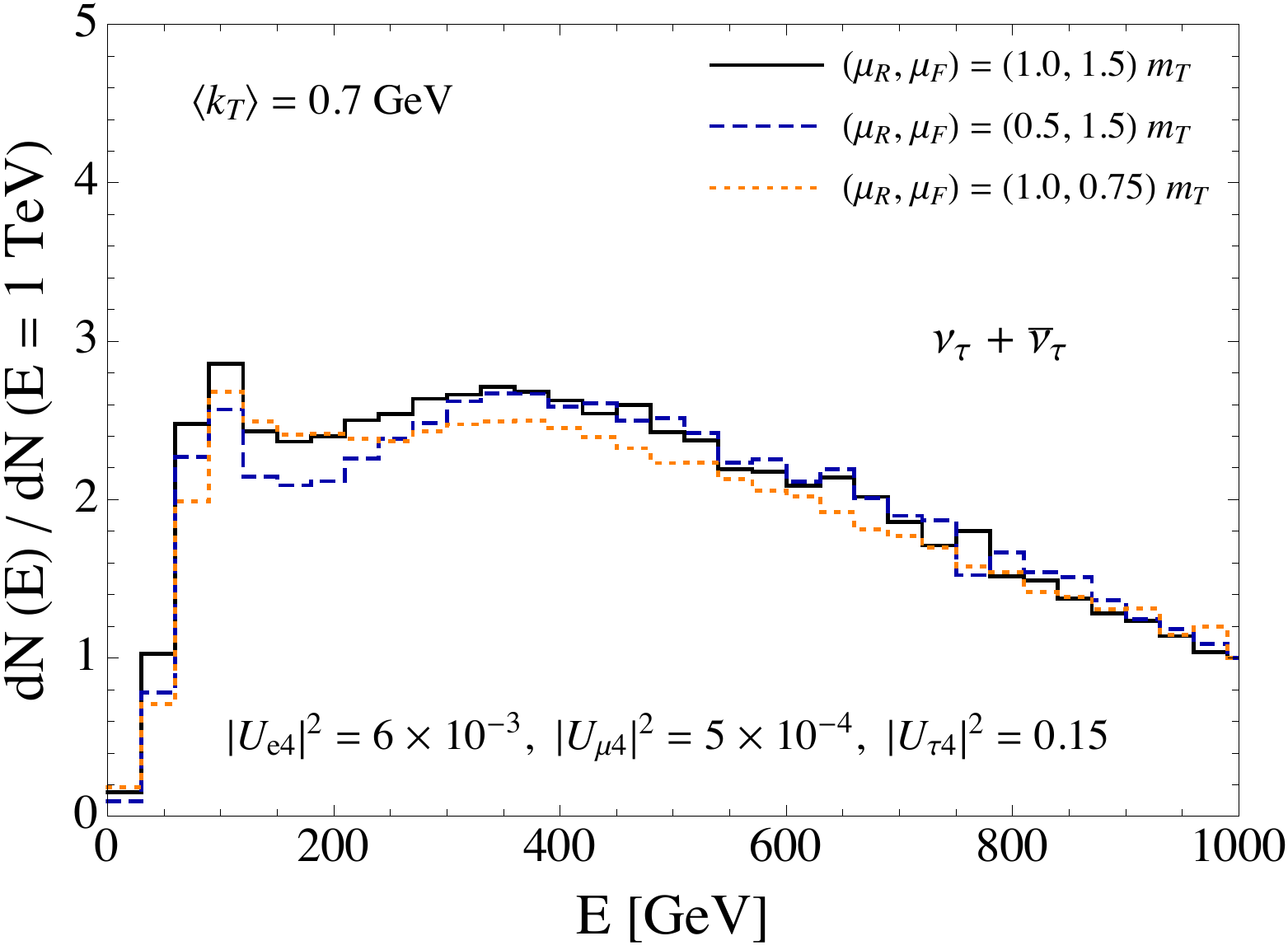}
\includegraphics[width=0.49\textwidth]{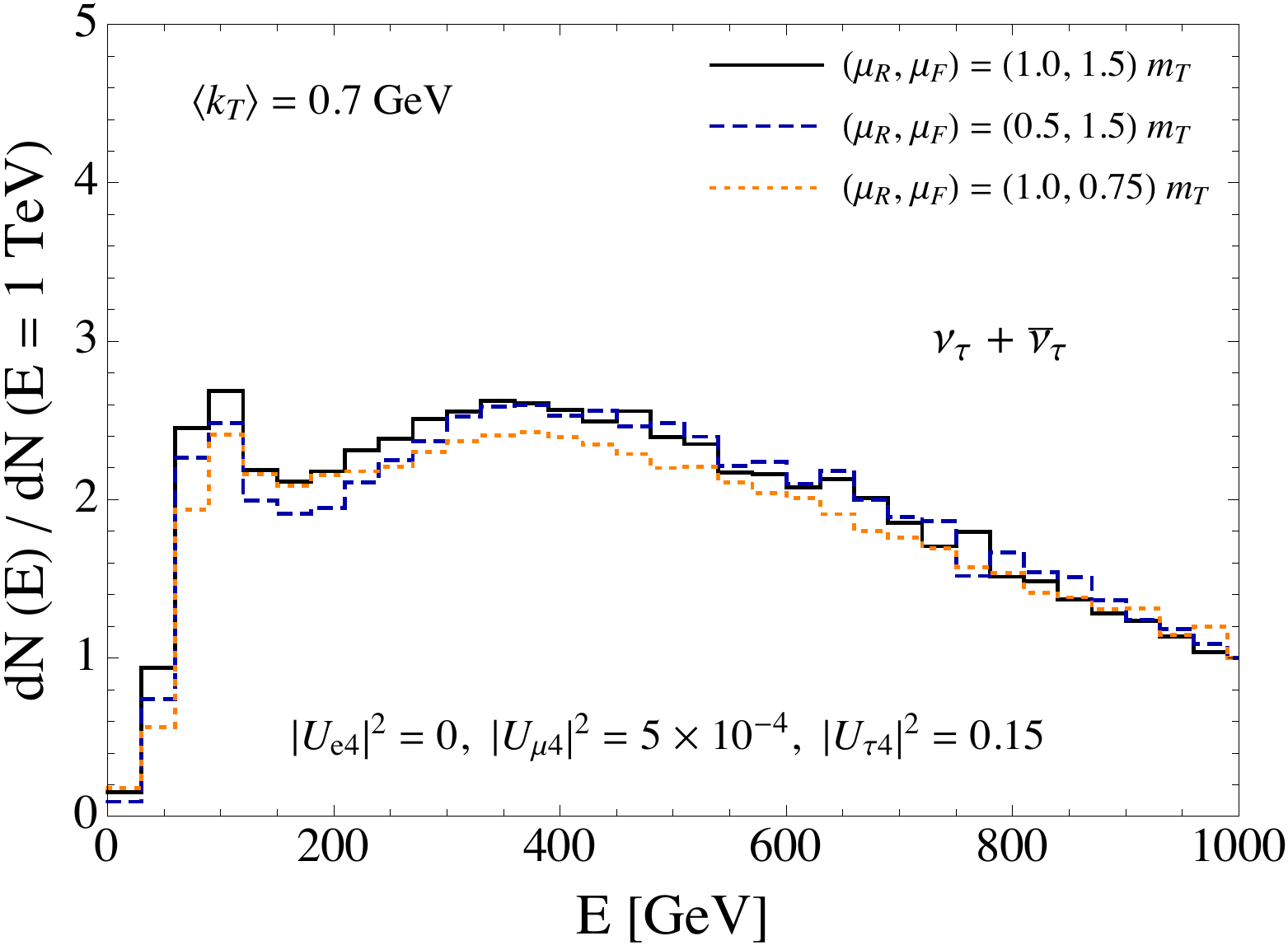}
\includegraphics[width=0.49\textwidth]{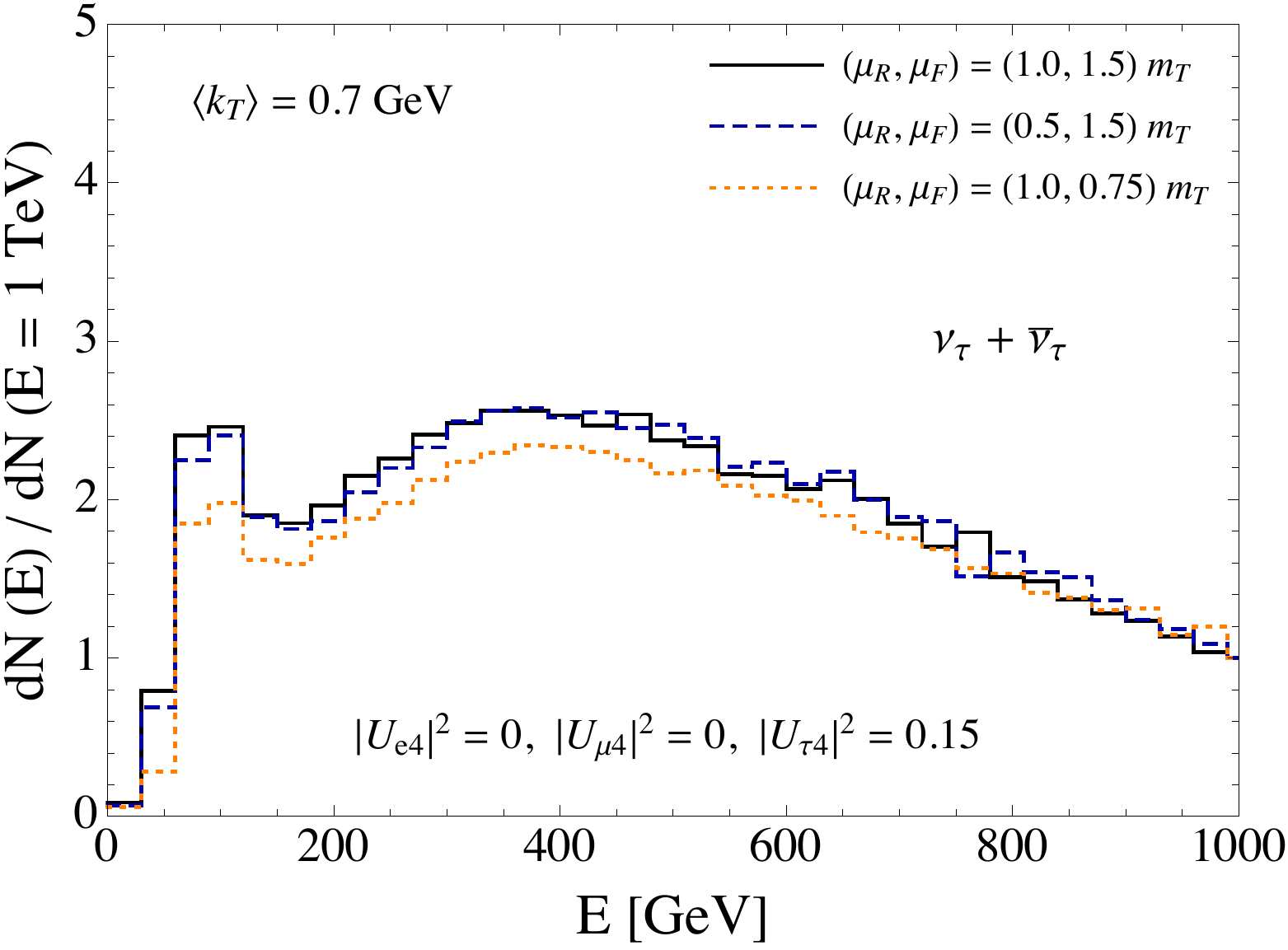}
\par\end{centering}
\caption{The ratio of the number of $\nu_\tau + \overline{\nu}_\tau$ charged-current interaction events and the number of events at 1 TeV in each case, as a function of neutrino energy, without oscillations (upper left plot) and in a 3+1 oscillation scenarios with different mixing parameters.
The numbers of events are evaluated for $\langle k_T \rangle$ = 0.7 GeV for three different ($\mu_R$, $\mu_F$) scale choices  that bracket the scale uncertainties.
}
\label{fig:Rnu_tau-scale}
\end{figure}

\begin{figure}
\begin{centering}
\includegraphics[width=0.49\textwidth]{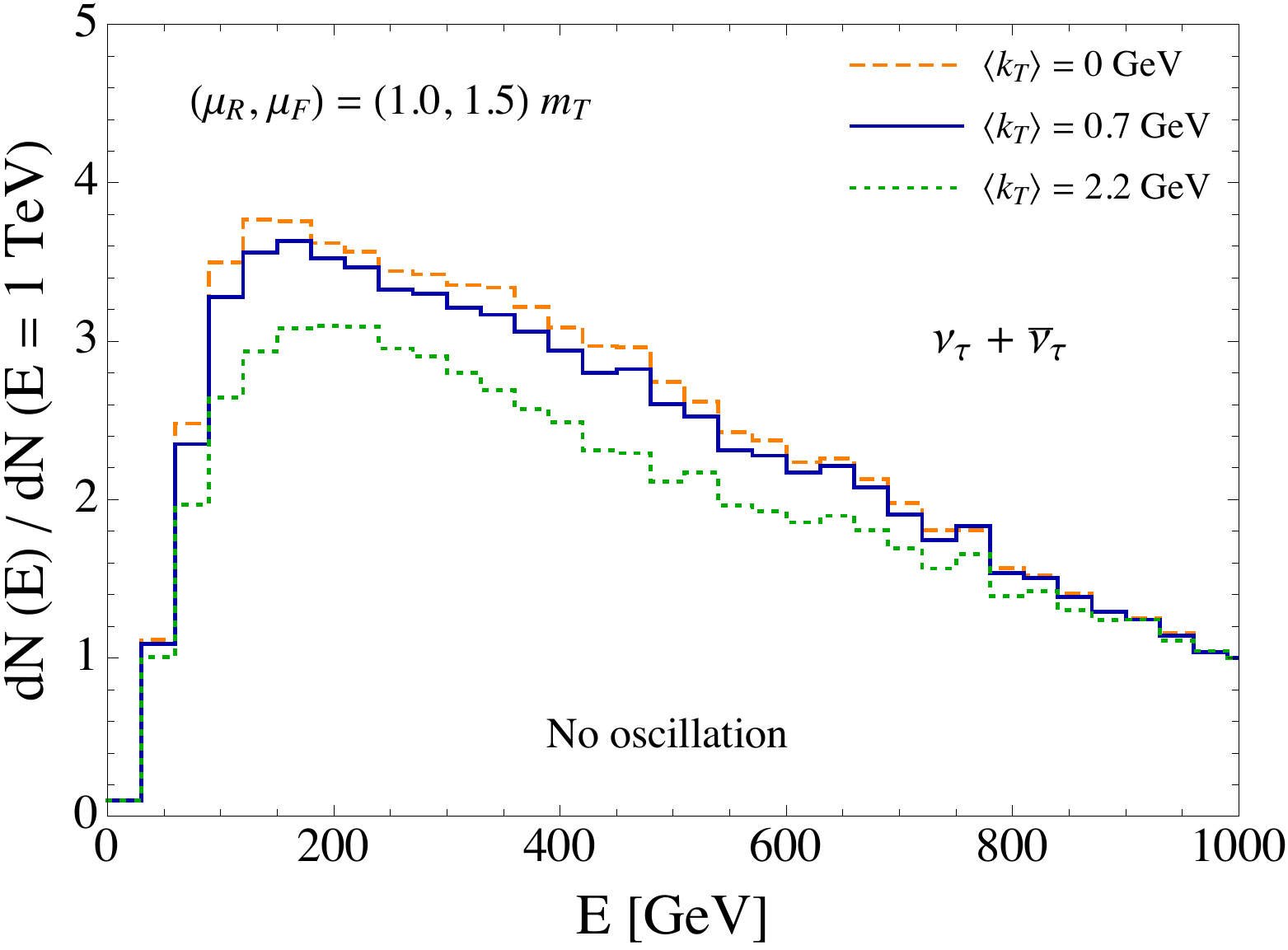}
\includegraphics[width=0.49\textwidth]{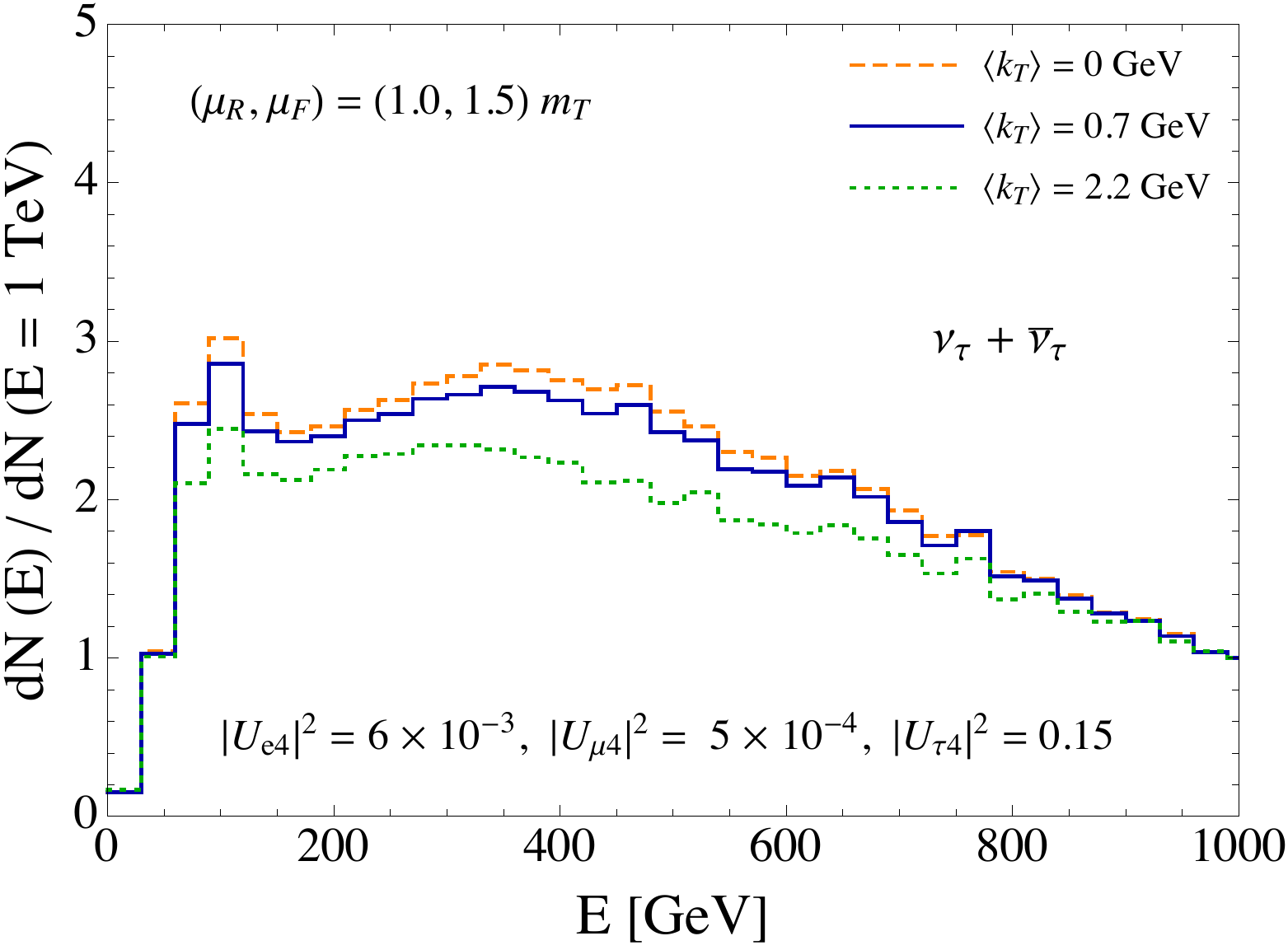}
\includegraphics[width=0.49\textwidth]{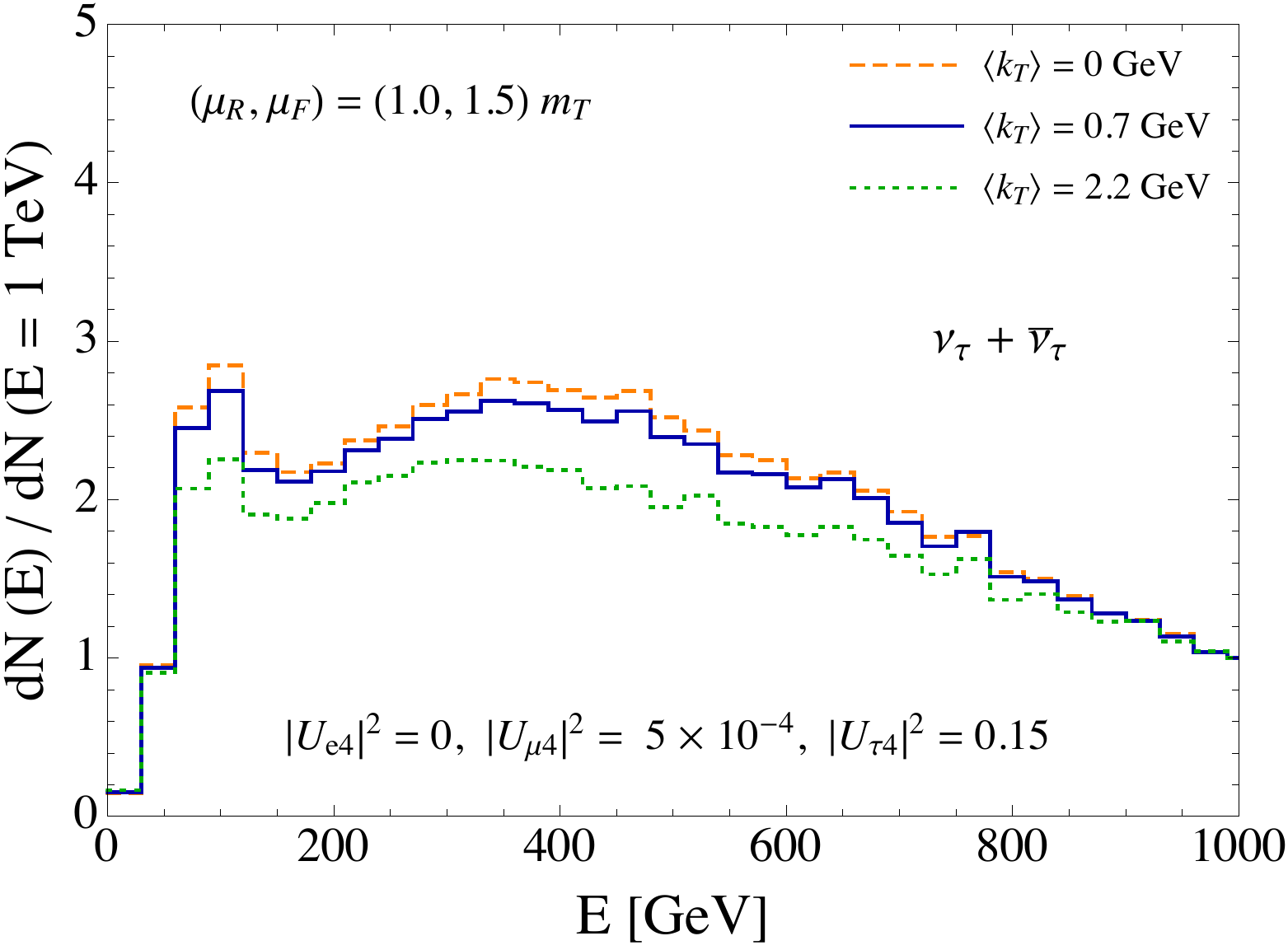}
\includegraphics[width=0.49\textwidth]{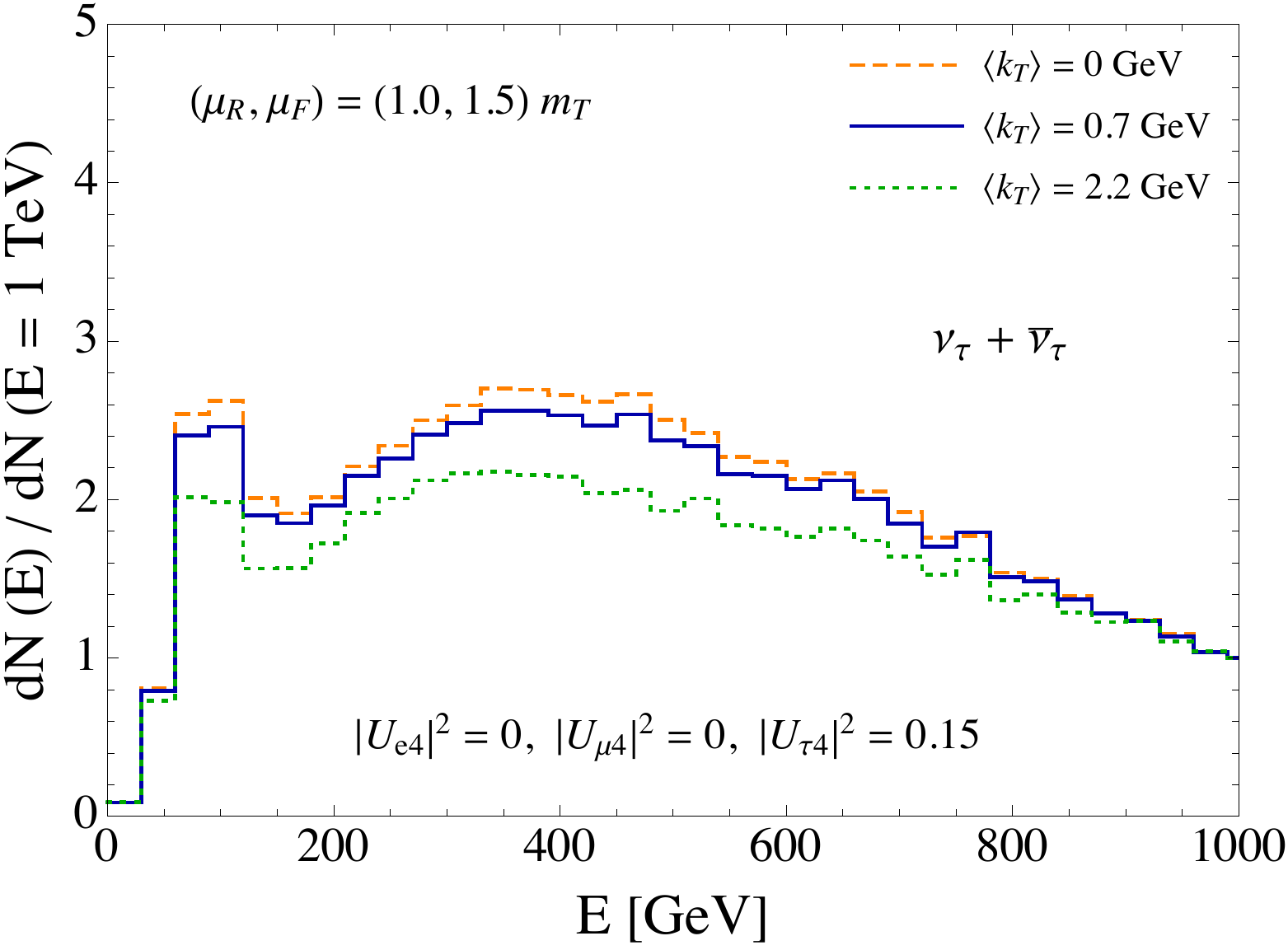}
\par\end{centering}
\caption{The same as figure \ref{fig:Rnu_tau-scale}, but considering different values of $\langle k_T \rangle$, for our default ($\mu_R$, $\mu_F$) scale choice.   
\label{fig:Rnu_tau-kT}}
\end{figure}

The oscillation dip may be partially filled in by $\nu_\mu\to \nu_\tau$ and $\nu_e\to \nu_\tau$ oscillations, where the $\nu_\mu$ and $\nu_e$ come from heavy-flavor decays and from $\pi^\pm,\ K^\pm$ and $K_L$ decays. The NOMAD experiment set the most stringent
limits on effective mixing angles with $\nu_\tau$ for 3+1 scenarios with $\Delta m^2\simeq m_4^2\gsim 30$ eV$^2$ 
\cite{Astier:2001yj,Astier:2003gs}, 
\begin{eqnarray}
\label{eq:etau}
\sin^2 2\theta_{e\tau} &\simeq & 4 |U_{e4}|^2 |U_{\tau 4}|^2 < 1.5\times 10^{-2}\\ 
\label{eq:mutau}
\sin^2 2\theta_{\mu\tau} &\simeq & 4 |U_{\mu 4}|^2 |U_{\tau 4}|^2 < 3.3\times 10^{-4}\\ 
\label{eq:emu}
\sin^2 2\theta_{e\mu} &\simeq & 4 |U_{e4}|^2 |U_{\mu 4}|^2 < 1.4\times 10^{-3}\ .
\end{eqnarray}
We use $|U_{\tau 4}|^2=0.15$ to illustrate sterile neutrino mixing effects.\footnote{ This is consistent with the 99\%CL limit from 
ref. \cite{Dentler:2018sju}, where results are shown only for $\Delta m_{41}^2\lsim 10$ eV$^2$.} For $|U_{\mu 4}|^2$ and $|U_{e4}|^2$, eqs. (\ref{eq:etau}-\ref{eq:emu}) must be satisfied. For illustration purposes, we take $|U_{\mu 4}|^2=5\times 10^{-4}$, a value close to the maximum mixing consistent with eq. (\ref{eq:mutau}). The Troisk tritium beta decay data can be used to
set limits on $|U_{e4}|^2$ for sterile neutrino masses in the range of $1-100$ eV
 \cite{Belesev:2012hx,Belesev:2013cba}.
For $m_4=20$~eV,
$|U_{e4}|^2\lsim 6.7\times 10^{-3}$ comes from the Troisk upper bound.
The other constraints, eqs. (\ref{eq:etau}) and (\ref{eq:emu}), are satisfied by this value of $|U_{e4}|^2$, considering our aforementioned choice for $|U_{\tau 4}|^2$. 
 We use parameters $|U_{\mu 4}|^2=5\times 10^{-4}$ and a slightly lower value of $|U_{e4}|^2 = 6\times 10^{-3}$, still satisfying the constraints discussed above, in the blue histogram in figure \ref{fig:nu_tau-nuOsci}. In the figure, $\nu_e$ and $\nu_\mu$ (and anti-neutrinos) from heavy-flavor decays and from $\pi^\pm,\ K^\pm$ and $K_L$ decays are included.

The spectral shape of the number of $\nu_\tau+\bar{\nu}_\tau$ events changes with different choices of mixing parameters.
If $|U_{e4}| ^2=|U_{\mu 4}| ^2=0$, spectral distortions will be conspicuous, even if $\langle k_T\rangle=2.2$ GeV (corresponding to a total smaller number of events than our default case $\langle k_T\rangle=0.7$ GeV), as can be seen in the right panel of figure \ref{fig:nu_tau-nuOsci}. Keeping the same value of $|U_{e 4}^2|$ and increasing the $|U_{\mu 4}|^2$ value to 
$|U_{\mu 4}|^2<10^{-3}$, the $\nu_\tau\to\nu_\tau$ oscillation dip in the energy distribution of the events is not significantly modified because muon mixing parameters are quite constrained
as discussed above. This effect is shown in the green histogram of figure \ref{fig:nu_tau-nuOsci}, obtained by setting $|U_{e4}| ^2=0$ and $|U_{\mu 4}|^2=5\times 10^{-4}$. When $|U_{e4}|^2\neq 0$, electron neutrinos coming primarily from $K^0_{e3}$ that oscillate to $\nu_\tau$ because of sterile neutrino mixing additionally fill in the spectral distortion of the $\nu_\tau+\bar{\nu}_\tau$
event number as shown in figure \ref{fig:nu_tau-nuOsci} with the solid blue histogram for $|U_{e4}| ^2=6\times 10^{-3}$.

Untangling the physics of a $3+1$ flavor oscillation scenario, from the standard 3-flavor oscillation scenario, considering the QCD theoretical uncertainties related to the choice of the scales and of the phenomenological $\langle k_T\rangle$ smearing parameter, will be difficult if the mixing parameters $|U_{e4}|^2$ and $|U_{\mu 4}|^2$ are close to $6 \times 10^{-3}$ and $5 \times 10^{-4}$, respectively. 
 Although the the values of $|U_{e4}|^2$ and $|U_{\mu4}|^2$ are small, the large number of $\nu_e$ and $\nu_\mu$ from light-meson production and decay means that $\nu_e$ and $\nu_\mu$ mixing with sterile neutrinos can have an non-negligible impact. 
The challenge is illustrated in figs. \ref{fig:Rnu_tau-scale} and \ref{fig:Rnu_tau-kT}. These figures show the numbers of tau neutrino plus antineutrino charged-current events, with each distribution normalized to the corresponding number of events at 1 TeV for that distribution, to highlight the effect of the scale dependence and of various values of $\langle k_T\rangle$ on the shape of the distribution. 
In figure \ref{fig:Rnu_tau-scale}, $\langle k_T\rangle
=0.7$ GeV is fixed and the scales are varied. In particular, although a peak is still visible, the case with all three elements $|U_{\ell 4}|^2\neq 0$ in the upper right panel presents a slightly shallower dip in the oscillated $3+1$ spectrum as compared to the case with $|U_{e4}|^2 = 0$ and $|U_{\mu 4}|^2 = 0$ shown in the lower right panel. The normalization of each distribution in figure \ref{fig:Rnu_tau-scale} to its own distribution at $E=1$ TeV collapses the scale uncertainty band of figure \ref{fig:NutauEventLHC-scale}.
Figure \ref{fig:Rnu_tau-kT} shows that a large transverse momentum smearing can somewhat obscure the spectral features of $3+1$ oscillations as well. 
 The effects of $\nu_e$ and $\nu_\mu$ mixing with sterile neutrinos are more important in each panel of figures \ref{fig:Rnu_tau-scale} and \ref{fig:Rnu_tau-kT}, namely, for $(\mu_R, \, \mu_F) = (1.0,\, 0.75)m_T$ in figure \ref{fig:Rnu_tau-scale} and $\langle k_T \rangle$ = 2.2 GeV in figure \ref{fig:Rnu_tau-kT}. 
The dip in the spectrum is least pronounced in the case of $(\mu_R, \, \mu_F) = (1.0,\, 0.75)m_T$ and $\langle k_T \rangle$ = 2.2 GeV.
A dedicated study of $3+1$ oscillation scenarios that includes detector resolution effects would be important to understand in detail the reach of a forward tau neutrino experiment along the LHC beamline.

%% file: conclusions.tex
\section{Conclusions}
\label{sec:conclusions}

Theoretical proposals to exploit collider production of tau neutrinos and antineutrinos via heavy-flavor decays in the far-forward region have a long history. Recent proposals of experiments to detect BSM particles with feeble interactions have made our evaluation timely. This work provides a first evaluation of the number of $\nu_\tau+\bar{\nu}_\tau$ charged-current interaction events in the very forward region, including QCD effects beyond the leading order/leading logarithmic 
accuracy considered in previous estimates and studying the effects of different sources of QCD uncertainties previously neglected. We focus on neutrinos with pseudorapidities $\eta>6.87$, consistently with the geometry of a 1 m radius cylindrical detector at a distance of 480 m from the interaction point \cite{Feng:2017uoz, Ariga:2018pin,Ariga:2019ufm}
to illustrate a number of effects.

Thousands of tau neutrino plus antineutrino events are predicted for 1 meter of lead (35.6 ton) target.
Theoretical uncertainties are quite large due to, in particular, the renormalization and factorization scale
variation in the heavy-flavour production cross-sections. We show results for both $(\mu_R,\mu_F)=(1.0,1.5)~m_T$ and $(\mu_R,\mu_F)=(1.0,1.0)~m_T$, to better match the LHCb data and for comparisons with the usual scale conventions adopted in QCD
studies of heavy-flavor production, respectively.
Seven-point scale variation 
around the central scale combinations $(\mu_R,\mu_F)=(1.0,1.0)~m_T$ and $(1.0,1.5)~m_T$ causes differences amounting to a factor of
$\sim 4 - 6$ between the upper and lower limit of the predicted event numbers, for the
intrinsic momentum parameter with which our NLO predictions convoluted with fragmentation functions approximately reproduce \textsc{Powheg~+~Pythia} results.

The introduction of the parameter $\langle k_T\rangle$ in the Gaussian smearing factor $f(\vec{k}_T)$ in eq. (3.1) distorts the far-forward tau neutrino and antineutrino spectra. This is also the case in fixed-target experiments like SHiP \cite{Bai:2018xum}. The collinear parton model is sufficient for central collisions, but for forward production, as we have shown, non-collinear effects have a significant impact on the number of events for forward neutrino production at the LHC.
Although the Gaussian form of $f(\vec{k}_T)$ is imperfect, it can approximate non-perturbative QCD effects and mimic part of the perturbative effects beyond fixed-order. 
A comparison with LHCb double-differential cross sections in $p_T$ and rapidity for $D_s$ production in $pp$ collisions at $\sqrt{s} = 13$ TeV, for $p_T$ $\in$ [0, 14] GeV in five rapidity bins in the range $y = 2.0 - 4.5$, shows that the experimental data are reasonably reproduced by theoretical predictions in a framework combining NLO pQCD corrections to the hard-scattering, 
$k_T$-smearing effects and phenomenological fragmentation functions.
Experiments that probe heavy-flavor physics in the very forward region (e.g., ref. \cite{Aoki:2019jry})
will help to better quantify and disentangle the need of accurate procedures jointly resumming different kinds of logarithms and that of more rigorous descriptions of 
non-perturbative QCD effects, as well as the limits of the collinear approximation. The inclusion of both small $x$ and $k_T$ effects in theoretical evaluations of heavy-flavor production at the 
Electron-Ion Collider (EIC) have implications for measurements (see, e.g., refs. \cite{Bacchetta:2018ivt,DAlesio:2019qpk}).
In another arena, predictions of the prompt atmospheric neutrino flux 
\cite{Bhattacharya:2015jpa,Bhattacharya:2016jce,Gauld:2015kvh,Garzelli:2016xmx,Benzke:2017yjn,Zenaiev:2019ktw} will also be constrained by
measurements of the tau neutrino energy spectrum in the far-forward region at the LHC.

For a baseline of $\sim$ 500 m, oscillations in the standard 3-flavor scenario are negligible at the energy scales of the tau neutrino beams in our setup. On the other hand, in a 3+1 oscillation scenario, including a sterile neutrino in the ${\cal O}$(20 eV) mass range with three active flavor eigenstates, oscillations could give rise to visible signals in the energy dependence of the tau neutrino events.
The contribution to $\nu_\tau$ events due to $\nu_\mu\to \nu_\tau$ oscillations is not important due to the small value of $|U_{\mu 4}|^2$.
This is the case even though there are $\sim 100$ times more $\nu_\mu+\bar{\nu}_\mu$ produced
from charged pion and kaon decays than 
from heavy-flavor decays. The number of
$\nu_\mu+\bar{\nu}_\mu$ from heavy flavor decays is larger than the number of $\nu_\tau+\bar{\nu}_\tau$ by a factor of $\sim$ 10. 
The  $\nu_e \to \nu_\tau$ oscillation effects are also small. 

Therefore 
a large number of $\nu_\tau+\bar{\nu}_\tau$ events in this energy range would provide an opportunity
to constrain $3+1$ oscillation models with a sterile neutrino in the 10's of eV mass range. In the case of $\nu_\tau$ disappearance, the location of a dip in the charged-current event distribution as a function of tau neutrino energy will constrain the mass of the fourth mass eigenstate (mostly sterile neutrino) $m_4$. The quantity $|U_{\tau 4}|^2$ is currently poorly constrained. We showed that, 
 as long as $\nu_e\to \nu_\tau$ appearance is suppressed,  
 for $|U_{\tau 4}|^2=0.15$ in 
 principle the oscillation effect would be unambiguous.
However, uncertainties in heavy-flavor production 
present challenges to precision constraints on the 3+1 sterile neutrino parameter space accessible to a far-forward neutrino experiment at the LHC. 
Practical aspects of tau neutrino detection in the high-luminosity environment will also be a challenge because of the muon neutrino background.

%% file: appendix.tex
\appendix
\label{sec:appendix}

\section{Decay Distributions}

The two-body decays of the $D_s$ in the $D_s$ rest frame come from energy and momentum conservation. Keeping the polarization of the $\tau$, the energy and angular distribution of the tau neutrino can also be obtained \cite{Barr:1988rb,Barr:1989,Gaisser:1990vg}. Detailed formulas for the direct $\nu_\tau$ and chain decay $\nu_\tau$ distributions appear in Appendix B of ref. \cite{Bai:2018xum}. Decays in the $D_s$ and $\tau$ rest frames are appropriatedly boosted to the collider frame, including full angular dependence. 

Neutrinos can be produced through the three-body semileptonic decay process 
of $D$ and $B$ mesons, $D(B) \to K (D)  l \nu_l \, ( l = e, \mu)$. The distribution of neutrinos from the decay in the rest frame is respectively given by 
\begin{eqnarray}
%
 \frac{{\rm d}\Gamma  (h_i \to \nu_l)}{{\rm d}x_\nu} 
&\sim& 
\begin{dcases}
 \frac{m_{D}^5 \, x_\nu^{2}(1-r_h-x_\nu)^{2}
[3+r_h(3-x_\nu)-5x_\nu+2x_\nu^{2}]}{(1-x_\nu)^{3}} \quad {\rm for} \quad h_i = D 
\\
 \frac{m_B^5 \, x_\nu^{2}(1-r_h-x_\nu)^{2}}{(1-x_\nu)} \hspace{13.5em} {\rm for} \quad h_i = B 
 \end{dcases}
%
\label{eq:dkdistm0}
\end{eqnarray}
where the fraction of energy transferred to the neutrino is $x_\nu = 2 E_\nu / m_i$ and the hadron mass fraction is $r_h = m_f^2/m_i^2$, with $m_i$ and $m_f$ being the hadron masses in the initial and final states, respectively. 
The corresponding cumulative distribution functions (CDF) are
\begin{eqnarray}
F(x_\nu) &=& \frac{1}{\Gamma} \int_0^{x_\nu} d x^\prime \frac{\textrm{d}\Gamma}{\textrm{d}x^\prime} \\  [1em]
 &=& 
 \begin{dcases}
 \frac{1}{D(r_h) (1 - x_\nu)^2} \biggl[ x_\nu 
 \Bigl( 2 r_h^3 x_\nu^2  - 6 \, r_h^2 (2 - x_\nu) (1 - x_\nu)  + (2 - x_\nu) (1 - x_\nu)^2 x_\nu^2    \nonumber \\
  \hspace{8.7em} 
-  2 r_h (1 - x_\nu)^2 x_\nu^2 \Bigr)   
-  12 r_h^2 (1- x_\nu)^2 \ln[1 - x_\nu] \biggr]  \quad {\rm for} \quad h_i = D    \\
  \frac{1}{D(r_h)} \biggl[  x_\nu \Bigl(x_\nu^2 (4 - 3 x_\nu) -8 r_h x_\nu^2 - 6 r_h^2 (2 + x_\nu)  \Bigr) 
 - 12 r_h^2 \ln[1 - x_\nu]  \biggr]   \hspace{1.2em}  {\rm for} \quad h_i = B  \nonumber
 \end{dcases}
\end{eqnarray}
where $D(r)=1-8r-12r^{2}\ln(r)+8r^{3}-r^{4}$.
The CDF is used to determine the neutrino energy in the heavy meson rest frame, and along with an isotropic decay distribution, the neutrino four-momentum in the rest frame is boosted to the collider frame, where the heavy meson $D(B)$ has four-momentum $p_{D(B)}$.

\begin{figure}
	\centering
	\includegraphics[width=0.45\textwidth]{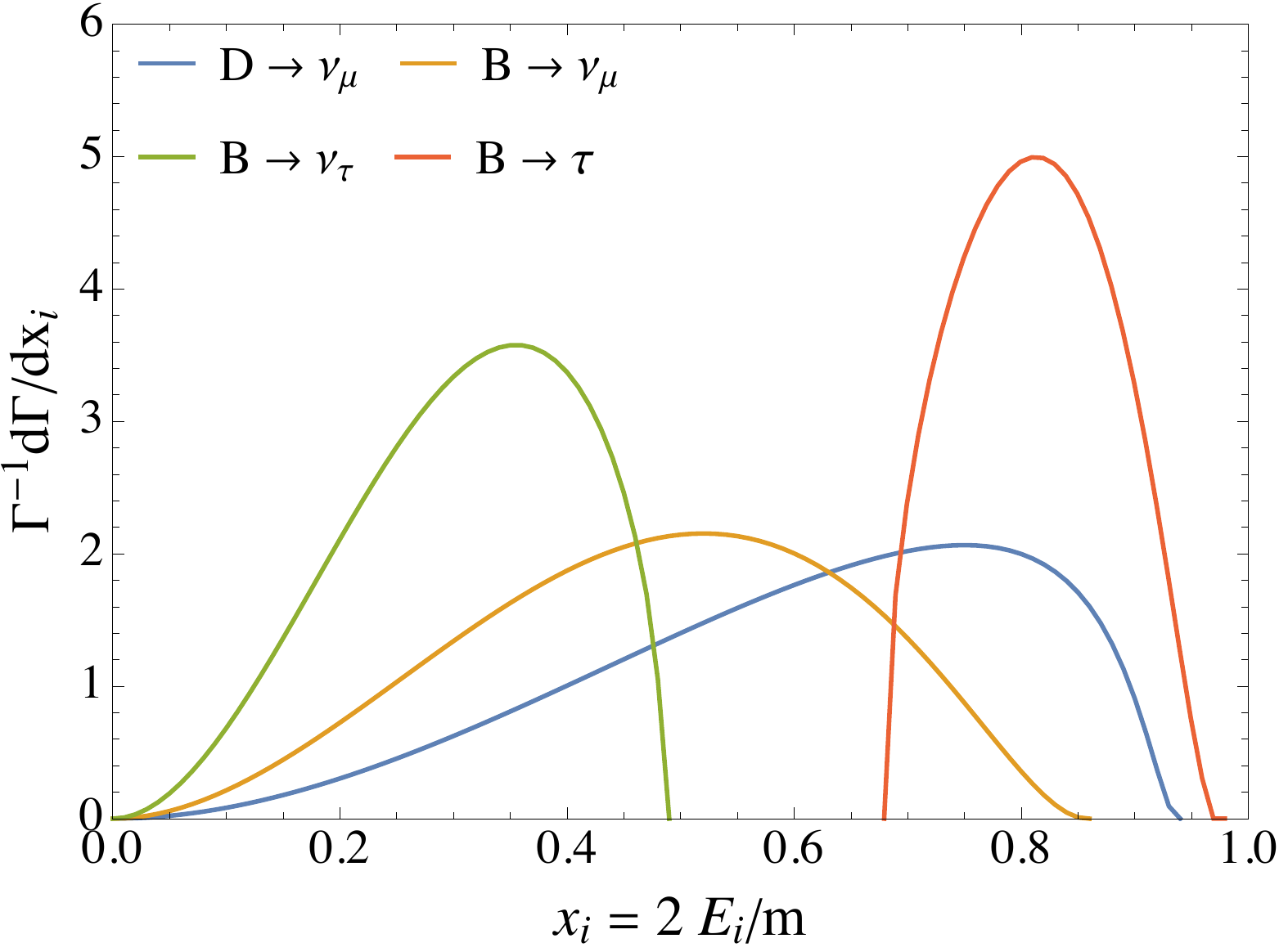}
	\includegraphics[width=0.45\textwidth]{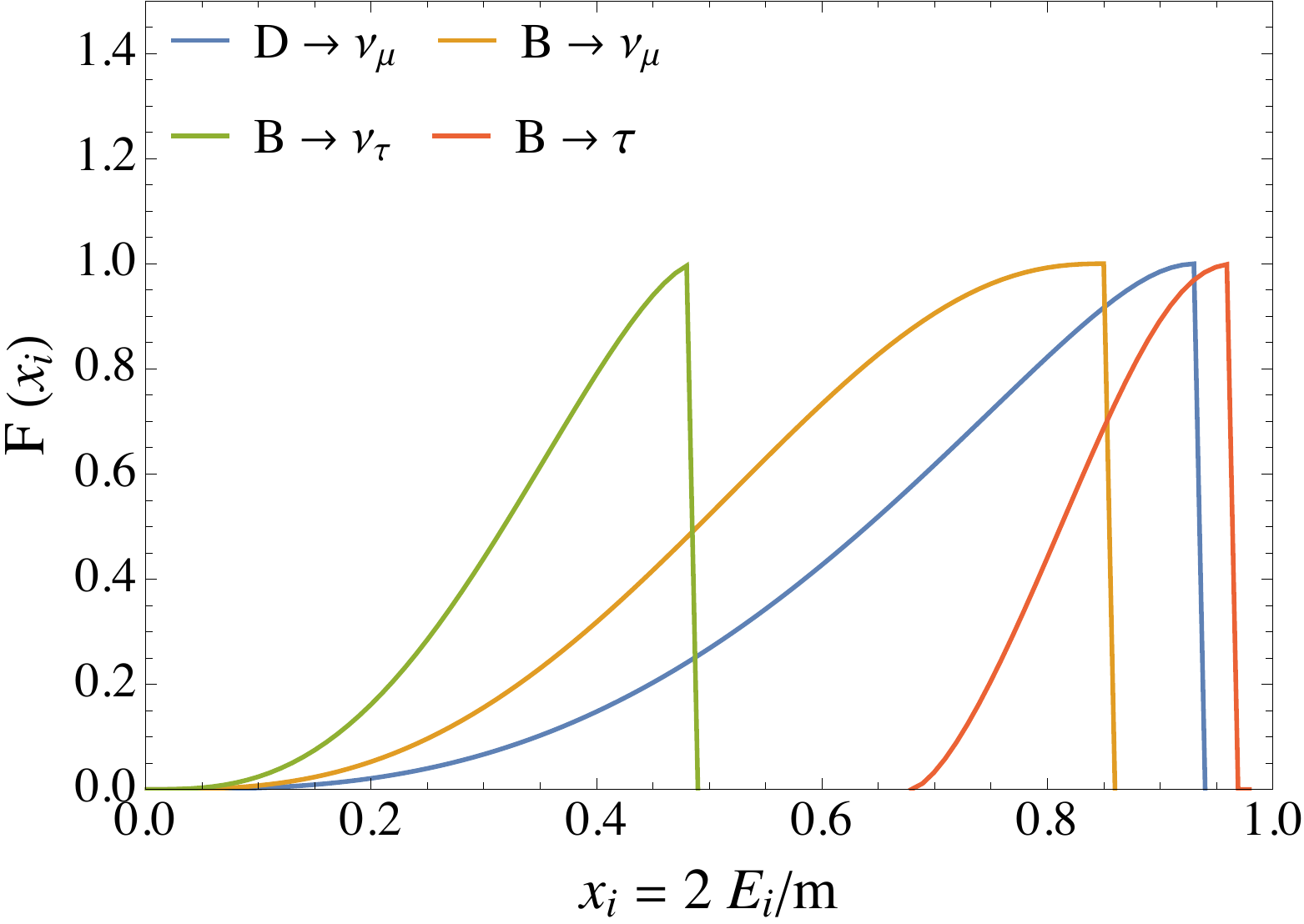}
	\caption{Left panel: The fractional energy distributions $x_i=E_i/(m/2)$ in three-body semi-leptonic decays in the rest frame of the decaying heavy particle with $m=m_{D_s}$ and $m=m_ B$ and $i=\nu_\mu,\ \nu_\tau$ and $\tau$. Right panel: The cumulative distribution function for each of the three-body decays shown in the left panel.
	} 
	\label{fig:dkdist}
\end{figure}

The $B$ meson also decays to tau neutrinos via $B \to D \tau  \nu_\tau$, and the $\tau$ lepton decays subsequently to $\nu_\tau$. 
In this case, there is an additional effect due to the non-negligible mass of $\tau$.   
With the additional mass term $r_\tau = m_\tau^2/m_B^2$, one obtains the distributions of $\tau$ and $\nu_\tau$ from $B$ decays:
\begin{eqnarray}
\frac{{\rm d}\Gamma (B\to\tau)}{{\rm d}x_\tau}
&\sim& \frac{m_B^5}{(1+r_\tau-x_\tau)^3}
 (1-x_\tau+r_\tau -r_h)^2 (x_\tau^2 - 4 r_\tau)^{1/2} 
 \biggl[  r_\tau^2 (3x_\tau-4 ) \\
&& 
+  ~ x \Bigl( 3+ 2x_\tau^2 - 5x_\tau +r_h(3-x_\tau) \Bigr)  
  - r_\tau \Bigl( 4+5x_\tau^2 - 10x_\tau + r_h(8-3x_\tau) \Bigr) \biggr], \nonumber \\ [1em]
%
\frac{{\rm d}\Gamma (B\to\nu_\tau)}{{\rm d}x_\nu}
&\sim &\frac{m_B^5}{1-x_\nu} \, x_\nu^2 (1-r_h -r_\tau-x_\nu)  
\Bigl(r_h^2 - 2 r_h(1+r_\tau-x_\nu)+(1-r_\tau-x_\nu)^2\Bigr)^{1/2}. \nonumber   
\end{eqnarray}
Here, $x_\tau=2E_\tau/m_i$ is the fraction of energy transferred to tau.   
The expression of the CDF for $\tau$ and $\nu_\tau$ from $B$ decays is too complicated to be presented here. Again, both energy and angular distributions are determined in the rest frame, then boosted to the collider frame. For reference, we show the fractional energy distribution (left) and CDFs (right) for $\nu_\mu$, $\nu_\tau$ and $\tau$ in the rest frame of the decaying $D$ and $B$ in Figure \ref{fig:dkdist} for the three-body decays.

The energy fractions $x$  span the following range: 
\begin{eqnarray}
0 < &x_{\nu_l} &< 1 - (\sqrt{r_l}+\sqrt{r_h})^2  \nonumber  \\
2 \sqrt{r_\tau} <& x_{\tau}& < 1 + r_\tau - r_h  \ .    
\end{eqnarray}
For electron neutrino ($\nu_e$) and muon neutrino ($\nu_\mu$), the maximum value of $x_\nu$ is simply $x_\nu^{\rm max} = 1-r_h$.

%% file: main.bbl
\providecommand{\href}[2]{#2}\begingroup\raggedright\begin{thebibliography}{10}

\bibitem{pdg:2016}
{\scshape Particle Data Group} collaboration, C.~Patrignani et~al.,
  \emph{{Review of Particle Physics}},
  \href{http://dx.doi.org/10.1088/1674-1137/40/10/100001}{\emph{Chin. Phys.}
  {\bf C40} (2016) 100001}.

\bibitem{Acciarri:2015uup}
{\scshape DUNE} collaboration, R.~Acciarri et~al., \emph{{Long-Baseline
  Neutrino Facility (LBNF) and Deep Underground Neutrino Experiment (DUNE)}},
  \href{http://arxiv.org/abs/1512.06148}{{\tt 1512.06148}}.

\bibitem{Acciarri:2016crz}
{\scshape DUNE} collaboration, R.~Acciarri et~al., \emph{{Long-Baseline
  Neutrino Facility (LBNF) and Deep Underground Neutrino Experiment (DUNE)}},
  \href{http://arxiv.org/abs/1601.05471}{{\tt 1601.05471}}.

\bibitem{Abi:2018dnh}
{\scshape DUNE} collaboration, B.~Abi et~al., \emph{{The DUNE Far Detector
  Interim Design Report Volume 1: Physics, Technology and Strategies}},
  \href{http://arxiv.org/abs/1807.10334}{{\tt 1807.10334}}.

\bibitem{Park:2011gh}
H.~Park, \emph{{The estimation of neutrino fluxes produced by proton-proton
  collisions at $\sqrt{s}=14$ TeV of the LHC}},
  \href{http://dx.doi.org/10.1007/JHEP10(2011)092}{\emph{JHEP} {\bf 10} (2011)
  092}, [\href{http://arxiv.org/abs/1110.1971}{{\tt 1110.1971}}].

\bibitem{Feng:2017uoz}
J.~L. Feng, I.~Galon, F.~Kling and S.~Trojanowski, \emph{{ForwArd Search
  ExpeRiment at the LHC}},
  \href{http://dx.doi.org/10.1103/PhysRevD.97.035001}{\emph{Phys. Rev.} {\bf
  D97} (2018) 035001}, [\href{http://arxiv.org/abs/1708.09389}{{\tt
  1708.09389}}].

\bibitem{Ariga:2018pin}
{\scshape FASER} collaboration, A.~Ariga et~al., \emph{{Technical Proposal for
  FASER: ForwArd Search ExpeRiment at the LHC}},
  \href{http://arxiv.org/abs/1812.09139}{{\tt 1812.09139}}.

\bibitem{Ariga:2019ufm}
{\scshape FASER} collaboration, A.~Ariga et~al., \emph{{FASER: ForwArd Search
  ExpeRiment at the LHC}},  \href{http://arxiv.org/abs/1901.04468}{{\tt
  1901.04468}}.

\bibitem{Buontempo:2018gta}
S.~Buontempo, G.~M. Dallavalle, G.~De~Lellis, D.~Lazic and F.~L. Navarria,
  \emph{{CMS-XSEN: LHC Neutrinos at CMS. Experiment Feasibility Study}},
  \href{http://arxiv.org/abs/1804.04413}{{\tt 1804.04413}}.

\bibitem{Beni:2019gxv}
N.~Beni et~al., \emph{{Physics Potential of an Experiment using LHC
  Neutrinos}}, \href{http://dx.doi.org/10.1088/1361-6471/ab3f7c}{\emph{J.
  Phys.} {\bf G46} (2019) 115008}, [\href{http://arxiv.org/abs/1903.06564}{{\tt
  1903.06564}}].

\bibitem{Beni:2019pyp}
{\scshape XSEN} collaboration, N.~Beni et~al., \emph{{XSEN: a $\nu$N Cross
  Section Measurement using High Energy Neutrinos from pp collisions at the
  LHC}},  \href{http://arxiv.org/abs/1910.11340}{{\tt 1910.11340}}.

\bibitem{DeRujula:1984pg}
A.~De~Rujula and R.~Ruckl, \emph{{Neutrino and muon physics in the collider
  mode of future accelerators}},  in \emph{{ECFA-CERN Workshop on large hadron
  collider in the LEP tunnel, Lausanne and CERN, Geneva, Switzerland, 21-27 Mar
  1984: Proceedings. 1.}}, pp.~571--596, 1984.
\newblock \href{http://dx.doi.org/10.5170/CERN-1984-010-V-2.571}{DOI}.

\bibitem{Winter:1990ry}
K.~Winter, \emph{{Detection of the tau-neutrino at the LHC}},  in \emph{{ECFA
  Large Hadron Collider Workshop, Aachen, Germany, 4-9 Oct 1990:
  Proceedings.2.}}, pp.~37--49, 1990.

\bibitem{DeRujula:1992sn}
A.~De~Rujula, E.~Fernandez and J.~J. Gomez-Cadenas, \emph{{Neutrino fluxes at
  future hadron colliders}},
  \href{http://dx.doi.org/10.1016/0550-3213(93)90427-Q}{\emph{Nucl. Phys.} {\bf
  B405} (1993) 80--108}.

\bibitem{Vannucci:1993ud}
F.~Vannucci, \emph{{Neutrino physics at LHC / SSC}},  in \emph{{5th
  International Symposium on Neutrino Telescopes Venice, Italy, March 2-4,
  1993}}, pp.~57--68, 1993.

\bibitem{Aaij:2018afd}
{\scshape LHCb} collaboration, R.~Aaij et~al., \emph{{Measurement of
  $D_s^{\pm}$ production asymmetry in $pp$ collisions at $\sqrt{s} =7$ and 8
  TeV}}, \href{http://dx.doi.org/10.1007/JHEP08(2018)008}{\emph{JHEP} {\bf 08}
  (2018) 008}, [\href{http://arxiv.org/abs/1805.09869}{{\tt 1805.09869}}].

\bibitem{Sjostrand:2019zhc}
T.~Sj{\"o}strand, \emph{{The PYTHIA Event Generator: Past, Present and
  Future}}, \href{http://dx.doi.org/10.1016/j.cpc.2019.106910}{\emph{Comput.
  Phys. Commun.} {\bf 246} (2020) 106910},
  [\href{http://arxiv.org/abs/1907.09874}{{\tt 1907.09874}}].

\bibitem{Abreu:2019yak}
{\scshape FASER} collaboration, H.~Abreu et~al., \emph{{Detecting and Studying
  High-Energy Collider Neutrinos with FASER at the LHC}},
  \href{http://arxiv.org/abs/1908.02310}{{\tt 1908.02310}}.

\bibitem{Abreu:2020ddv}
{\scshape FASER} collaboration, H.~Abreu et~al., \emph{{Technical Proposal:
  FASERnu}},  \href{http://arxiv.org/abs/2001.03073}{{\tt 2001.03073}}.

\bibitem{Sjostrand:2014zea}
T.~Sj{\"o}strand, S.~Ask, J.~R. Christiansen, R.~Corke, N.~Desai, P.~Ilten
  et~al., \emph{{An Introduction to PYTHIA 8.2}},
  \href{http://dx.doi.org/10.1016/j.cpc.2015.01.024}{\emph{Comput. Phys.
  Commun.} {\bf 191} (2015) 159--177},
  [\href{http://arxiv.org/abs/1410.3012}{{\tt 1410.3012}}].

\bibitem{Ahdida:2020evc}
{\scshape SHiP} collaboration, C.~Ahdida et~al., \emph{{SND@LHC}},
  \href{http://arxiv.org/abs/2002.08722}{{\tt 2002.08722}}.

\bibitem{Kodama:2000mp}
{\scshape DONUT} collaboration, K.~Kodama et~al., \emph{{Observation of tau
  neutrino interactions}},
  \href{http://dx.doi.org/10.1016/S0370-2693(01)00307-0}{\emph{Phys. Lett.}
  {\bf B504} (2001) 218--224}, [\href{http://arxiv.org/abs/hep-ex/0012035}{{\tt
  hep-ex/0012035}}].

\bibitem{Kodama:2007aa}
{\scshape DONuT} collaboration, K.~Kodama et~al., \emph{{Final tau-neutrino
  results from the DONuT experiment}},
  \href{http://dx.doi.org/10.1103/PhysRevD.78.052002}{\emph{Phys. Rev.} {\bf
  D78} (2008) 052002}, [\href{http://arxiv.org/abs/0711.0728}{{\tt
  0711.0728}}].

\bibitem{Pupilli:2016zrb}
{\scshape OPERA} collaboration, F.~Pupilli, \emph{{Recent results of the OPERA
  experiment}}, \href{http://dx.doi.org/10.1063/1.4953321}{\emph{AIP Conf.
  Proc.} {\bf 1743} (2016) 060004}.

\bibitem{Agafonova:2018auq}
{\scshape OPERA} collaboration, N.~Agafonova et~al., \emph{{Final Results of
  the OPERA Experiment on $\nu_\tau$ Appearance in the CNGS Neutrino Beam}},
  \href{http://dx.doi.org/10.1103/PhysRevLett.121.139901,
  10.1103/PhysRevLett.120.211801}{\emph{Phys. Rev. Lett.} {\bf 120} (2018)
  211801}, [\href{http://arxiv.org/abs/1804.04912}{{\tt 1804.04912}}].

\bibitem{Li:2017dbe}
{\scshape Super-Kamiokande} collaboration, Z.~Li et~al., \emph{{Measurement of
  the tau neutrino cross section in atmospheric neutrino oscillations with
  Super-Kamiokande}},
  \href{http://dx.doi.org/10.1103/PhysRevD.98.052006}{\emph{Phys. Rev.} {\bf
  D98} (2018) 052006}, [\href{http://arxiv.org/abs/1711.09436}{{\tt
  1711.09436}}].

\bibitem{Aartsen:2019tjl}
{\scshape IceCube} collaboration, M.~G. Aartsen et~al., \emph{{Measurement of
  Atmospheric Tau Neutrino Appearance with IceCube DeepCore}},
  \href{http://dx.doi.org/10.1103/PhysRevD.99.032007}{\emph{Phys. Rev.} {\bf
  D99} (2019) 032007}, [\href{http://arxiv.org/abs/1901.05366}{{\tt
  1901.05366}}].

\bibitem{Aartsen:2017kpd}
{\scshape IceCube} collaboration, M.~G. Aartsen et~al., \emph{{Measurement of
  the multi-TeV neutrino cross section with IceCube using Earth absorption}},
  \href{http://dx.doi.org/10.1038/nature24459}{\emph{Nature} {\bf 551} (2017)
  596--600}, [\href{http://arxiv.org/abs/1711.08119}{{\tt 1711.08119}}].

\bibitem{Bustamante:2017xuy}
M.~Bustamante and A.~Connolly, \emph{{Extracting the Energy-Dependent
  Neutrino-Nucleon Cross Section above 10 TeV Using IceCube Showers}},
  \href{http://dx.doi.org/10.1103/PhysRevLett.122.041101}{\emph{Phys. Rev.
  Lett.} {\bf 122} (2019) 041101}, [\href{http://arxiv.org/abs/1711.11043}{{\tt
  1711.11043}}].

\bibitem{Nason:1987xz}
P.~Nason, S.~Dawson and R.~K. Ellis, \emph{{The Total Cross-Section for the
  Production of Heavy Quarks in Hadronic Collisions}},
  \href{http://dx.doi.org/10.1016/0550-3213(88)90422-1}{\emph{Nucl. Phys.} {\bf
  B303} (1988) 607--633}.

\bibitem{Nason:1989zy}
P.~Nason, S.~Dawson and R.~K. Ellis, \emph{{The One Particle Inclusive
  Differential Cross-Section for Heavy Quark Production in Hadronic
  Collisions}}, \href{http://dx.doi.org/10.1016/0550-3213(90)90180-L,
  10.1016/0550-3213(89)90286-1}{\emph{Nucl. Phys.} {\bf B327} (1989) 49--92}.

\bibitem{Mangano:1991jk}
M.~L. Mangano, P.~Nason and G.~Ridolfi, \emph{{Heavy quark correlations in
  hadron collisions at next-to-leading order}},
  \href{http://dx.doi.org/10.1016/0550-3213(92)90435-E}{\emph{Nucl. Phys.} {\bf
  B373} (1992) 295--345}.

\bibitem{Aaij:2015bpa}
{\scshape LHCb} collaboration, R.~Aaij et~al., \emph{{Measurements of prompt
  charm production cross-sections in $pp$ collisions at $ \sqrt{s}=13 $ TeV}},
  \href{http://dx.doi.org/10.1007/JHEP03(2016)159, 10.1007/JHEP09(2016)013,
  10.1007/JHEP05(2017)074}{\emph{JHEP} {\bf 03} (2016) 159},
  [\href{http://arxiv.org/abs/1510.01707}{{\tt 1510.01707}}].

\bibitem{Koers:2006dd}
H.~B.~J. Koers, A.~Pe'er and R.~A. M.~J. Wijers, \emph{{Parameterization of the
  energy and rapidity distributions of secondary pions and kaons produced in
  energetic proton-proton collisions}}, {\emph{Submitted to: Phys. Rev. D}
  (2006) }, [\href{http://arxiv.org/abs/hep-ph/0611219}{{\tt hep-ph/0611219}}].

\bibitem{Bruning:2004ej}
O.~S. Bruning, P.~Collier, P.~Lebrun, S.~Myers, R.~Ostojic, J.~Poole et~al.,
  \emph{{LHC Design Report Vol.1: The LHC Main Ring}}, {\emph{CERN-2004-003-V1}
  (2004) }.

\bibitem{Apanasevich:1998ki}
L.~Apanasevich et~al., \emph{{$k_{T}$ effects in direct photon production}},
  \href{http://dx.doi.org/10.1103/PhysRevD.59.074007}{\emph{Phys. Rev.} {\bf
  D59} (1999) 074007}, [\href{http://arxiv.org/abs/hep-ph/9808467}{{\tt
  hep-ph/9808467}}].

\bibitem{Mangano:1998oia}
M.~L. Mangano, \emph{{Two lectures on heavy quark production in hadronic
  collisions}},
  \href{http://dx.doi.org/10.3254/978-1-61499-222-6-95}{\emph{Proc.\ Int.\
  Sch.\ Phys.\ Fermi} {\bf 137} (1998) 95--137},
  [\href{http://arxiv.org/abs/hep-ph/9711337}{{\tt hep-ph/9711337}}].

\bibitem{Miu:1998ju}
G.~Miu and T.~Sjostrand, \emph{{$W$ production in an improved parton shower
  approach}},
  \href{http://dx.doi.org/10.1016/S0370-2693(99)00068-4}{\emph{Phys. Lett.}
  {\bf B449} (1999) 313--320}, [\href{http://arxiv.org/abs/hep-ph/9812455}{{\tt
  hep-ph/9812455}}].

\bibitem{Balazs:2000sz}
C.~Balazs, J.~Huston and I.~Puljak, \emph{{Higgs production: A Comparison of
  parton showers and resummation}},
  \href{http://dx.doi.org/10.1103/PhysRevD.63.014021}{\emph{Phys.\ Rev.\ D}
  {\bf 63} (2001) 014021}, [\href{http://arxiv.org/abs/hep-ph/0002032}{{\tt
  hep-ph/0002032}}].

\bibitem{Skands:2010ak}
P.~Z. Skands, \emph{{Tuning Monte Carlo Generators: The Perugia Tunes}},
  \href{http://dx.doi.org/10.1103/PhysRevD.82.074018}{\emph{Phys. Rev.} {\bf
  D82} (2010) 074018}, [\href{http://arxiv.org/abs/1005.3457}{{\tt
  1005.3457}}].

\bibitem{Catani:1990eg}
S.~Catani, M.~Ciafaloni and F.~Hautmann, \emph{{High-energy factorization and
  small x heavy flavor production}},
  \href{http://dx.doi.org/10.1016/0550-3213(91)90055-3}{\emph{Nucl. Phys.} {\bf
  B366} (1991) 135--188}.

\bibitem{Collins:1991ty}
J.~C. Collins and R.~K. Ellis, \emph{{Heavy quark production in very
  high-energy hadron collisions}},
  \href{http://dx.doi.org/10.1016/0550-3213(91)90288-9}{\emph{Nucl. Phys.} {\bf
  B360} (1991) 3--30}.

\bibitem{Peterson:1982ak}
C.~Peterson, D.~Schlatter, I.~Schmitt and P.~M. Zerwas, \emph{{Scaling
  Violations in Inclusive $e^+ e^-$ Annihilation Spectra}},
  \href{http://dx.doi.org/10.1103/PhysRevD.27.105}{\emph{Phys. Rev.} {\bf D27}
  (1983) 105}.

\bibitem{Lisovyi:2015uqa}
M.~Lisovyi, A.~Verbytskyi and O.~Zenaiev, \emph{{Combined analysis of
  charm-quark fragmentation-fraction measurements}},
  \href{http://dx.doi.org/10.1140/epjc/s10052-016-4246-y}{\emph{Eur. Phys. J.}
  {\bf C76} (2016) 397}, [\href{http://arxiv.org/abs/1509.01061}{{\tt
  1509.01061}}].

\bibitem{pdg:2019}
{\scshape Particle Data Group} collaboration, M.~Tanabashi et~al.,
  \emph{{Review of Particle Physics}},
  \href{http://dx.doi.org/10.1103/PhysRevD.98.030001}{\emph{Phys. Rev.} {\bf
  D98} (2018) 030001}.

\bibitem{Aaij:2019pqz}
{\scshape LHCb} collaboration, R.~Aaij et~al., \emph{{Measurement of $b$-hadron
  fractions in 13 TeV $pp$ collisions}},
  \href{http://arxiv.org/abs/1902.06794}{{\tt 1902.06794}}.

\bibitem{Kovarik:2015cma}
K.~Kovarik et~al., \emph{{nCTEQ15 - Global analysis of nuclear parton
  distributions with uncertainties in the CTEQ framework}},
  \href{http://dx.doi.org/10.1103/PhysRevD.93.085037}{\emph{Phys. Rev.} {\bf
  D93} (2016) 085037}, [\href{http://arxiv.org/abs/1509.00792}{{\tt
  1509.00792}}].

\bibitem{Cacciari:2012ny}
M.~Cacciari, S.~Frixione, N.~Houdeau, M.~L. Mangano, P.~Nason and G.~Ridolfi,
  \emph{{Theoretical predictions for charm and bottom production at the LHC}},
  \href{http://dx.doi.org/10.1007/JHEP10(2012)137}{\emph{JHEP} {\bf 10} (2012)
  137}, [\href{http://arxiv.org/abs/1205.6344}{{\tt 1205.6344}}].

\bibitem{Benzke:2017yjn}
M.~Benzke, M.~V. Garzelli, B.~Kniehl, G.~Kramer, S.~Moch and G.~Sigl,
  \emph{{Prompt neutrinos from atmospheric charm in the general-mass
  variable-flavor-number scheme}},
  \href{http://dx.doi.org/10.1007/JHEP12(2017)021}{\emph{JHEP} {\bf 12} (2017)
  021}, [\href{http://arxiv.org/abs/1705.10386}{{\tt 1705.10386}}].

\bibitem{Zenaiev:2019ktw}
{\scshape PROSA} collaboration, O.~Zenaiev, M.~V. Garzelli, K.~Lipka, S.~O.
  Moch, A.~Cooper-Sarkar, F.~Olness et~al., \emph{{Improved constraints on
  parton distributions using LHCb, ALICE and HERA heavy-flavour measurements
  and implications for the predictions for prompt atmospheric-neutrino
  fluxes}},  \href{http://arxiv.org/abs/1911.13164}{{\tt 1911.13164}}.

\bibitem{Aaij:2017qml}
{\scshape LHCb} collaboration, R.~Aaij et~al., \emph{{Measurement of the
  $B^{\pm}$ production cross-section in pp collisions at $\sqrt{s} =$ 7 and 13
  TeV}}, \href{http://dx.doi.org/10.1007/JHEP12(2017)026}{\emph{JHEP} {\bf 12}
  (2017) 026}, [\href{http://arxiv.org/abs/1710.04921}{{\tt 1710.04921}}].

\bibitem{underway}
W.~Bai, M.~Diwan, M.~V. Garzelli, Y.~S. Jeong and M.~H. Reno{\emph{\ in
  preparation} }.

\bibitem{DAlesio:2017rzj}
U.~D'Alesio, F.~Murgia, C.~Pisano and P.~Taels, \emph{{Probing the gluon Sivers
  function in $p^\uparrow p\to J/\psi\,X$ and $p^\uparrow p \to D\,X$}},
  \href{http://dx.doi.org/10.1103/PhysRevD.96.036011}{\emph{Phys. Rev.} {\bf
  D96} (2017) 036011}, [\href{http://arxiv.org/abs/1705.04169}{{\tt
  1705.04169}}].

\bibitem{Aaij:2016bqq}
{\scshape LHCb} collaboration, R.~Aaij et~al., \emph{{Measurement of the
  J/$\psi$ pair production cross-section in pp collisions at $ \sqrt{s}=13 $
  TeV}}, \href{http://dx.doi.org/10.1007/JHEP06(2017)047,
  10.1007/JHEP10(2017)068}{\emph{JHEP} {\bf 06} (2017) 047},
  [\href{http://arxiv.org/abs/1612.07451}{{\tt 1612.07451}}].

\bibitem{Lansberg:2017dzg}
J.-P. Lansberg, C.~Pisano, F.~Scarpa and M.~Schlegel, \emph{{Pinning down the
  linearly-polarised gluons inside unpolarised protons using quarkonium-pair
  production at the LHC}},
  \href{http://dx.doi.org/10.1016/j.physletb.2018.08.004,
  10.1016/j.physletb.2019.01.057}{\emph{Phys. Lett.} {\bf B784} (2018)
  217--222}, [\href{http://arxiv.org/abs/1710.01684}{{\tt 1710.01684}}].

\bibitem{Bacchetta:2018lna}
A.~Bacchetta, G.~Bozzi, M.~Radici, M.~Ritzmann and A.~Signori, \emph{{Effect of
  Flavor-Dependent Partonic Transverse Momentum on the Determination of the $W$
  Boson Mass in Hadronic Collisions}},
  \href{http://dx.doi.org/10.1016/j.physletb.2018.11.002}{\emph{Phys. Lett.}
  {\bf B788} (2019) 542--545}, [\href{http://arxiv.org/abs/1807.02101}{{\tt
  1807.02101}}].

\bibitem{Frixione:2007vw}
S.~Frixione, P.~Nason and C.~Oleari, \emph{{Matching NLO QCD computations with
  Parton Shower simulations: the POWHEG method}},
  \href{http://dx.doi.org/10.1088/1126-6708/2007/11/070}{\emph{JHEP} {\bf 11}
  (2007) 070}, [\href{http://arxiv.org/abs/0709.2092}{{\tt 0709.2092}}].

\bibitem{Frixione:2007nw}
S.~Frixione, P.~Nason and G.~Ridolfi, \emph{{A Positive-weight
  next-to-leading-order Monte Carlo for heavy flavour hadroproduction}},
  \href{http://dx.doi.org/10.1088/1126-6708/2007/09/126}{\emph{JHEP} {\bf 09}
  (2007) 126}, [\href{http://arxiv.org/abs/0707.3088}{{\tt 0707.3088}}].

\bibitem{Barr:1988rb}
S.~M. Barr, T.~K. Gaisser, P.~Lipari and S.~Tilav, \emph{{Ratio of $\nu_e /
  \nu_\mu$ in Atmospheric Neutrinos}},
  \href{http://dx.doi.org/10.1016/0370-2693(88)90468-6}{\emph{Phys. Lett.} {\bf
  B214} (1988) 147--150}.

\bibitem{Pasquali:1998xf}
L.~Pasquali and M.~H. Reno, \emph{{Tau-neutrino fluxes from atmospheric
  charm}}, \href{http://dx.doi.org/10.1103/PhysRevD.59.093003}{\emph{Phys.
  Rev.} {\bf D59} (1999) 093003},
  [\href{http://arxiv.org/abs/hep-ph/9811268}{{\tt hep-ph/9811268}}].

\bibitem{Bhattacharya:2016jce}
A.~Bhattacharya, R.~Enberg, Y.~S. Jeong, C.~S. Kim, M.~H. Reno, I.~Sarcevic
  et~al., \emph{{Prompt atmospheric neutrino fluxes: perturbative QCD models
  and nuclear effects}},
  \href{http://dx.doi.org/10.1007/JHEP11(2016)167}{\emph{JHEP} {\bf 11} (2016)
  167}, [\href{http://arxiv.org/abs/1607.00193}{{\tt 1607.00193}}].

\bibitem{Bai:2018xum}
W.~Bai and M.~H. Reno, \emph{{Prompt neutrinos and intrinsic charm at SHiP}},
  \href{http://dx.doi.org/10.1007/JHEP02(2019)077}{\emph{JHEP} {\bf 02} (2019)
  077}, [\href{http://arxiv.org/abs/1807.02746}{{\tt 1807.02746}}].

\bibitem{Kretzer:2002fr}
S.~Kretzer and M.~H. Reno, \emph{{Tau neutrino deep inelastic charged current
  interactions}},
  \href{http://dx.doi.org/10.1103/PhysRevD.66.113007}{\emph{Phys. Rev.} {\bf
  D66} (2002) 113007}, [\href{http://arxiv.org/abs/hep-ph/0208187}{{\tt
  hep-ph/0208187}}].

\bibitem{Kretzer:2003iu}
S.~Kretzer and M.~H. Reno, \emph{{Target mass corrections to electroweak
  structure functions and perturbative neutrino cross-sections}},
  \href{http://dx.doi.org/10.1103/PhysRevD.69.034002}{\emph{Phys. Rev.} {\bf
  D69} (2004) 034002}, [\href{http://arxiv.org/abs/hep-ph/0307023}{{\tt
  hep-ph/0307023}}].

\bibitem{Jeong:2010za}
Y.~S. Jeong and M.~H. Reno, \emph{{Quark mass effects in high energy neutrino
  nucleon scattering}},
  \href{http://dx.doi.org/10.1103/PhysRevD.81.114012}{\emph{Phys. Rev.} {\bf
  D81} (2010) 114012}, [\href{http://arxiv.org/abs/1001.4175}{{\tt
  1001.4175}}].

\bibitem{Jeong:2010nt}
Y.~S. Jeong and M.~H. Reno, \emph{{Tau neutrino and antineutrino cross
  sections}}, \href{http://dx.doi.org/10.1103/PhysRevD.82.033010}{\emph{Phys.
  Rev.} {\bf D82} (2010) 033010}, [\href{http://arxiv.org/abs/1007.1966}{{\tt
  1007.1966}}].

\bibitem{Reno:2006hj}
M.~H. Reno, \emph{{Electromagnetic structure functions and neutrino nucleon
  scattering}}, \href{http://dx.doi.org/10.1103/PhysRevD.74.033001}{\emph{Phys.
  Rev.} {\bf D74} (2006) 033001},
  [\href{http://arxiv.org/abs/hep-ph/0605295}{{\tt hep-ph/0605295}}].

\bibitem{Lipari:1994pz}
P.~Lipari, M.~Lusignoli and F.~Sartogo, \emph{{The Neutrino cross-section and
  upward going muons}},
  \href{http://dx.doi.org/10.1103/PhysRevLett.74.4384}{\emph{Phys. Rev. Lett.}
  {\bf 74} (1995) 4384--4387}, [\href{http://arxiv.org/abs/hep-ph/9411341}{{\tt
  hep-ph/9411341}}].

\bibitem{Kretzer:2004wk}
S.~Kretzer and M.~H. Reno, \emph{{sigma DIS (nu N), NLO perturbative QCD and
  O(1 GeV) mass corrections}},
  \href{http://dx.doi.org/10.1016/j.nuclphysbps.2004.11.237}{\emph{Nucl. Phys.
  Proc. Suppl.} {\bf 139} (2005) 134--139},
  [\href{http://arxiv.org/abs/hep-ph/0410184}{{\tt hep-ph/0410184}}].

\bibitem{crmc}
``Cosmic ray monte carlo package.''
  \url{https://web.ikp.kit.edu/rulrich/crmc.html}.

\bibitem{Pierog:2013ria}
T.~Pierog, I.~Karpenko, J.~M. Katzy, E.~Yatsenko and K.~Werner, \emph{{EPOS
  LHC: Test of collective hadronization with data measured at the CERN Large
  Hadron Collider}},
  \href{http://dx.doi.org/10.1103/PhysRevC.92.034906}{\emph{Phys. Rev.} {\bf
  C92} (2015) 034906}, [\href{http://arxiv.org/abs/1306.0121}{{\tt
  1306.0121}}].

\bibitem{Ostapchenko:2010vb}
S.~Ostapchenko, \emph{{Monte Carlo treatment of hadronic interactions in
  enhanced Pomeron scheme: I. QGSJET-II model}},
  \href{http://dx.doi.org/10.1103/PhysRevD.83.014018}{\emph{Phys. Rev.} {\bf
  D83} (2011) 014018}, [\href{http://arxiv.org/abs/1010.1869}{{\tt
  1010.1869}}].

\bibitem{Engel:2019dsg}
R.~Engel, A.~Fedynitch, T.~K. Gaisser, F.~Riehn and T.~Stanev, \emph{{The
  hadronic interaction model Sibyll 2.3c and extensive air showers}},
  \href{http://arxiv.org/abs/1912.03300}{{\tt 1912.03300}}.

\bibitem{Fedynitch:2018cbl}
A.~Fedynitch, F.~Riehn, R.~Engel, T.~K. Gaisser and T.~Stanev, \emph{{Hadronic
  interaction model sibyll 2.3c and inclusive lepton fluxes}},
  \href{http://dx.doi.org/10.1103/PhysRevD.100.103018}{\emph{Phys. Rev.} {\bf
  D100} (2019) 103018}, [\href{http://arxiv.org/abs/1806.04140}{{\tt
  1806.04140}}].

\bibitem{Riehn:2017mfm}
F.~Riehn, H.~P. Dembinski, R.~Engel, A.~Fedynitch, T.~K. Gaisser and T.~Stanev,
  \emph{{The hadronic interaction model SIBYLL 2.3c and Feynman scaling}},
  \href{http://dx.doi.org/10.22323/1.301.0301}{\emph{PoS} {\bf ICRC2017} (2018)
  301}, [\href{http://arxiv.org/abs/1709.07227}{{\tt 1709.07227}}].

\bibitem{Cirigliano:2001mk}
V.~Cirigliano, M.~Knecht, H.~Neufeld, H.~Rupertsberger and P.~Talavera,
  \emph{{Radiative corrections to K(l3) decays}},
  \href{http://dx.doi.org/10.1007/s100520100825}{\emph{Eur. Phys. J.} {\bf C23}
  (2002) 121--133}, [\href{http://arxiv.org/abs/hep-ph/0110153}{{\tt
  hep-ph/0110153}}].

\bibitem{Adriani:2015iwv}
{\scshape LHCf} collaboration, O.~Adriani et~al., \emph{{Measurements of
  longitudinal and transverse momentum distributions for neutral pions in the
  forward-rapidity region with the LHCf detector}},
  \href{http://dx.doi.org/10.1103/PhysRevD.94.032007}{\emph{Phys. Rev.} {\bf
  D94} (2016) 032007}, [\href{http://arxiv.org/abs/1507.08764}{{\tt
  1507.08764}}].

\bibitem{Antchev:2014lez}
{\scshape TOTEM} collaboration, G.~Antchev et~al., \emph{{Measurement of the
  forward charged particle pseudorapidity density in pp collisions at $\sqrt{s}
  = 8$ TeV using a displaced interaction point}},
  \href{http://dx.doi.org/10.1140/epjc/s10052-015-3343-7}{\emph{Eur. Phys. J.}
  {\bf C75} (2015) 126}, [\href{http://arxiv.org/abs/1411.4963}{{\tt
  1411.4963}}].

\bibitem{Sirunyan:2017nsj}
{\scshape CMS} collaboration, A.~M. Sirunyan et~al., \emph{{Measurement of the
  inclusive energy spectrum in the very forward direction in proton-proton
  collisions at $ \sqrt{s}=13 $ TeV}},
  \href{http://dx.doi.org/10.1007/JHEP08(2017)046}{\emph{JHEP} {\bf 08} (2017)
  046}, [\href{http://arxiv.org/abs/1701.08695}{{\tt 1701.08695}}].

\bibitem{Esteban:2018azc}
I.~Esteban, M.~C. Gonzalez-Garcia, A.~Hernandez-Cabezudo, M.~Maltoni and
  T.~Schwetz, \emph{{Global analysis of three-flavour neutrino oscillations:
  synergies and tensions in the determination of $\theta_{23}$, $\delta_{CP}$,
  and the mass ordering}},
  \href{http://dx.doi.org/10.1007/JHEP01(2019)106}{\emph{JHEP} {\bf 01} (2019)
  106}, [\href{http://arxiv.org/abs/1811.05487}{{\tt 1811.05487}}].

\bibitem{Giunti:2019aiy}
C.~Giunti and T.~Lasserre, \emph{{eV-scale Sterile Neutrinos}},
  \href{http://dx.doi.org/10.1146/annurev-nucl-101918-023755}{\emph{Ann. Rev.
  Nucl. Part. Sci.} {\bf 69} (2019) 163--190},
  [\href{http://arxiv.org/abs/1901.08330}{{\tt 1901.08330}}].

\bibitem{Jones:2019nix}
{\scshape IceCube} collaboration, B.~J.~P. Jones, \emph{{IceCube Sterile
  Neutrino Searches}},
  \href{http://dx.doi.org/10.1051/epjconf/201920704005}{\emph{EPJ Web Conf.}
  {\bf 207} (2019) 04005}, [\href{http://arxiv.org/abs/1902.06185}{{\tt
  1902.06185}}].

\bibitem{Blennow:2018hto}
M.~Blennow, E.~Fernandez-Martinez, J.~Gehrlein, J.~Hernandez-Garcia and
  J.~Salvado, \emph{{IceCube bounds on sterile neutrinos above 10 eV}},
  \href{http://dx.doi.org/10.1140/epjc/s10052-018-6282-2}{\emph{Eur. Phys. J.}
  {\bf C78} (2018) 807}, [\href{http://arxiv.org/abs/1803.02362}{{\tt
  1803.02362}}].

\bibitem{Ohlsson:1999xb}
T.~Ohlsson and H.~Snellman, \emph{{Three flavor neutrino oscillations in
  matter}}, \href{http://dx.doi.org/10.1063/1.533270}{\emph{J. Math. Phys.}
  {\bf 41} (2000) 2768--2788}, [\href{http://arxiv.org/abs/hep-ph/9910546}{{\tt
  hep-ph/9910546}}].

\bibitem{Li:2018ezt}
W.~Li, J.~Ling, F.~Xu and B.~Yue, \emph{{Matter Effect of Light Sterile
  Neutrino: An Exact Analytical Approach}},
  \href{http://dx.doi.org/10.1007/JHEP10(2018)021}{\emph{JHEP} {\bf 10} (2018)
  021}, [\href{http://arxiv.org/abs/1808.03985}{{\tt 1808.03985}}].

\bibitem{Astier:2001yj}
{\scshape NOMAD} collaboration, P.~Astier et~al., \emph{{Final NOMAD results on
  muon-neutrino to tau-neutrino and electron-neutrino to tau-neutrino
  oscillations including a new search for tau-neutrino appearance using
  hadronic tau decays}},
  \href{http://dx.doi.org/10.1016/S0550-3213(01)00339-X}{\emph{Nucl. Phys.}
  {\bf B611} (2001) 3--39}, [\href{http://arxiv.org/abs/hep-ex/0106102}{{\tt
  hep-ex/0106102}}].

\bibitem{Astier:2003gs}
{\scshape NOMAD} collaboration, P.~Astier et~al., \emph{{Search for $\nu_\mu
  \to \nu_e$ oscillations in the NOMAD experiment}},
  \href{http://dx.doi.org/10.1016/j.physletb.2003.07.029}{\emph{Phys. Lett.}
  {\bf B570} (2003) 19--31}, [\href{http://arxiv.org/abs/hep-ex/0306037}{{\tt
  hep-ex/0306037}}].

\bibitem{Dentler:2018sju}
M.~Dentler, A.~Hern\'andez-Cabezudo, J.~Kopp, P.~A.~N. Machado, M.~Maltoni,
  I.~Martinez-Soler et~al., \emph{{Updated Global Analysis of Neutrino
  Oscillations in the Presence of eV-Scale Sterile Neutrinos}},
  \href{http://dx.doi.org/10.1007/JHEP08(2018)010}{\emph{JHEP} {\bf 08} (2018)
  010}, [\href{http://arxiv.org/abs/1803.10661}{{\tt 1803.10661}}].

\bibitem{Belesev:2012hx}
A.~I. Belesev, A.~I. Berlev, E.~V. Geraskin, A.~A. Golubev, N.~A. Likhovid,
  A.~A. Nozik et~al., \emph{{An upper limit on additional neutrino mass
  eigenstate in 2 to 100 eV region from 'Troitsk nu-mass' data}},
  \href{http://dx.doi.org/10.1134/S0021364013020033}{\emph{JETP Lett.} {\bf 97}
  (2013) 67--69}, [\href{http://arxiv.org/abs/1211.7193}{{\tt 1211.7193}}].

\bibitem{Belesev:2013cba}
A.~I. Belesev, A.~I. Berlev, E.~V. Geraskin, A.~A. Golubev, N.~A. Likhovid,
  A.~A. Nozik et~al., \emph{{The search for an additional neutrino mass
  eigenstate in the 2–100 eV region from ‘Troitsk nu-mass’ data: a
  detailed analysis}},
  \href{http://dx.doi.org/10.1088/0954-3899/41/1/015001}{\emph{J. Phys.} {\bf
  G41} (2014) 015001}, [\href{http://arxiv.org/abs/1307.5687}{{\tt
  1307.5687}}].

\bibitem{Aoki:2019jry}
{\scshape DsTau} collaboration, S.~Aoki et~al., \emph{{DsTau: Study of tau
  neutrino production with 400 GeV protons from the CERN-SPS}},
  \href{http://arxiv.org/abs/1906.03487}{{\tt 1906.03487}}.

\bibitem{Bacchetta:2018ivt}
A.~Bacchetta, D.~Boer, C.~Pisano and P.~Taels, \emph{{Gluon TMDs and NRQCD
  matrix elements in $J/\psi$ production at an EIC}},
  \href{http://dx.doi.org/10.1140/epjc/s10052-020-7620-8}{\emph{Eur. Phys. J.}
  {\bf C80} (2020) 72}, [\href{http://arxiv.org/abs/1809.02056}{{\tt
  1809.02056}}].

\bibitem{DAlesio:2019qpk}
U.~D'Alesio, F.~Murgia, C.~Pisano and P.~Taels, \emph{{Azimuthal asymmetries in
  semi-inclusive $J/\psi\,+\,\mathrm{jet}$ production at an EIC}},
  \href{http://dx.doi.org/10.1103/PhysRevD.100.094016}{\emph{Phys. Rev.} {\bf
  D100} (2019) 094016}, [\href{http://arxiv.org/abs/1908.00446}{{\tt
  1908.00446}}].

\bibitem{Bhattacharya:2015jpa}
A.~Bhattacharya, R.~Enberg, M.~H. Reno, I.~Sarcevic and A.~Stasto,
  \emph{{Perturbative charm production and the prompt atmospheric neutrino flux
  in light of RHIC and LHC}},
  \href{http://dx.doi.org/10.1007/JHEP06(2015)110}{\emph{JHEP} {\bf 06} (2015)
  110}, [\href{http://arxiv.org/abs/1502.01076}{{\tt 1502.01076}}].

\bibitem{Gauld:2015kvh}
R.~Gauld, J.~Rojo, L.~Rottoli, S.~Sarkar and J.~Talbert, \emph{{The prompt
  atmospheric neutrino flux in the light of LHCb}},
  \href{http://dx.doi.org/10.1007/JHEP02(2016)130}{\emph{JHEP} {\bf 02} (2016)
  130}, [\href{http://arxiv.org/abs/1511.06346}{{\tt 1511.06346}}].

\bibitem{Garzelli:2016xmx}
{\scshape PROSA} collaboration, M.~V. Garzelli, S.~Moch, O.~Zenaiev,
  A.~Cooper-Sarkar, A.~Geiser, K.~Lipka et~al., \emph{{Prompt neutrino fluxes
  in the atmosphere with PROSA parton distribution functions}},
  \href{http://dx.doi.org/10.1007/JHEP05(2017)004}{\emph{JHEP} {\bf 05} (2017)
  004}, [\href{http://arxiv.org/abs/1611.03815}{{\tt 1611.03815}}].

\bibitem{Barr:1989}
G.~Barr, T.~K. Gaisser and T.~Stanev, \emph{Flux of atmospheric neutrinos},
  \href{http://dx.doi.org/10.1103/PhysRevD.39.3532}{\emph{Phys. Rev. D} {\bf
  39} (Jun, 1989) 3532--3534}.

\bibitem{Gaisser:1990vg}
T.~Gaisser, \emph{{Cosmic rays and particle physics}}.
\newblock Cambridge University Press, 1990.

\end{thebibliography}\endgroup
